\pdfoutput=1
\documentclass[11pt]{article}

\usepackage[utf8]{inputenc}
\usepackage[T1]{fontenc}
\usepackage{newtxtext}
\usepackage[margin=1in]{geometry}
\usepackage{amsmath,amssymb,amsthm,mathtools}
\usepackage{newtxmath}
\usepackage{bm}
\usepackage{setspace}
\usepackage[longnamesfirst,sort&compress]{natbib}
\usepackage{booktabs}
\usepackage{enumitem}
\usepackage{titlesec}
\usepackage{titling}
\usepackage[dvipsnames]{xcolor}
\usepackage{etoolbox}
\usepackage{graphicx}
\usepackage{subcaption}
\usepackage{adjustbox}
\usepackage{rotating}
\usepackage[hidelinks,breaklinks=true]{hyperref}
\hypersetup{colorlinks=true,
	linkcolor=blue!55!black,
	citecolor=blue!55!black,
	urlcolor=blue!55!black}

\bibpunct{(}{)}{;}{a}{,}{,}

\titleformat{\section}
{\normalfont\Large\scshape\filcenter}{\thesection.}{0.7em}{}
\titlespacing*{\section}{0pt}{2.2ex plus 1ex minus .2ex}{1.4ex plus .2ex}

\titleformat{\subsection}
{\normalfont\large\itshape}{\thesubsection.}{0.6em}{}
\titlespacing*{\subsection}{0pt}{1.8ex plus .6ex minus .2ex}{1.0ex plus .2ex}

\titleformat{\subsubsection}
{\normalfont\normalsize\itshape}{\thesubsubsection.}{0.6em}{}
\pretitle{\centering\normalfont\LARGE\scshape}
\posttitle{\par\vskip 0.6em}

\preauthor{\centering\normalfont\normalsize
	\textsc{By}\vskip 0.3em}
\postauthor{\par\vskip 2em}

\predate{\centering\normalfont\small}
\postdate{\par}

\theoremstyle{plain}
\newtheorem{proposition}{\scshape Proposition}

\newtheorem{lemma}{\scshape Lemma}
\newtheorem{corollary}{\scshape Corollary}

\theoremstyle{definition}
\newtheorem{assumption}{\scshape Assumption}

\theoremstyle{remark}
\newtheorem{remark}{\scshape Remark}

\usepackage{algorithm}
\usepackage{algpseudocode}
\usepackage{tabularx}

\makeatletter
\renewenvironment{proof}[1][\textsc{Proof}]{%
	\par\pushQED{\qed}\normalfont\topsep6\p@\@plus6\p@\relax
	\trivlist\item[\hskip\labelsep #1:]\ignorespaces}%
{\popQED\endtrivlist\@endpefalse}
\makeatother

\newcommand{\E}{\mathbb{E}}
\newcommand{\Var}{\operatorname{Var}}
\newcommand{\Cov}{\operatorname{Cov}}

\newcommand{\sig}{s_{ig}}
\newcommand{\sir}{s_{ir}}
\newcommand{\uisr}{u_{i}^{rg}}

\DeclareMathOperator{\rank}{rank}
\DeclareMathOperator{\col}{col}

\title{Hedging out the HEDGE Bias:\\
	Identifying Dynamic Partial-Equilibrium Effects with Panel LP-IV}

\author{%
	\begin{tabular}{c@{\hspace{1em}}c}
		Endong Wang\footnotemark[1] & Ren Zhang\footnotemark[2]\\[2pt]
		{\small \textit{Dept.\ of Economics, University of Mannheim}} &
		{\small \textit{McCoy College of Business, Texas State Univ.}}
	\end{tabular}%
}

\date{This version: November 9, 2026}

\usepackage{placeins}

\newcommand{\R}{\mathbb R}
\newcommand{\ve}{\operatorname{vec}}
\newcommand{\vh}{\operatorname{vech}}
\newcommand{\LRV}{\operatorname{LRV}}
\begin{document}
	\renewcommand{\thefootnote}{\fnsymbol{footnote}}
	\maketitle

	\footnotetext[1]{Department of Economics, University of Mannheim, L7, 3--5, Mannheim, Germany, 68161.
		TEL: +49 621-181-1879; Email: \href{mailto:endong.wang@uni-mannheim.de}{endong.wang@uni-mannheim.de}.}

	\footnotetext[2]{Department of Finance and Economics, McCoy College of Business,
		Texas State University, 601 University Drive, San Marcos, TX 78666.
		Email: \href{mailto:r_z79@txstate.edu}{\nolinkurl{r_z79@txstate.edu}}.}

	\renewcommand{\thefootnote}{\arabic{footnote}}
	\setcounter{footnote}{0}

	\begin{center}
		\begin{minipage}{0.88\textwidth}
			\small
			\singlespacing
			\noindent

Cross-sectional designs are widely used to estimate dynamic
partial-equilibrium effects of an aggregate treatment. We show analytically
that these estimates are biased when exposures to the treatment and to
aggregate confounders are correlated or the confounders feed back into the
treatment. This paper develops the heterogeneous-exposure local-projection
instrumental variables (HELP-IV) framework, which compares units while holding
the confounder path fixed by instrumenting the treatment and this path with
exposure-weighted shock proxies. We prove that HELP-IV identifies the
partial-equilibrium effect. We further extend the framework, with the
corresponding inference, to multiple exposure margins and confounders,
forward-looking outcomes, and cumulative multipliers. Finally, we apply HELP-IV
to state fiscal multipliers and regional stock-wealth effects, where it yields
partial-equilibrium effects that are smaller than conventional estimates and
deliver different implications when ported to macroeconomic models.

			\medskip
	\noindent\textsc{Keywords:} Cross-sectional identification, panel local
	projections, partial-equilibrium effects, portable elasticities, shift-share
	designs, two-way fixed effects.

			\noindent\textsc{JEL Classification:} C13, C26, C33, E52, E62.
		\end{minipage}
	\end{center}

	\vspace{1em}
	\newpage
	\section{Introduction}
	\label{sec:intro}

A growing body of literature relies on cross-sectional identification to recover
dynamic partial-equilibrium effects, which measure the direct responses of unit
outcomes to a treatment driven by an aggregate shock when the paths of the other
aggregates that move with the shock (the confounders) are held
fixed.\footnote{In macroeconomics, applications include
	\citet{huber2018,berajafusterhurstvavra2019,chodorowreich2019,guren2021,jeenas2026}.
	Outside macroeconomics, the same design is used to study trade shocks
	\citep{autordornhanson2013,acemogluetal2016} and, in finance and the broader
	business literature, the effects of common price, policy, or technology shocks
	on firms and markets that differ in their exposure.} Estimating these effects
is of particular importance because, as portable statistics in the sense of
\citet{nakamurasteinsson2018}, the estimates discipline and test structural
models. Specifically, these studies use panel data on
units that differ in their exposure to the shock and estimate two-way
fixed-effects instrumental-variables (TWFE-IV) local projections, which include
unit and time effects. The design presumes that the time effects control
for the influence of the confounders, which are common to all units, so that the
estimated relative responses recover the dynamic partial-equilibrium effects.

The time effects absorb only the part of the confounders' influence that is
common to all units. A confounder may therefore bias the TWFE-IV estimate when
the confounder responds to the treatment and units that are more exposed to the
treatment are also more or less sensitive to the confounder
\citep{chodorowreichnenovsimsek2021,guren2021,arkhangelskykorovkin2024}.
A recent paper by \citet{donaldson2026} terms this bias the
heterogeneous-exposure-to-dynamic-general-equilibrium (HEDGE) bias. The central
contribution of this paper is to characterize the HEDGE bias under general
aggregate dynamics and to develop a regression-based framework,
heterogeneous-exposure local-projection instrumental variables (HELP-IV), which
identifies the dynamic partial-equilibrium effects across a broad range of
environments. We summarize these results in turn.

We start by deriving an explicit formula for the dynamic partial-equilibrium
effect in a backward-looking environment and by characterizing in closed form the
HEDGE bias of the TWFE-IV estimate, which contains an exposure-correlation
component and an aggregate-feedback component. Intuitively, the effect is the
outcome response divided by the impact response of the treatment, in the
experiment that holds the confounder path fixed. The exposure-correlation
component arises when exposures to the treatment and to the confounder are
correlated and the treatment shock moves the confounder, so that the numerator of
the TWFE-IV estimate also reflects the influence of the confounder. The
aggregate-feedback component arises when the confounder shock also moves the
treatment (two-way general equilibrium), so that holding the confounder path
fixed changes the treatment path, and possibly its impact response in the
denominator, whereas the TWFE-IV estimate follows the unrestricted treatment
path. Thus, the TWFE-IV estimate is generally biased whenever either component
is present.

To hedge out the HEDGE bias, we develop for this environment the benchmark
HELP-IV design, which augments the TWFE-IV local projection with the path
controls and instruments boxed below. At each horizon $h$, HELP-IV regresses the
outcome $Y_{i,t+h}$ of unit $i$ on the treatment $G_t$ and on the confounder path
$R_t,\ldots,R_{t+h}$, each interacted with the exposure $\sig$ of the unit to the
treatment, together with unit and time effects,
\begin{align}
	Y_{i,t+h}
	&=
	b_h^{\mathrm{HELP}}\sig G_t
	+
	\boxed{\sum_{s=0}^{h}c_{h,s}\sig R_{t+s}}
	+
	\mu_{i,h}
	+
	\lambda_{t,h}
	+
	\xi_{i,t,h}.
	\label{eq:intro-help}
\end{align}
HELP-IV instruments these regressors with
$\sig\bigl(z_{g,t},\boxed{z_{r,t},\ldots,z_{r,t+h}}\bigr)$, where $z_{g,t}$ is a
proxy for the treatment shock and $z_{r,t+s}$ is a proxy for the confounder shock
dated $t+s$. We prove that, with valid instruments, $b_h^{\mathrm{HELP}}$ identifies the dynamic
partial-equilibrium effect. Intuitively, the regression identifies its $h+2$
coefficients through $h+2$ moment conditions, one for each instrument, and
identifying $b_h^{\mathrm{HELP}}$ is equivalent to taking a linear combination of these
conditions that eliminates the coefficients on the path controls. This
elimination is possible exactly when the corresponding combination of
instruments is uncorrelated with every value of the confounder path, so the
combination mimics a treatment shock in the experiment that holds the confounder
path fixed. This linear combination then expresses $b_h^{\mathrm{HELP}}$ as the response of the
outcome to the combined instruments divided by the response of the treatment to
them. The numerator and the denominator of this ratio therefore correspond to
those that define the dynamic partial-equilibrium effect, so HELP-IV removes both
components of the HEDGE bias simultaneously.

We then extend the benchmark HELP-IV design to four more general environments
that arise frequently in empirical applications.
First, when units are exposed to the treatment along multiple margins, we
identify a separate partial-equilibrium effect for each margin by interacting the
treatment, the path controls, and the instruments with each margin. Second, when
multiple confounders are present, we develop three designs, each suited to a
different class of applications. These designs, respectively, (i) instrument each
value on every confounder path with its own proxy, (ii) combine all proxies into one
instrument that leaves the confounder paths unchanged (composite HELP-IV), and (iii) combine the time-series identification of HELP-IV with the
cross-sectional targeting rule of \citet{manuel2026} to accommodate confounders
that lack proxies but do not move the treatment. Third, when outcomes also
depend on expected future aggregates, we build the composite instrument with
news shocks about future confounders and, when such news shocks are few,
identify an information-constrained effect.
Fourth, when the target is defined on unit-level treatments and outcomes that
may span multiple dates, many applications instrument the treatment with the
exposure-weighted aggregate treatment itself rather than its innovation. We show that this instrument
introduces an omitted-history bias in addition to the HEDGE bias, because it also
reflects shocks that predate the treatment shock. HELP-IV, which instruments with
shock proxies, removes both biases. Although the four environments
differ, the HELP-IV framework can be adapted to all of them because every
adapted design retains the common logic that a combination of instruments
mimics a treatment shock holding the relevant confounder paths fixed.

The HELP-IV method has the advantage that inference for these designs builds on
standard panel IV methods. With available proxies, we follow
\citet{almuzarasancibrian2025} and compute standard errors that are robust to the
correlation that common shocks induce across units and over time. We also adapt
this inference where our designs go beyond this setting. First, the instruments
may be estimated in a first step, as in composite HELP-IV or when shocks from a
structural vector autoregression (SVAR) replace unavailable proxies. We then
stack the first-step moments with the IV moment, so that the standard errors
also reflect the first-step estimation error. Second, under the targeting rule
of \citet{manuel2026}, the instruments are uncorrelated with exposure to the
confounders without proxies only in the population, so their sample correlation
adds variance across units. We characterize this variance as a
cross-sectional term that adds to the time-series variance and that can be
estimated when the exposures to these confounders are identified.

To assess the performance of HELP-IV in finite samples and its practical
importance, we conduct a set of Monte Carlo simulations and two empirical
applications. The simulations, which cover all our designs and environments, show
that HELP-IV recovers the target effect, with confidence intervals whose coverage
is generally close to the nominal level.
Our first application re-estimates the state-level fiscal multiplier of
\citet{nakamurasteinsson2014} under different assumptions about the environment.
When the interest rate is the only confounder, HELP-IV lowers the two-year
cumulative multiplier from the conventional TWFE-IV estimate of 1.45 to 1.00 in
the backward-looking environment and to 0.90 in the forward-looking one. When we
allow for multiple confounders through the targeting rule, the two-year
multiplier falls to 0.35, comparable to the estimates of \citet{manuel2026}. We
thus conclude that the TWFE-IV estimate is biased upward, whereas the
partial-equilibrium multiplier itself depends on the assumptions about the
environment and on which confounders are held fixed.
Our second application re-estimates the effect of stock wealth on local
employment and payroll in \citet{chodorowreichnenovsimsek2021}. They address four
aggregate confounders through a control-function approach that interacts
exposure with these confounders, and they report that their estimates are robust
to dropping these interactions. Applying composite HELP-IV, we find that the
two-year response of nontradable payroll falls from their estimate of 3.24 to
0.83 basis points. When we use this portable estimate in their structural model,
the implied marginal propensity to consume out of stock wealth and the aggregate
labor-market effects are substantially lower. In contrast to their robustness
claim, HELP-IV reveals that the confounders bias the TWFE-IV estimate upward,
even though this estimate moves only from 2.62 to 3.24 basis points when
\citet{chodorowreichnenovsimsek2021} add these interactions as controls.

\subsection{Related literature}
\label{sec:literature}

Our paper contributes mainly to the literature that estimates dynamic partial-equilibrium effects with panel local projections, which combine cross-sectional differences in exposure to aggregate shocks with two-way fixed effects \citep{nakamurasteinsson2014,ottonellowinberry2020,jeenas2026}. Much of this literature abstracts from aggregate confounders that move with the treatment and to which units respond differently. \citet{chodorowreichnenovsimsek2021} and \citet{guren2021} are notable exceptions. They recognize that time fixed effects do not absorb such confounders, and they confront this identification threat by interacting exposure with the suspected confounders. These studies, however, neither derive the resulting bias nor formally prove that their interactions correct for it. We contribute to this literature by deriving the HEDGE and omitted-history biases of the TWFE-IV estimators used in this literature in closed form at all horizons, and by developing a new regression-based framework that corrects for them.

Our paper also contributes to a recently developed literature that proposes corrections for the HEDGE bias by exploiting either cross-sectional or time-series variation. On the cross-sectional side, \citet{arkhangelskykorovkin2024} reweight units to balance their exposure to unobserved confounders, but only in a static setting. \citet{manuel2026} instead controls for an observed targeting rule, the characteristics that govern the assignment of exposure. Although he targets a different object, we show that his method also identifies our partial-equilibrium effect when the confounders do not feed back into the treatment. By contrast, our framework allows for two-way confounders and requires no targeting rule. When such a rule is available, however, HELP-IV can complement his method by controlling for the rule, which combines cross-sectional and time-series identification in a single regression.

On the time-series side, \citet{donaldson2026} develops a multi-step method that decomposes the two-way fixed-effects estimate into the partial-equilibrium effect and a bias term, which she estimates from a time-series policy counterfactual. Her application, however, interacts exposure with the aggregate change in spending rather than with a shock, so her estimate retains the omitted-history bias. HELP-IV differs from the decomposition method in two further respects. On the one hand, HELP-IV applies to more general environments. The benchmark design allows the confounders to feed back into the treatment, whereas the closed-form analysis in the main text of \citet{donaldson2026} assumes one-way confounders.\footnote{\citet{donaldson2026} shows that the decomposition method recovers the partial-equilibrium effect in two examples with two-way feedback, in which the treatment shock is observed and moves the treatment one for one when the confounder path is held fixed. In more general environments, however, the decomposition method no longer recovers this effect, as we show formally in Section~\ref{sec:donaldson-connection}.} The extension of HELP-IV to multiple exposure margins, multiple confounders, and transformed treatments and outcomes broadens this scope further. %
On the other hand, most HELP-IV designs are one-step LP-IV regressions, so they can be estimated with standard statistical packages, and inference builds on standard IV methods. The benchmark design is therefore accessible to researchers who are not familiar with constructing policy counterfactuals. Taken together, these two differences make HELP-IV a general and accessible framework for estimating dynamic partial-equilibrium effects across economics, finance, and the broader business and social sciences.

Our paper also speaks to three further strands of research, namely policy counterfactuals, shift-share designs, and the empirical questions that our applications address. First, a growing body of work constructs policy counterfactuals directly from impulse responses in semi-structural time-series models \citep{simszha2006md,mckaywolf2023,barnichonmesters2023,caravellomckaywolf2024,wang2026}. Our framework extends these counterfactuals from the time series to cross-sectional data through a panel local-projection design. Second, our paper relates to the literature on shift-share designs \citep{goldsmithpinkhametal2020,adaokolesarmorales2019,borusyakhulljaravel2022}, in particular to the recentered instruments of \citet{borusyakhull2023}, which purge a shift-share instrument of its expected value. The omitted-history bias in our paper is the time-series counterpart of the bias that recentering removes \citep{jaegerruiststuhler2018,borusyakhulljaravel2025}. HELP-IV corrects this bias together with the HEDGE bias. Third, our applications speak to the literatures on cross-sectional fiscal multipliers \citep{nakamurasteinsson2014,duporguerrero2017,pennings2021,chodorowreich2019} and on stock wealth effects \citep{zhoucarroll2012,dimaggiokermanimajlesi2020,chodorowreichnenovsimsek2021}. \citet{matthesetal2025} revisit the same two applications to estimate the aggregate effects that time fixed effects absorb, building on the cross-sectional identification of the original studies, whereas we re-examine that identification itself. The resulting estimates are portable and offer structural models in their respective fields a firmer empirical anchor for calibration.

\medskip
The rest of the paper is organized as follows. Section~\ref{sec:portable-markov}
defines the partial-equilibrium effect, derives the HEDGE bias, and establishes
identification by HELP-IV. Section~\ref{sec:confounders} extends HELP-IV to more
general environments, and Section~\ref{sec:help-inference} develops its
inference. Section~\ref{sec:montecarlo} evaluates the designs in Monte Carlo
simulations, Section~\ref{sec:applications} presents the two applications, and
Section~\ref{sec:conclusion} concludes.

	\section{Backward-looking HELP-IV}
	\label{sec:portable-markov}

	This section studies the partial-equilibrium effect in a \emph{backward-looking environment} where unit outcomes respond only to current and past aggregates. This isolates the identification problem in its simplest form. We first describe the model environment and define the partial-equilibrium effect we seek to identify (Section~\ref{sec:dyn-general}). We then show that the standard TWFE-IV design does not recover this effect and characterize the resulting bias in closed form (Section~\ref{sec:twfe-hedge}). We next introduce a novel local-projection design, HELP-IV, and establish that it identifies the partial-equilibrium effect (Section~\ref{sec:dyn-identification}). Finally, in Section~\ref{sec:donaldson-connection}, we compare HELP-IV with the decomposition method of \citet{donaldson2026}.

	\subsection{Environment and the identification target}
	\label{sec:dyn-general}

	The object of interest in this section is the partial-equilibrium effect: the response of a unit-level outcome to a one-unit autonomous movement in the aggregate \emph{treatment}, holding the path of the aggregate \emph{confounder} fixed over the response horizon. For example, when a researcher studies how regional output responds to government spending, the outcome is regional output, the treatment is government spending, and a candidate confounder is the interest rate, which moves with spending and also affects output directly.

	\subsubsection{Model setup}
	\label{sec:dyn-modelsetup}

	The model specifies the dependence of unit outcomes on the aggregate variables, the joint process of those aggregates, and the cross-sectional relation between the exposures to them. In the backward-looking environment, the outcome of unit $i$, denoted $Y_{it}$, depends on its own lag $Y_{i,t-1}$ and on the current and lagged values of the treatment $G_t$ and the confounder $R_t$:
	\begin{align}\label{eq:outcome-dyn}
		Y_{it}&=\mu_i+\lambda_t+\rho Y_{i,t-1}+\sum_{\ell=0}^{p}\left(\beta_\ell\sig G_{t-\ell}+\gamma_\ell\sir R_{t-\ell}\right)+u_{it}.
	\end{align} Here $\mu_i$ and $\lambda_t$ are unit and time effects, and $\rho$, with $|\rho|<1$, measures own persistence.\footnote{A single own lag keeps the notation light. An own-lag polynomial of order $q$ may change the closed-form expression for the identification target, but leaves the identification propositions in this paper unchanged.} The coefficients $\beta_\ell$ and $\gamma_\ell$ measure the dependence of the outcome on the two aggregates, $\sig$ and $\sir$ are unit $i$'s exposures to them, and $u_{it}$ is an idiosyncratic disturbance. The two exposures are predetermined, and their cross-sectional relation is

	\begin{align}\label{eq:exposure-dyn}
		\sir&=\phi\sig+\uisr.
	\end{align}
	We maintain that $\sig$ and $\uisr$ are independent, $\E[\uisr]=0$, and
	$(\mu_i,\sig,\uisr,\{u_{it}\})$ is independent of the aggregate process. The coefficient $\phi$ measures exposure correlation, and $\phi\neq0$ challenges the identification of the partial-equilibrium effect, because units with greater exposure to $G_t$ then also have systematically different exposure to $R_t$. We restrict attention to a single confounder and extend the framework to multiple confounders in the next section.

	We assume that the two aggregates admit the stable structural
	moving-average representation
	\begin{align}
		\mathbf W_t
		&=
		\sum_{j=0}^{\infty}
		\Psi_j\boldsymbol\epsilon_{t-j},
		\label{eq:svma-dyn}
	\end{align}
	where $\mathbf W_t=(G_t,R_t)'$ and
	$\boldsymbol\epsilon_t=(\epsilon_{g,t},\epsilon_{r,t})'$. The sequence
	$\{\Psi_j\}_{j\geq0}$ is absolutely summable. The two structural shocks, which we call the
	\emph{treatment shock} $\epsilon_{g,t}$ and the \emph{confounder shock} $\epsilon_{r,t}$, have mean
	zero, are serially uncorrelated, and are mutually orthogonal. We leave the scale of these shocks
	unrestricted, because
	the effects below are ratios of responses and do not depend on it. Each aggregate can respond to the other's shock, so the representation permits
	feedback in both directions between $G_t$ and $R_t$, a case we refer to as \emph{two-way general
	equilibrium}. When $R_t$ responds to $\epsilon_{g,t}$ while $G_t$ does not respond to
	$\epsilon_{r,t}$ at any horizon, the feedback runs in one direction, and we refer to this special
	case as \emph{one-way general equilibrium}.

	\subsubsection{The partial-equilibrium effect}
	\label{sec:dyn-pe-effect}

	To formalize the definition of the partial-equilibrium effect, let
	$\breve G_\ell$ denote the controlled response of $G_{t+\ell}$ to
	$\epsilon_{g,t}$ when the path $R_t,\ldots,R_{t+h}$ is held fixed, and let
	$\breve Y_{i,h}$ denote the corresponding response of the outcome. The path is
	held fixed by confounder shocks dated $t$ through $t+h$, referred to as the
	offsets \citep{simszha2006md}. The confounder path does not change, so its
	direct contribution to the outcome response is zero, and applying
	equation~\eqref{eq:outcome-dyn} under this experiment gives
	\begin{align}
		\breve Y_{i,h}
		&=
		\sig
		\sum_{\ell=0}^{\min\{p,h\}}
		\beta_\ell\breve G_{h-\ell}
		+
		\rho \breve Y_{i,h-1}.
		\label{eq:pe-outcome-controlled}
	\end{align}

Normalizing by the impact movement in $G_t$ that this experiment generates gives the partial-equilibrium effect $\beta_h^P=\breve Y_{i,h}/(\sig\breve G_0)$, which satisfies the recursion

	\begin{align}
		\beta_h^P
		&=
		\frac{
			\displaystyle
			\sum_{\ell=0}^{\min\{p,h\}}
			\beta_\ell\breve G_{h-\ell}
		}{
			\breve G_0
		}
		+
		\rho\beta_{h-1}^P,
		\qquad
		\beta_{-1}^P=0.
		\label{eq:portable-elasticity-normalized}
	\end{align}
	The initial condition holds because outcomes do not respond before the
	shock, so that $\beta_0^P=\beta_0$. Both the numerator and the denominator of
	$\beta_h^P$ are measured in the same fixed-path experiment, so any structural
	model that runs this experiment yields a direct counterpart of $\beta_h^P$. In
	this sense, $\beta_h^P$ is a portable statistic
	\citep{nakamurasteinsson2018}. Its value, however, varies with the controlled
	treatment path $\breve G_0,\ldots,\breve G_h$, so a model reproduces
	$\beta_h^P$ only if it generates the same controlled path.

	\subsection{The HEDGE bias of the TWFE-IV design}
	\label{sec:twfe-hedge}

	A large empirical literature targets the partial-equilibrium effect with a
	TWFE-IV design, for example in fiscal policy
	\citep{nakamurasteinsson2014,chodorowreich2019}, monetary policy and firm
	investment \citep{ottonellowinberry2020}, and household wealth
	\citep{miansufi2014,guren2021,chodorowreichnenovsimsek2021}. Specifications differ in the
	definition of the treatment, in the horizon, and in the controls, but they
	share the interaction of predetermined exposure with the treatment. We write the design as the two-way fixed-effects local projection
	\begin{align}
		\label{eq:lp-twfe}
		Y_{i,t+h}
		&=
		b_h^{\mathrm{TWFE}}\sig G_t
		+\mu_{i,h}
		+\lambda_{t,h}
		+v_{i,t,h}.
	\end{align}
	The exposure-weighted treatment $\sig G_t$ is instrumented by $\sig z_{g,t}$, where the treatment
	proxy $z_{g,t}$ is
	\begin{align}
		\label{eq:g-proxy}
		z_{g,t}
		&=
		\kappa_g\epsilon_{g,t}+\eta_{g,t},
	\end{align}
	for $\kappa_g\neq0$, and the measurement error $\eta_{g,t}$ has mean zero and is orthogonal to
	the structural shocks at all leads and lags. We also maintain that the exposure-weighted proxies are uncorrelated with the unit-level disturbances at all leads and lags. We refer to this condition as panel exclusion. It must be imposed separately, because orthogonality of the proxy to the aggregate shocks does not imply it.

		Proposition~\ref{lemma:hedge-bias} provides a closed-form expression for the wedge between the TWFE-IV coefficient and $\beta_h^P$.

	\begin{proposition}[HEDGE bias of TWFE-IV]
		\label{lemma:hedge-bias}
		Suppose \eqref{eq:outcome-dyn}--\eqref{eq:svma-dyn} and
		\eqref{eq:g-proxy} hold, the proxy satisfies the panel exclusion,
		$\Var(\sig)>0$, $G_0^g\neq0$, and $\breve G_0\neq0$. Then
		the population TWFE-IV coefficient in
		\eqref{eq:lp-twfe}
		satisfies
		\begin{align}
			b_h^{\mathrm{TWFE}}-\beta_h^P
			&=
			\underbrace{
				\left(
				(G_0^g)^{-1}\sum_{j=0}^{h}\varphi_jG_{h-j}^g
				-
				\breve G_0^{-1}\sum_{j=0}^{h}\varphi_j\breve G_{h-j}
				\right)
			}_{\text{aggregate feedback}}
			+
			\underbrace{
				\phi\,(G_0^g)^{-1}
				\sum_{j=0}^{h}\bar\gamma_jR_{h-j}^g
			}_{\text{exposure correlation}},
			\label{eq:hedge-bias-lemma}
		\end{align}
		where
		\begin{align*}
			G_h^g=\frac{\E[G_{t+h}\epsilon_{g,t}]}{\E[\epsilon_{g,t}^2]},
			\qquad
			\varphi_j=\sum_{\ell=0}^{\min\{p,j\}}\rho^{j-\ell}\beta_\ell,
			\qquad
			\bar\gamma_j=\sum_{\ell=0}^{\min\{p,j\}}\rho^{j-\ell}\gamma_\ell,
		\end{align*}
		and $G_h^r$, $R_h^g$, and $R_h^r$ are defined analogously to $G_h^g$.
	\end{proposition}

\noindent The proof of Proposition~\ref{lemma:hedge-bias} is in
Appendix~\ref{proof:hedge-bias}. Proposition~\ref{lemma:hedge-bias} shows that the HEDGE bias has two components. The
first is the \emph{aggregate-feedback component}. TWFE-IV measures the outcome response
along the unrestricted treatment path per unit of its impact $G_0^g$. The
target $\beta_h^P$ instead measures it along the controlled path per unit of
$\breve G_0$. The two normalized responses both equal $\beta_0$ at $h=0$,
and at longer horizons they can differ only if confounder shocks feed back
into the treatment. This component is therefore zero under one-way general equilibrium,
where $G_\ell^r=0$ for all $\ell\geq0$ and hence $\breve G_\ell=G_\ell^g$.
The second is the \emph{exposure-correlation component}. When $\phi\neq0$, the
treatment shock generates cross-sectional variation through both $\sig G_{t+\ell}$ and
$\sir R_{t+\ell}$, $\ell=0,\ldots,h$, and time effects remove only their common components, so the coefficient on $\sig G_t$ also
picks up the response through $\sir R_{t+\ell}$. This component is therefore zero when $\phi=0$ or when the confounder does not respond to the treatment
shock at any horizon up to $h$.

	\subsection{Identification by the HELP-IV design}
	\label{sec:dyn-identification}

Both components of the HEDGE bias arise because the TWFE-IV design leaves the
confounder path unrestricted. To identify the partial-equilibrium effect, we
design a heterogeneous-exposure local-projection instrumental variables
(HELP-IV) regression that holds this path fixed. It adds the realized values of
the confounder path to the TWFE-IV regression as path controls and instruments
these controls with dated proxies for the confounder shocks.

In particular, we estimate
\begin{align}
	Y_{i,t+h}
	&=
	b_h^{\mathrm{HELP}}\sig G_t
	+
	\sum_{s=0}^{h}c_{h,s}\sig R_{t+s}
	+
	\mu_{i,h}
	+
	\lambda_{t,h}
	+
	\xi_{i,t,h},
	\qquad
	h\geq0.
	\label{eq:lpiv-dyn}
\end{align}
We instrument the exposure-weighted treatment $\sig G_t$ and the path controls
$\sig R_t,\ldots,\sig R_{t+h}$ with the exposure-weighted proxies
\begin{align}
	\mathbf Z_{it,h}
	&=
	\sig
	\left(
	z_{g,t},
	z_{r,t},
	z_{r,t+1},
	\ldots,
	z_{r,t+h}
	\right)',
	\label{eq:iv-dyn}
\end{align}
where $z_{r,t}$ is a proxy for the confounder shock,
\begin{align}
	z_{r,t}
	&=
	\kappa_r\epsilon_{r,t}
	+
	\eta_{r,t},
	\qquad
	\kappa_r\neq0,
	\label{eq:r-proxy}
\end{align}
and $\eta_{r,t}$ has mean zero and is orthogonal to the structural shocks at all
leads and lags.\footnote{The regression can also include controls determined
before date $t$, such as the lagged outcome, without affecting identification,
because the proxies are orthogonal to them.}

\subsubsection{Why HELP-IV identifies the partial-equilibrium effect}
\label{sec:help-why}

We start from the identifying moment conditions. Exogeneity of the proxies and
panel exclusion imply that the coefficients of the HELP-IV regression solve the
population moment system
\begin{align}
	\E
	\left[
	\widetilde{\mathbf Z}_{it,h}
	\left(
	\widetilde Y_{i,t+h}
	-
	\widetilde{\boldsymbol X}_{it,h}'
	\boldsymbol\vartheta_h
	\right)
	\right]
	&=
	\mathbf 0,
	\label{eq:help-population-system}
\end{align}
where
\begin{align*}
	\boldsymbol X_{it,h}
	&=
	\sig\left(G_t,R_t,\ldots,R_{t+h}\right)',
	\qquad
	\boldsymbol\vartheta_h
	=
	\left(b_h^{\mathrm{HELP}},c_{h,0},\ldots,c_{h,h}\right)'
\end{align*}
collect the endogenous regressors and their coefficients, and a tilde on panel
variables denotes population residualization with respect to the unit and time
effects. The HELP-IV regression is exactly identified, because the moment system
has $h+2$ equations, one for each instrument, in the $h+2$ elements of
$\boldsymbol\vartheta_h$.

To show how $b_h^{\mathrm{HELP}}$ identifies the partial-equilibrium effect, we
solve the moment system \eqref{eq:help-population-system} for
$b_h^{\mathrm{HELP}}$ through a linear combination of its equations from which the
path coefficients $c_{h,0},\ldots,c_{h,h}$ drop out.
Premultiplying system \eqref{eq:help-population-system} by a row vector
$(1,w_0,\ldots,w_h)$ adds the equations of the confounder proxies, weighted by
the offset weights $w_0,\ldots,w_h$, to the equation of the treatment proxy. The combined equation
is the moment condition of a single instrument, the exposure-weighted combined
proxy $\sig(z_{g,t}+\sum_{j=0}^{h}w_jz_{r,t+j})$. Up to the common factor
$\Var(\sig)$, each coefficient in this equation is multiplied by the covariance
of the combined proxy with the aggregate part of its regressor, which is $G_t$
for $b_h^{\mathrm{HELP}}$ and $R_{t+s}$ for $c_{h,s}$. The path coefficients
therefore drop out of the combined
equation exactly when the combined proxy is uncorrelated with every value of the
confounder path,
\begin{align}
	\E\Bigl[R_{t+\ell}\Bigl(z_{g,t}+\sum_{j=0}^{h}w_jz_{r,t+j}\Bigr)\Bigr]
	&=
	0,
	\qquad
	\ell=0,\ldots,h,
	\label{eq:help-weights-path}
\end{align}
in which case the combined equation reduces to a single equation in
$b_h^{\mathrm{HELP}}$,
\begin{align}
	\E\Bigl[\dot s_{ig}\Bigl(z_{g,t}+\sum_{j=0}^{h}w_jz_{r,t+j}\Bigr)\,\widetilde Y_{i,t+h}\Bigr]
	&=
	\Var(\sig)\,b_h^{\mathrm{HELP}}\,
	\E\Bigl[G_t\Bigl(z_{g,t}+\sum_{j=0}^{h}w_jz_{r,t+j}\Bigr)\Bigr],
	\label{eq:help-weights-first-stage}
\end{align}
where $\dot s_{ig}=\sig-\E[\sig]$ denotes demeaned exposure.

Condition \eqref{eq:help-weights-path} holds the confounder path fixed. Since
each proxy covaries only with the shock of its own type and date, condition
\eqref{eq:help-weights-path} is equivalent to
$R_\ell^g+\sum_{j=0}^{\ell}e_jR_{\ell-j}^r=0$ for $\ell=0,\ldots,h$, in the
notation of Proposition~\ref{lemma:hedge-bias}, where $e_j$ is proportional to
$w_j$. Each such sum is the response of $R_{t+\ell}$ to a unit treatment shock
accompanied by confounder shocks of sizes $e_0,\ldots,e_h$ at dates $t$ through
$t+h$. Because the path is held fixed, the confounder channel is shut down in the
reduced form on the left side of condition \eqref{eq:help-weights-first-stage},
and the exposure-correlation component is removed.

Condition \eqref{eq:help-weights-first-stage} equates the covariance of the
exposure-weighted combined proxy with the outcome, the reduced form, to
$b_h^{\mathrm{HELP}}$ times its covariance with the treatment, the first stage.
Under condition \eqref{eq:help-weights-path} and up to the common factor
$\Var(\sig)\E[z_{g,t}\epsilon_{g,t}]$, the first stage is $\breve G_0$, the impact
response of $G_t$ in the fixed-path experiment, and the reduced form is the outcome response
per unit of exposure, which equals $\breve G_0\beta_h^P$ by the definition of
$\beta_h^P$. Condition \eqref{eq:help-weights-first-stage} therefore reads
$\breve G_0b_h^{\mathrm{HELP}}=\breve G_0\beta_h^P$, so that
$b_h^{\mathrm{HELP}}=\beta_h^P$ whenever $\breve G_0\neq0$. HELP-IV thus divides
the controlled reduced form $\breve G_0\beta_h^P$ by the controlled first stage
$\breve G_0$, whereas TWFE-IV divides the unrestricted reduced form by the
unrestricted first stage $G_0^g$. Because the numerator and the denominator of
HELP-IV are measured in the same fixed-path experiment, the aggregate-feedback
component is removed as well.\footnote{To see the logic in
the simplest case, take $h=0$. Condition \eqref{eq:help-weights-path} then gives
$w_0=-\E[R_tz_{g,t}]/\E[R_tz_{r,t}]$ when $\E[R_tz_{r,t}]\neq0$. The combined proxy
$z_{g,t}+w_0z_{r,t}$ thus moves $G_t$ only through the treatment shock accompanied
by the confounder shock that keeps $R_t$ unchanged, so its covariance with $G_t$
is proportional to $\breve G_0$, and condition
\eqref{eq:help-weights-first-stage} yields $b_0^{\mathrm{HELP}}=\beta_0^P=\beta_0$.}

The following proposition formally establishes that HELP-IV identifies the
partial-equilibrium effect.

\begin{proposition}[HELP-IV identification]
	\label{prop:help-identification}
	Suppose \eqref{eq:outcome-dyn}--\eqref{eq:svma-dyn}, \eqref{eq:g-proxy},
	\eqref{eq:iv-dyn}, and \eqref{eq:r-proxy} hold, the exposure-weighted
	proxies satisfy the panel exclusion restrictions, $\Var(\sig)>0$,
	$\E[R_tz_{r,t}]\neq0$, and $\breve G_0\neq0$. Then
	\eqref{eq:help-population-system} has a unique solution whose first
	element satisfies
	\begin{align}
		b_h^{\mathrm{HELP}}
		&=
		\beta_h^P.
		\label{eq:help-identifies}
	\end{align}
\end{proposition}

\noindent The proof is in Appendix~\ref{app:proof-help-identification}. The two
relevance conditions play distinct roles. The condition $\E[R_tz_{r,t}]\neq0$
requires the confounder proxy to move the confounder on impact, which guarantees
that weights satisfying condition \eqref{eq:help-weights-path} exist and are
unique. The condition $\breve G_0\neq0$ requires the treatment to move in the
fixed-path experiment, which allows condition
\eqref{eq:help-weights-first-stage} to determine $b_h^{\mathrm{HELP}}$.

\subsubsection{Why ordinary controls are insufficient}
\label{sec:why-instrument}

Prior studies address aggregate confounders by interacting exposure with the
suspected confounders and adding these interactions to the TWFE-IV regression as
ordinary controls \citep[e.g.,][]{chodorowreichnenovsimsek2021,guren2021}. A
control for the current confounder $\sig R_t$, however, falls short of the
fixed-path experiment for two reasons. First, it
restricts the confounder only at date $t$, although the outcome at date $t+h$
responds to the entire path $R_t,\ldots,R_{t+h}$. Second, it restricts the
realized value $R_t$, which reflects all aggregate shocks rather than the
treatment shock alone, and therefore removes only the component of the later path
that $R_t$ predicts through the unconditional autocovariance of the confounder.
Controlling only for the current confounder can thus still bias the estimate of
the partial-equilibrium effect.

Controlling for the entire realized path $\sig R_t,\ldots,\sig R_{t+h}$ resolves
the first problem but not the second, since these controls are themselves
realized values. The realized path responds also to shocks dated before $t$ and
to subsequent treatment shocks, which affect the outcome through terms omitted
from the regression, namely the lags in equation \eqref{eq:outcome-dyn} and later
values of the treatment. By the Frisch--Waugh--Lovell theorem, the effective
instrument for $\sig G_t$ is the exposure-weighted treatment proxy net of its
projection on the realized path. Since the treatment shock moves the confounder,
this projection is nonzero and transmits the omitted shocks to the instrument, so
that the realized path becomes an endogenous control. Instrumenting the path
controls is therefore necessary, and the dated confounder proxies are valid
instruments for this purpose, as each is correlated only with the confounder
shock of its date.

	\subsection{Identification comparison with the decomposition of
		\texorpdfstring{\citet{donaldson2026}}{Donaldson (2026)}}
	\label{sec:donaldson-connection}

\citet{donaldson2026} also studies the HEDGE bias of the TWFE-IV design and proposes a decomposition method that targets the same partial-equilibrium effect as HELP-IV. This subsection compares the two methods. We first show that, under the
	backward-looking data-generating process of this section, the
	decomposition method removes the exposure-correlation component of the
	HEDGE bias but differs from the partial-equilibrium effect when the offsets
	change the treatment response used for normalization. We then discuss two further advantages of HELP-IV.

	The decomposition method proceeds in three steps. The cross-sectional step
	regresses the outcome on the exposure-weighted proxies of both shocks,
	\begin{align}
		Y_{i,t+h}
		&=
		b_h^{\mathrm{Don}}\sig z_{g,t}
		+
		c_h^{\mathrm{Don}}\sig z_{r,t}
		+
		\mu_{i,h}
		+
		\lambda_{t,h}
		+
		e_{i,t,h}.
		\label{eq:don-cs-pop}
	\end{align}
	The coefficient $b_h^{\mathrm{Don}}$ is the unrestricted reduced-form response of the outcome to the treatment
	proxy, which contains the HEDGE bias because the treatment shock also moves the
	confounder path. The coefficient $c_h^{\mathrm{Don}}$ is the reduced-form response of the outcome to the confounder
	proxy. The time-series step, following \citet{simszha2006md}, finds the
	sequence of confounder proxies that reproduces the confounder path generated by the
	treatment proxy. It chooses $\{\nu_{r,k}\}_{k=0}^{h}$ such that
	\begin{align}
		\pi_{R,\ell}^{g}
		&=
		\sum_{k=0}^{\ell}
		\pi_{R,\ell-k}^{r}\nu_{r,k},
		\qquad
		\ell=0,\ldots,h,
		\label{eq:don-pop-recursion}
	\end{align}
	where $\pi_{X,\ell}^{j}=\E[X_{t+\ell}z_{j,t}]/\E[z_{j,t}^2]$ denotes the
	response of $X_{t+\ell}$ to the proxy $z_{j,t}$, for $X\in\{G,R\}$ and
	$j\in\{g,r\}$, so that $\ell=0$ gives the impact response at date $t$.
	The weight $\nu_{r,k}$ applies to the confounder proxy dated $t+k$. The final
	step subtracts the outcome effect of this sequence,
	$\Delta_h=\sum_{k=0}^{h}c_{h-k}^{\mathrm{Don}}\nu_{r,k}$, from $b_h^{\mathrm{Don}}$, using the
	coefficients $c_0^{\mathrm{Don}},\ldots,c_h^{\mathrm{Don}}$ from the regressions at each horizon. It then
	divides the result by the first-stage coefficient $\pi_{G,0}^{g}$, which
	removes the unknown scale of the proxy,
	\begin{align}
		\beta_h^{\mathrm{Don}}
		&=
		\left(\pi_{G,0}^{g}\right)^{-1}
		\left(b_h^{\mathrm{Don}}-\Delta_h\right).
		\label{eq:don-statistic}
	\end{align}
	The difference $b_h^{\mathrm{Don}}-\Delta_h$ is thus proportional to the controlled reduced form
of Section~\ref{sec:help-why}, and the decomposition method normalizes it by the
unrestricted first stage $\pi_{G,0}^{g}$ instead of the controlled one.

	The partial-equilibrium effect of Section~\ref{sec:dyn-pe-effect} instead divides
	the same outcome response by the impact movement of $G_t$ under the
	experiment. Because offsetting confounder shocks dated after $t$ cannot move $G_t$,
	this movement is the unrestricted impact response less the impact movement
	induced by the offsetting proxy dated $t$. Hence,
	\begin{align}
		\beta_h^P
		&=
		\left(\pi_{G,0}^{g}-\nu_{r,0}\pi_{G,0}^{r}\right)^{-1}
		\left(b_h^{\mathrm{Don}}-\Delta_h\right).
		\label{eq:portable-response-form}
	\end{align}
	The statistics in \eqref{eq:don-statistic} and
	\eqref{eq:portable-response-form} share the numerator. Their denominators
	differ by $\nu_{r,0}\pi_{G,0}^{r}$, the impact response of $G_t$ to the
	offsetting confounder proxy. Under one-way general equilibrium, the treatment does not respond to confounder
shocks, so this term vanishes and the decomposition method recovers	the partial-equilibrium effect. Under two-way general equilibrium, the term is
	nonzero when the confounder shock moves $G_t$ on impact and the treatment
	shock moves $R_t$ on impact. When either response arises only with a lag,
	the denominators of equations~\eqref{eq:don-statistic}
	and~\eqref{eq:portable-response-form} coincide for the date-$t$ treatment of
	this section. When the term is nonzero, the
	decomposition method removes the
	exposure-correlation component of Proposition~\ref{lemma:hedge-bias} and
	the part of the aggregate-feedback component that operates through the
	outcome response.\footnote{Under two-way general equilibrium, the HEDGE bias given in Proposition~\ref{lemma:hedge-bias} also differs from its counterpart in \citet{donaldson2026}. It is derived in a more general environment, with an aggregate block that has an infinite-order moving-average representation and distributed lags in the outcome equation.} It retains, however, the part that operates through the
	normalization, so the statistic in \eqref{eq:don-statistic} is biased for
	the partial-equilibrium effect.\footnote{Appendices~A.1 and~A.3.3 of \citet{donaldson2026} show that the
		decomposition method recovers the partial-equilibrium effect in specific
		two-way examples. There, the treatment shock enters the treatment
		equation with a unit coefficient, so a unit shock moves $G_t$ by one when
		the confounder path is held fixed, and the corrected reduced form
		$b_h^{\mathrm{Don}}-\Delta_h$ needs no normalization. Her application instead divides by the
		horizon-$h$ first stage of cumulative local defense spending, whereas the
		relevant denominator is the response of cumulative spending through $t+h$
		to the proxy and the entire offsetting sequence, which lagged feedback
		also changes (Section~\ref{sec:generalized-target}).}

	HELP-IV has two further advantages over the decomposition method. First,
	HELP-IV imposes the fixed-path experiment and the normalization within a
	single IV regression, whereas the decomposition method combines a
	cross-sectional regression with a time-series step. This single regression can
	be estimated with standard statistical packages and delivers both the
	estimate and its standard error (Section~\ref{sec:help-inference}).
	Second, HELP-IV can be extended to more general environments with multiple
exposure margins (Section~\ref{sec:multiple-exposure-margins}), multiple
confounders (Section~\ref{sec:multiple-aggregate-confounders}), forward-looking
outcomes (Section~\ref{sec:forward-portable-section}), or transformed treatments
and outcomes (Section~\ref{sec:generalized-target}). Together, these advantages
	widen the range of applications of HELP-IV.

	\section{HELP-IV in more general environments}

	\label{sec:backward-extensions}
	\label{sec:confounders}

Many empirical applications target the partial-equilibrium effect in
environments more general than that of Section~\ref{sec:portable-markov}. This
section extends the HELP-IV design to multiple exposure margins, multiple
confounders, outcomes that depend on expected future aggregates, and transformed
treatments and outcomes, so that it can be implemented in such applications.

	\subsection{Multiple exposure margins}
	\label{sec:multiple-exposure-margins}

	Units are often exposed to the same aggregate treatment along multiple margins. For example, in estimating local spending
	multipliers, \citet{parkzhouzubairy2026} distinguish counties' exposure to
	prime-contract spending from their exposure to subcontract spending. They
	estimate separate prime-contract and subcontract multipliers, which
	capture economically distinct sources of local fiscal transmission. Motivated by this application, we extend HELP-IV to identify a
	separate partial-equilibrium effect for each exposure margin.

	Consider an environment in which the aggregate treatment affects unit $i$
	through $M$ predetermined exposure margins:
	\begin{align}
		Y_{it}
		&=
		\mu_i+\lambda_t+\rho Y_{i,t-1}
		+
		\sum_{\ell=0}^{p}
		\sum_{m=1}^{M}
		\beta_{m,\ell}s_{ig,m}G_{t-\ell}
		+
		\sum_{\ell=0}^{p}
		\gamma_\ell\sir R_{t-\ell}
		+
		u_{it}.
		\label{eq:outcome-margins-main}
	\end{align}
	Here $s_{ig,m}$ denotes exposure along margin $m$, and
	$\mathbf s_{ig}=(s_{ig,1},\ldots,s_{ig,M})'$. Confounder exposure may be
	correlated with each treatment-exposure margin:
	$\sir=\sum_{m=1}^{M}\phi_m s_{ig,m}+\uisr$, where
	$\mathbf s_{ig}$ is independent of $\uisr$ and $\E[\uisr]=0$. The aggregate
	dynamics are unchanged from Section~\ref{sec:portable-markov}. For each margin $m$, the
	partial-equilibrium effect $\beta_{m,h}^{P}$ is the response of $Y_{i,t+h}$ per unit of
	exposure along margin $m$, holding exposure along the other margins fixed. As in
	Section~\ref{sec:dyn-pe-effect}, this response is measured in the fixed-path experiment
	and normalized by the impact movement $\breve G_0$, so that
	\begin{align}
		\beta_{m,h}^{P}
		&=
		\breve G_0^{-1}
		\sum_{\ell=0}^{\min\{p,h\}}
		\beta_{m,\ell}\breve G_{h-\ell}
		+
		\rho\beta_{m,h-1}^{P},
		\qquad
		\beta_{m,-1}^{P}=0.
		\label{eq:portable-elasticity-margins-main}
	\end{align}

	Because the confounder exposure loads on every margin, the path controls must be
	interacted with each margin. HELP-IV with multiple exposure margins therefore
	estimates the regression
	\begin{align}
		Y_{i,t+h}
		&=
		\sum_{m=1}^{M}
		b_{m,h}^{\mathrm{HELP}}s_{ig,m}G_t
		+
		\sum_{s=0}^{h}
		\sum_{m=1}^{M}
		c_{m,h,s}s_{ig,m}R_{t+s}
		+
		\mu_{i,h}
		+
		\lambda_{t,h}
		+
		\xi_{i,t,h}.
		\label{eq:lpiv-margins-main}
	\end{align}
	The endogenous variables are instrumented using
	$\mathbf s_{ig}\otimes
	(z_{g,t},z_{r,t},z_{r,t+1},\ldots,z_{r,t+h})'$.

	\begin{corollary}[Identification with multiple exposure margins]
		\label{cor:margins}
		Suppose the conditions of Proposition~\ref{prop:help-identification} hold
		for the model and instruments of this subsection, with $\Var(\sig)>0$
		replaced by nonsingularity of $\Var(\mathbf s_{ig})$. Then the population coefficients
		in \eqref{eq:lpiv-margins-main} satisfy
		$b_{m,h}^{\mathrm{HELP}}=\beta_{m,h}^{P}$ for $m=1,\ldots,M$.
	\end{corollary}

	The proof is in Appendix~\ref{app:margins}. Corollary~\ref{cor:margins}
	also applies to exposure transmitted through production or spatial networks. For example, \citet{acemogluetal2016} measure
	exposure to import competition both directly and through input--output
	linkages, while \citet{adaoarkolakisesposito2019} measure exposure to trade
	shocks through linkages across locations. One may then set
	$s_{ig,1}=\sig$ for own exposure,
	$s_{ig,2}=\sum_j a^U_{ij}s_{jg}$ for upstream exposure, and
	$s_{ig,3}=\sum_j a^D_{ij}s_{jg}$ for downstream exposure, where $a^U$ and
	$a^D$ are input--output or spatial linkage weights. The corollary then identifies one
	partial-equilibrium effect for each margin.

	\subsection{Multiple aggregate confounders}
	\label{sec:multiple-aggregate-confounders}

		In many applications, multiple confounders are present. For example, \citet{acemoglurestrepo2020} estimate the effect of robot adoption on local labor markets and treat four aggregates as potential confounders: Chinese imports, Mexican imports, offshoring, and the decline in routine tasks. If commuting zones more exposed to robots are also more exposed to these aggregates, and their paths move with robot adoption, each aggregate generates HEDGE bias in the TWFE-IV design, which control variables alone do not remove (Section~\ref{sec:why-instrument}).  This subsection therefore extends HELP-IV to hold all confounder paths fixed jointly.

	We extend the data-generating process to $K$ confounders. The outcome
	satisfies
	\begin{align}
		Y_{it}
		&=
		\mu_i+\lambda_t+\rho Y_{i,t-1}
		+
		\sum_{\ell=0}^{p}
		\left(
		\beta_\ell\sig G_{t-\ell}
		+
		\sum_{q=1}^{K}
		\gamma_{q,\ell}s_{ir_q}R_{q,t-\ell}
		\right)
		+
		u_{it}.
		\label{eq:outcome-conf}
	\end{align}
	Here $R_{1t},\ldots,R_{Kt}$ are the confounders and $s_{ir_q}$
	is unit $i$'s exposure to $R_{qt}$. Exposure to each confounder may be
	correlated with treatment exposure,
	\begin{align}
		s_{ir_q}
		&=
		\phi_q\sig+u_i^{r_q},
		\qquad
		q=1,\ldots,K,
		\label{eq:exposure-conf}
	\end{align}
	where $\phi_q$ measures that exposure correlation, and $\sig$ and
	$u_i^{r_q}$ satisfy the conditions maintained for
	equation~\eqref{eq:exposure-dyn}. Let $\mathbf R_t=(R_{1t},\ldots,R_{Kt})'$ and
	$\boldsymbol\epsilon_{r,t}=(\epsilon_{r_1,t},\ldots,\epsilon_{r_K,t})'$. The
	aggregate dynamics follow equation~\eqref{eq:svma-dyn}, with $R_t$ and
	$\epsilon_{r,t}$ replaced by $\mathbf R_t$ and $\boldsymbol\epsilon_{r,t}$,
	and with the same stability and orthogonality conditions.

	The target is the response to a movement in $G_t$ caused by $\epsilon_{g,t}$
	when $\mathbf R_t,\ldots,\mathbf R_{t+h}$ are held fixed jointly. The
	partial-equilibrium effect therefore retains the form in
	equation~\eqref{eq:portable-elasticity-normalized}, with
	$\{\breve G_\ell\}$ now denoting the treatment path under this joint
	fixed-path experiment. We consider three designs:
	multiple-confounder HELP-IV, composite HELP-IV, and targeting-rule
	HELP-IV. All three identify
	the same target under their own conditions, but each is suited to a different empirical application.

	\subsubsection{Multiple-confounder HELP-IV}
	\label{sec:vec-conf}

	The direct extension of HELP-IV adds the full path of every confounder as
	path controls and instruments these controls jointly with the corresponding
	proxies. We refer to this design as multiple-confounder HELP-IV. At horizon
	$h$, consider
	\begin{align}
		Y_{i,t+h}
		&=
		b_h^{\mathrm{HELP}}\sig G_t
		+
		\sum_{q=1}^{K}
		\sum_{s=0}^{h}
		c_{q,h,s}\sig R_{q,t+s}
		+
		\mu_{i,h}
		+
		\lambda_{t,h}
		+
		\xi_{i,t,h}.
		\label{eq:lpiv-vec}
	\end{align}

	The endogenous variables $\sig G_t$ and $\sig R_{q,t+s}$, for
	$q=1,\ldots,K$ and $s=0,\ldots,h$, are instrumented by
	\begin{align}
		\mathbf Z_{it,h}
		&=
		\sig
		\left(
		z_{g,t},
		\boldsymbol z_{r,t}',
		\boldsymbol z_{r,t+1}',
		\ldots,
		\boldsymbol z_{r,t+h}'
		\right)',
		\label{eq:iv-vec}
	\end{align}
	where $\boldsymbol z_{r,t}=(z_{r_1,t},\ldots,z_{r_K,t})'$ collects the
	proxies for the confounder shocks, each defined as in
	equation~\eqref{eq:r-proxy}. The system is exactly identified, with $1+K(h+1)$
	instruments for as many endogenous variables.

	Let $\Sigma_{rz,0}=\E[\mathbf R_t\boldsymbol z_{r,t}']$. Joint relevance requires
	$\Sigma_{rz,0}$ to be nonsingular. Because a confounder proxy dated $t+s$ does
	not move the confounders dated before $t+s$, the first-stage system for
	$\mathbf R_t,\ldots,\mathbf R_{t+h}$ is block triangular with $\Sigma_{rz,0}$
	on each diagonal block, so the confounder-path block has full rank, while
	$\breve G_0\neq0$ supplies the remaining rank condition for the
	treatment.

	\begin{corollary}[Identification with multiple confounders]
		\label{cor:vec-conf}
		Suppose \eqref{eq:svma-dyn}, \eqref{eq:g-proxy}, \eqref{eq:r-proxy},
		\eqref{eq:outcome-conf}, \eqref{eq:exposure-conf}, and \eqref{eq:iv-vec} hold, the exposure-weighted
		proxies satisfy the panel exclusion restrictions, $\Var(\sig)>0$,
		$\Sigma_{rz,0}$ is nonsingular, and $\breve G_0\neq0$. Then the
		population coefficient on
		$\sig G_t$ in \eqref{eq:lpiv-vec} satisfies $b_h^{\mathrm{HELP}}=\beta_h^P$.
	\end{corollary}

	The proof is in Appendix~\ref{app:confounders}. Although multiple-confounder
	HELP-IV identifies $\beta_h^P$ at any horizon and for any number of
	confounders, it uses $K(h+1)$ path controls, a count that grows with both.
	Each path control is instrumented by its own proxy, so the estimate is
	precise only if the proxies separate the treatment from every path value.
	These controls are aggregate series interacted with a common exposure, so the
	variation that separates them is time-series variation. When the proxies
	measure the confounder shocks with substantial error, the matrix of sample
	cross moments between the instruments and the regressors can be nearly
	singular in a given sample, and this risk rises with the number of path
	controls. This motivates an alternative design that does not need to separate
	the path values one by one.

	\subsubsection{Composite HELP-IV}
	\label{sec:additional-composite-help}

	We construct this alternative by replacing the instrumented control for each
	confounder at each date with a single instrument that moves $G_t$ but leaves
	the confounder path unchanged. The idea follows \citet{simszha2006md}, who hold
	a policy variable on a given path through a sequence of policy shocks. Here the
	instrument combines the treatment proxy with dated confounder proxies that
	offset every movement the treatment shock causes in the confounder path.
	Because the path does not move with the instrument, the first stage and the
	reduced form are the movements in $G_t$ and in $Y_{i,t+h}$ along the fixed
	path, so the regression needs no path control. Their ratio is therefore
	$\beta_h^P$. We refer to this design as composite HELP-IV, because its instrument
	is a composite of the treatment proxy and the dated confounder proxies.

	To implement this design, we form, for each horizon $h$, the instrument
	\begin{align}
		\mathrm z_{g,t,h}^C
		&=
		z_{g,t}
		+
		\sum_{s=0}^{h}
		\boldsymbol w_s'\boldsymbol z_{r,t+s},
		\label{eq:joint-composite-instrument}
	\end{align}
	where $z_{g,t}$ is the treatment proxy of equation~\eqref{eq:g-proxy},
	$\boldsymbol z_{r,t+s}$ collects the confounder proxies of
	equation~\eqref{eq:r-proxy} dated $t+s$, and $\boldsymbol w_s$ is the $K\times1$
	vector of offset weights on them. The weights are chosen so that the instrument
	leaves the joint confounder path
	$\mathbf R_{t,h}=(\mathbf R_t',\ldots,\mathbf R_{t+h}')'$ unchanged,
	\begin{align}
		\E\left[\mathbf R_{t,h}\mathrm z_{g,t,h}^C\right]
		&=
		\mathbf0.
		\label{eq:joint-composite-orthogonality}
	\end{align}
	With a single confounder, this instrument is the exposure-weighted combined
	proxy of Section~\ref{sec:help-why} without the factor $\sig$, and requirement
	\eqref{eq:joint-composite-orthogonality} is condition
	\eqref{eq:help-weights-path}.
	Since each confounder proxy dated $t+s$ covaries only with the confounder
	shock dated $t+s$, and that shock does not move the confounders realized before
	$t+s$, requirement \eqref{eq:joint-composite-orthogonality} is a triangular
	linear system in the weights. Specifically, its $K$ equations for date
	$t+\ell$ involve only $\boldsymbol w_0,\ldots,\boldsymbol w_\ell$, and the
	weights $\boldsymbol w_\ell$ on the proxies dated $t+\ell$ enter them only
	through the impact covariance $\Sigma_{rz,0}$. Because $\Sigma_{rz,0}$ is
	nonsingular, this system has a unique solution, obtained recursively from
	the impact horizon,
	\begin{align}
		\boldsymbol w_\ell
		&=
		-\Sigma_{rz,0}^{-1}
		\left(
		\mathbf v_\ell
		+
		\sum_{s=0}^{\ell-1}
		\Sigma_{rz,\ell-s}\boldsymbol w_s
		\right),
		\qquad
		\ell=0,\ldots,h,
		\label{eq:composite-weights}
	\end{align}
	where $\mathbf v_\ell=\E[\mathbf R_{t+\ell}z_{g,t}]$ is the covariance of the
	confounders at horizon $\ell$ with the treatment proxy and
	$\Sigma_{rz,\ell}=\E[\mathbf R_{t+\ell}\boldsymbol z_{r,t}']$ is their covariance with
	their own proxies.

	The weight $\boldsymbol w_\ell$ therefore offsets, across all
	$K$ confounders at once, the movement in $\mathbf R_{t+\ell}$ that remains after
	the treatment proxy and the offsets at earlier dates. With these weights, composite HELP-IV estimates
	\begin{align}
		Y_{i,t+h}
		&=
		b_h^{\mathrm{HELP\text{-}C}}\sig G_t
		+
		\mu_{i,h}
		+
		\lambda_{t,h}
		+
		\xi_{i,t,h}^{C},
		\label{eq:joint-composite-help}
	\end{align}
	using $\sig\mathrm z_{g,t,h}^C$ as the instrument for $\sig G_t$. The
	weights are determined by the moments $\mathbf v_\ell$ and $\Sigma_{rz,\ell}$, which can be
	estimated in two ways. With proxies, these moments are the sample cross moments of the
	aggregates and the proxies. When proxies are unavailable, shocks estimated from an SVAR
	replace the proxies in instrument~\eqref{eq:joint-composite-instrument}, and these moments
	are implied by the estimated impulse responses and shock variances. In both cases, the
	estimated weights define the instrument $\widehat{\mathrm z}_{g,t,h}^C$.
	Section~\ref{sec:composite-estimation} derives its inference with proxies, and
	Appendix~\ref{app:svar-composite} derives it with SVAR shocks.

	The following proposition establishes that
	$b_h^{\mathrm{HELP\text{-}C}}$ identifies the partial-equilibrium effect.

	\begin{proposition}[Composite HELP-IV]
		\label{prop:composite-conf}
		Suppose the conditions of Proposition~\ref{prop:help-identification} hold
		for \eqref{eq:outcome-conf} and \eqref{eq:exposure-conf}, with $\E[R_tz_{r,t}]\neq0$ replaced by
		$\det\Sigma_{rz,0}\neq0$. Then the weights satisfying
		\eqref{eq:joint-composite-orthogonality} exist and are unique. If, in
		addition, $\E[G_t\mathrm z_{g,t,h}^C]\neq0$, the population coefficient on
		$\sig G_t$ in \eqref{eq:joint-composite-help} satisfies
		\begin{align}
			\label{eq:joint-composite-identifies}
			b_h^{\mathrm{HELP\text{-}C}}
			&=
			\beta_h^P.
		\end{align}
	\end{proposition}

	The proof is in Appendix~\ref{app:composite-conf}. The condition
	$\E[G_t\mathrm z_{g,t,h}^C]\neq0$ in the proposition is implied by
	$\breve G_0\neq0$, because this covariance is proportional to $\breve G_0$. Composite HELP-IV and
	multiple-confounder HELP-IV use the same weights in population and differ only in
	how they estimate them. The argument of Section~\ref{sec:help-why} extends to $K$
	confounders and applies equally to sample moments, so the sample coefficient of
	multiple-confounder HELP-IV equals the IV coefficient obtained with $\sig$ times an
	instrument of the form in equation~\eqref{eq:joint-composite-instrument}. The weights of this
	instrument set its sample cross moment with every path control to zero, after the
	unit and time effects are partialed out, and in population they solve requirement
	\eqref{eq:joint-composite-orthogonality}.

	The two designs differ in finite samples. Multiple-confounder HELP-IV performs
	the offsetting implicitly, estimating the weights jointly with the effect of
	interest, so its instrument is exactly uncorrelated with the path controls in the
	sample. Composite HELP-IV instead estimates the weights in a separate first step,
	from recursion \eqref{eq:composite-weights}. This recursion imposes the causality
	restrictions $\E[\mathbf R_{t+\ell}\boldsymbol z_{r,t+s}']=\mathbf0$ for $s>\ell$
	in place of their sample counterparts, so the composite instrument remains
	correlated with the path controls in the sample. The composite estimate therefore
	absorbs part of the dependence of the outcome on the confounder path, in proportion
	to the path coefficients $c_{q,h,s}$ and to this remaining correlation. This part
	vanishes as the number of periods grows.

	Each design has its own advantages. Composite HELP-IV is more stable when the
	proxies are noisy and the number of path controls $K(h+1)$ is large. Because
	composite HELP-IV estimates the offset weights in a separate first step, it keeps
	a single endogenous regressor for any $K$ and $h$ and inverts only the $K\times K$
	impact covariance $\Sigma_{rz,0}$ in recursion \eqref{eq:composite-weights}.
	Multiple-confounder HELP-IV instead inverts the $[1+K(h+1)]\times[1+K(h+1)]$
	matrix of cross moments between its instruments and its regressors, which can then
	be nearly singular, so that some estimates take extreme values. By contrast, when
	the proxies are accurate and this matrix is well conditioned, multiple-confounder
	HELP-IV is more precise. Multiple-confounder HELP-IV sets the sample cross moments
	of its instrument with the path controls exactly to zero and thus avoids the
	finite-sample error that the composite estimate incurs by absorbing part of the
	dependence of the outcome on the confounder path. Multiple-confounder HELP-IV is
	also a standard one-step IV regression. Therefore, the two designs complement each
	other.

	\subsubsection{Targeting-rule HELP-IV}
	\label{sec:combined-conf}
	\label{sec:additional-one-way}

Multiple-confounder HELP-IV and composite HELP-IV hold the path of every
confounder fixed and therefore require a proxy for each confounder shock. This
requirement is difficult to meet in some applications. This
	subsection relaxes it by requiring shock proxies only for the two-way
	confounders, whose shocks move the treatment. The one-way confounders,
	whose shocks move neither the treatment nor a two-way confounder
	(Assumption~\ref{ass:one-way-block} below), require no proxies and can remain unspecified. By
	Proposition~\ref{lemma:hedge-bias}, a one-way confounder can generate only
	the exposure-correlation component of the HEDGE bias, which arises when
	exposure to the confounder is correlated with exposure to the treatment and
	the treatment shock moves the confounder. To correct this component, we draw
	on the targeting-rule approach of \citet{manuel2026}, which removes the first
	of these two conditions and was originally developed for general-equilibrium
	biases in shift-share regressions.\footnote{The target
		of \citet{manuel2026} is the direct effect of the treatment along its
		equilibrium path, an object that is well defined whether the
		general-equilibrium aggregates are one-way or two-way. Our target
		$\beta_h^P$ is instead evaluated along the treatment path obtained when
		the confounder paths are held fixed. The two treatment paths coincide for
		one-way confounders but generally differ for two-way ones, so the targeting
		rule serves our target only for one-way confounders.}

	We partition the confounders into a two-way block and a one-way
	block. We index the two-way confounders by
	$q=1,\ldots,K_1$ and the one-way confounders by $q=K_1+1,\ldots,K$.
	Equation~\eqref{eq:exposure-conf} is maintained only for the two-way block,
	and we impose two assumptions, the first on the one-way block.

	\begin{assumption}[One-way confounders]
		\label{ass:one-way-block}
		For $q>K_1$, $j\leq K_1$, and $\ell\geq0$,
		\[
		G_\ell^{r_q}=0,\qquad R_{j,\ell}^{r_q}=0,
		\]
		where $G_\ell^{r_q}=\E[G_{t+\ell}\epsilon_{r_q,t}]/\E[\epsilon_{r_q,t}^2]$
		and $R_{j,\ell}^{r_q}=\E[R_{j,t+\ell}\epsilon_{r_q,t}]/\E[\epsilon_{r_q,t}^2]$.
		The matrix $(R_{q',0}^{r_q})_{q,q'>K_1}$ is nonsingular.
	\end{assumption}

	In words, the shocks of the one-way block affect neither the treatment nor any
	two-way confounder at any horizon, so offsetting them cannot change the
	controlled treatment path, either directly or through the two-way confounders.
	Because the impact matrix of this block is nonsingular, its paths can be offset
	without changing the two-way block.

	The second assumption is a targeting rule, which governs the
	cross-sectional allocation of treatment exposure.

	\begin{assumption}[Targeting rule]
		\label{ass:targeting}
		Let $\mathbf a_i$ be a vector of observed, predetermined, time-invariant
		characteristics that includes a constant and is independent of the
		aggregate process, with $\E[\mathbf a_i\mathbf a_i']$
		nonsingular, and let
		$\sig^\perp=\sig-\mathbf a_i'\E[\mathbf a_i\mathbf a_i']^{-1}\E[\mathbf a_i\sig]$.
		The residual satisfies $\Var(\sig^\perp)>0$ and $\E[\sig^\perp s_{ir_q}]=0$ for
		$q=K_1+1,\ldots,K$.
	\end{assumption}

	Adapting the identifying condition of \citet{manuel2026}, we obtain a
	sufficient condition for Assumption~\ref{ass:targeting}. Consider the
	conditional expectation of treatment exposure given $\mathbf a_i$ and the
	unit characteristics that determine exposure to the confounders. If this
	expectation is linear in $\mathbf a_i$ and does not depend on the other
	characteristics, then $\sig^\perp$ is orthogonal to
	any function of those characteristics, including exposure to each one-way
	confounder. The assumption therefore holds for all one-way confounders
	simultaneously, without requiring the researcher to specify them.

	Building on these assumptions, we identify $\beta_h^P$ by combining HELP-IV
	for the two-way block with the targeting rule of \citet{manuel2026} for the
	one-way block. At horizon $h$, we estimate
	\begin{align}
		Y_{i,t+h}
		&=
		b_h^{\mathrm{HELP\text{-}T}}\sig G_t
		+
		\sum_{q=1}^{K_1}
		\sum_{s=0}^{h}
		c_{q,h,s}\sig R_{q,t+s}
		+
		\mu_{i,h}
		+
		\lambda_{t,h}
		+
		\xi_{i,t,h}^{\mathrm T},
		\label{eq:targeted-one-way-regression}
	\end{align}
	and instrument $\sig G_t$ and
	$\{\sig R_{q,t+s}\}_{q\leq K_1,\;s=0,\ldots,h}$ by $\sig^\perp z_{g,t}$ and
	$\{\sig^\perp z_{r_q,t+s}\}_{q\leq K_1,\;s=0,\ldots,h}$, respectively. We refer to this design as targeting-rule HELP-IV, which requires no confounder proxy when
	$K_1=0$, because the path-control block is then empty. Because the
	instruments use only $\sig^\perp$, Assumption~\ref{ass:targeting} removes
	the one-way confounders from every IV moment. The design thus reduces to
	multiple-confounder HELP-IV for the two-way block, with $\sig^\perp$ in
	place of $\sig$ in the instruments, and its identification follows from
	Corollary~\ref{cor:vec-conf}. Assumption~\ref{ass:one-way-block}, in
	turn, ensures that holding the paths of the one-way confounders fixed would
	not alter the treatment path, so the reduced design still identifies
	$\beta_h^P$. The following corollary formalizes this result, and its proof is in
	Appendix~\ref{app:target-identification}.

	\begin{corollary}[Targeting-rule HELP-IV]
		\label{cor:combined-conf}
		Suppose \eqref{eq:svma-dyn} and \eqref{eq:outcome-conf} hold for all $K$
		confounders and the remaining conditions of Corollary~\ref{cor:vec-conf}
		hold for the first $K_1$ confounders, with $\sig^\perp$ in place of $\sig$
		in the instruments, and the exposures $s_{ir_q}$, $q>K_1$, are
		predetermined and independent of the aggregate process. Under Assumptions~\ref{ass:one-way-block}
		and~\ref{ass:targeting}, the population IV coefficient in
		\eqref{eq:targeted-one-way-regression} satisfies
		$b_h^{\mathrm{HELP\text{-}T}}=\beta_h^P$.
	\end{corollary}

	Using $\sig^\perp$ in the instruments is equivalent in population to
	controlling for $\mathbf a_i$ interacted with time effects. With $K_1=0$,
	targeting-rule HELP-IV is therefore the method of \citet{manuel2026}, applied
	with a treatment shock proxy as the instrument, and
	Corollary~\ref{cor:combined-conf} shows that this method identifies
	$\beta_h^P$ when no confounder feeds back into the treatment, because the
	equilibrium treatment path then coincides with the controlled path.

	\subsection{Forward-looking environment}
	\label{sec:forward-portable-section}
	\label{sec:forward-portable-target}

	The preceding sections maintain a backward-looking outcome equation in the
	data-generating process, in which the outcome depends only on current and
	lagged aggregates. In many macroeconomic and finance models, however,
	current outcomes can depend on expected future aggregates. Identifying the
	partial-equilibrium effect then requires holding the confounder path fixed both in expectation and ex post. In this subsection, we develop
	this extension for a single exposure margin and a single confounder, which
	isolates the role of expectations.\footnote{The method
		extends to multiple confounders by stacking their paths and their
		contemporaneous and news shocks. It extends to multiple exposure margins
		by interacting the composite instrument introduced below with each
		margin, since the offset is common to all margins.} We first state how the data-generating process changes and define the
	forward-looking partial-equilibrium effect. We then show how composite
	HELP-IV can be extended to this environment.

	The forward-looking environment differs from the backward-looking
	environment of Section~\ref{sec:dyn-general} in the outcome equation and
	in the evolution of the aggregate process, while the exposure restriction
	in equation~\eqref{eq:exposure-dyn}, which relates confounder exposure $\sir$ to
	treatment exposure $\sig$, is unchanged. First, the outcome also responds
	to anticipated future values of both aggregates,
	\begin{align}
		\label{eq:outcome-forward}
		Y_{it}
		&=
		\mu_i+\lambda_t+\rho Y_{i,t-1}
		+\sig\left(
		\sum_{\ell=0}^{p}\beta_\ell G_{t-\ell}
		+\sum_{f=1}^{F}\zeta_f^{g}\E_tG_{t+f}
		\right)
		\nonumber\\
		&\quad
		+\sir\left(
		\sum_{\ell=0}^{p}\gamma_\ell R_{t-\ell}
		+\sum_{f=1}^{F}\zeta_f^{r}\E_tR_{t+f}
		\right)
		+u_{it},
		\qquad
		|\rho|<1,
	\end{align}
	where $\E_t[\cdot]$ denotes the expectation conditional on aggregate
	information at date $t$. Second, the aggregates are also driven by news
	shocks about the confounder, so that $\mathbf W_t=(G_t,R_t)'$ satisfies
	\begin{align}
		\label{eq:svma-forward}
		\mathbf W_t
		&=
		\sum_{j=0}^{\infty}
		\Psi_j\boldsymbol\epsilon_{t-j},
		\qquad
		\boldsymbol\epsilon_t
		=
		\left(
		\epsilon_{g,t},
		\boldsymbol\epsilon_{r,t}^{E\prime},
		\boldsymbol\epsilon_{o,t}'
		\right)',
	\end{align}
	where $\boldsymbol\epsilon_{r,t}^{E}
	=(\epsilon_{r,t},\epsilon_{r,t,1}^{E},\ldots,\epsilon_{r,t,d}^{E})'$
	collects the contemporaneous shock $\epsilon_{r,t}$ and $d$ news shocks
	for which proxies are available, and $\boldsymbol\epsilon_{o,t}$
	collects the remaining shocks. The news shocks are realized at date $t$ and
	move expectations of future values of $R$. The
	shocks satisfy the conditions of equation~\eqref{eq:svma-dyn}, with $G_0^g=1$ and
	each element of $\boldsymbol\epsilon_{r,t}^{E}$ normalized to move $R_t$
	one for one on impact or, for a news shock that does not move $R_t$ on impact, to move
	contemporaneously an observed variable one for one.

The forward-looking partial-equilibrium effect is the response of $Y_{i,t+h}$ to a movement in $G_t$ caused by $\epsilon_{g,t}$ when the confounder path is held fixed both ex post and in the expectations formed at dates $t$ through $t+h$. As in equation~\eqref{eq:portable-response-form} of Section~\ref{sec:donaldson-connection}, we express this effect in terms of shock responses. It divides the outcome response to an experiment that combines the treatment shock with offsetting shocks to $R$ by the impact movement of $G_t$ under the same experiment. Relative to the backward-looking case, the experiment must satisfy two additional requirements. First, because $Y_{i,t+h}$ depends on expectations formed at date $t+h$, the offset must cover $R_t,\ldots,R_{t+H}$ for $H\geq\bar h+F$, where $\bar h$ is the largest horizon of interest. Second, because shocks realized after $t$ cannot alter the expectations formed at $t$, the offsetting shocks must be dated $t$, as in \citet{mckaywolf2023}. A single contemporaneous shock moves the path along one impulse response and
therefore cannot offset an arbitrary path response over $H+1$ dates. The
experiment therefore also requires news shocks, which are realized at date $t$
but move the expected confounder at later dates.
Let $\mathbf r_H^g$ and $\mathbf R_H^E$ denote the responses of the path $R_t,\ldots,R_{t+H}$ to $\epsilon_{g,t}$ and to the elements of $\boldsymbol\epsilon_{r,t}^{E}$. The offsetting combination $\mathbf e_H^E$ of the date-$t$ shocks then holds this path fixed, $\mathbf r_H^g+\mathbf R_H^E\mathbf e_H^E=\mathbf0$, when such a combination exists. Otherwise, it minimizes the remaining movement of the path, as we make precise below.

	Under this experiment, the outcome moves by
	$b_h+(\mathbf e_H^E)'\mathbf c_h^E$, where $b_h$ and $\mathbf c_h^E$ are the
	outcome responses to $\epsilon_{g,t}$ and to
	$\boldsymbol\epsilon_{r,t}^{E}$, defined by the population projection
	\begin{align}
		\label{eq:forward-decomposed-regression}
		Y_{i,t+h}
		&=
		b_h\sig\epsilon_{g,t}
		+
		\left(\mathbf c_h^E\right)'
		\sig\boldsymbol\epsilon_{r,t}^{E}
		+
		\mu_{i,h}
		+
		\lambda_{t,h}
		+
		e_{i,t,h}.
	\end{align}
	The treatment moves on impact by
	$\breve G_{0,H}^{E}=1+(\mathbf g_0^E)'\mathbf e_H^E$, where
	$\mathbf g_0^E$ collects the impact responses of $G_t$ to
	$\boldsymbol\epsilon_{r,t}^{E}$. For given weights $\mathbf e_H^E$, the experiment therefore yields the ratio of the outcome response
	to the impact movement of the treatment,
	\begin{align}
		\label{eq:forward-portable-vector-form}
		\beta_h^{E}
		&=
		\left(\breve G_{0,H}^{E}\right)^{-1}
		\left[
		b_h+\left(\mathbf e_H^E\right)'\mathbf c_h^E
		\right].
	\end{align}

	We extend composite HELP-IV of
	Section~\ref{sec:additional-composite-help} to identify $\beta_h^E$. For
	each $h=0,\ldots,\bar h$, the extension estimates the same regression as in
	that section,
	\begin{align}
		\label{eq:forward-composite-help}
		Y_{i,t+h}
		&=
		b_h^{\mathrm{E\text{-}HELP}}\sig G_t
		+
		\mu_{i,h}
		+
		\lambda_{t,h}
		+
		\xi_{i,t,h}^{E},
	\end{align}
	and differs only in the instrument for $\sig G_t$. Let $\mathrm z_{g,t}$
	be a proxy for the treatment shock $\epsilon_{g,t}$, and let
	$\boldsymbol{\mathrm z}_{r,t}^{E}$ collect proxies for the
	contemporaneous shock and the $d$ news confounder shocks in
	$\boldsymbol\epsilon_{r,t}^{E}$. Each proxy is relevant for its own shock and
	exogenous, in the sense that it is orthogonal to all other shocks at all
	leads and lags, as in equations~\eqref{eq:g-proxy} and~\eqref{eq:r-proxy}
	and analogously for the news proxies. We write these proxies upright,
	because each is also normalized to have unit covariance with its own shock,
	by dividing it by its covariance with the variable that its shock moves one
	for one on impact. The
	composite instrument weights these proxies by the offsetting combination,
	\begin{align}
		\label{eq:forward-composite-instrument}
		\mathrm z_{g,t}^{E}
		&=
		\mathrm z_{g,t}
		+
		\left(\mathbf e_H^E\right)'
		\boldsymbol{\mathrm z}_{r,t}^{E},
	\end{align}
	and regression~\eqref{eq:forward-composite-help} instruments $\sig G_t$ by
	$\sig\mathrm z_{g,t}^{E}$. Because each proxy has unit covariance with its
	own shock, $\E[(R_t,\ldots,R_{t+H})'\mathrm z_{g,t}^{E}]=\mathbf r_H^g+\mathbf
	R_H^E\mathbf e_H^E$. Holding the path fixed is therefore the forward-looking
	counterpart of requirement~\eqref{eq:joint-composite-orthogonality}. This
	instrument reflects both requirements of the target experiment. First, its
	weights offset the confounder path through $t+H$. Second, it combines only
	proxies dated $t$, including news-shock proxies. Unlike the
	backward-looking composite instrument, which holds the path fixed only ex
	post, this instrument also offsets the path in the expectations
	formed at $t$. We refer to this design as E-HELP-IV.

The composite instrument in equation~\eqref{eq:forward-composite-instrument} is fully
specified once its weights $\mathbf e_H^E$ are determined. Holding the path
fixed requires $\mathbf r_H^g+\mathbf R_H^E\mathbf e_H^E=\mathbf0$, a system of
$H+1$ equations in $d+1$ weights, so the solution depends on the number of
available news shocks. If this number equals the number of future dates to be
offset, $d=H$, and $\mathbf R_H^E$ is nonsingular, the system has the unique
solution $\mathbf e_H^E=-(\mathbf R_H^E)^{-1}\mathbf r_H^g$. In most
applications, however, the news shocks are fewer, $d<H$, and the system has
more equations than weights. It then generically has no solution,
so we instead take the weights that minimize the remaining movement of the
path, given by the Moore--Penrose solution \citep{auclertetal2021},
\begin{align}
	\label{eq:forward-fixing-mp}
	\mathbf e_H^E
	&=
	-\left(\mathbf R_H^E\right)^+\mathbf r_H^g,
\end{align}
which reduces to the unique solution when $\mathbf R_H^E$ is nonsingular. In
practice, $\mathbf r_H^g$ and $\mathbf R_H^E$ are estimated, either from sample
cross moments of the aggregates with the proxies or from an SVAR, and rescaling
a proxy only rescales its weight. Composite HELP-IV treats the cases $d=H$ and $d<H$ within one framework. The benchmark HELP-IV design of Section~\ref{sec:dyn-identification} could also handle the case $d=H$, with the path controls $\sig R_t,\ldots,\sig R_{t+H}$ instrumented by the date-$t$ proxies, and is then exactly identified. With $d<H$, however, that design has fewer instruments than endogenous regressors and is under-identified. A pseudo-inverse solution of its moment conditions then depends on the units of the path controls. We therefore refer to the case $d=H$ with $\mathbf R_H^E$ nonsingular and the case $d<H$ as the exactly identified and the under-identified case of the offset.

As in the backward-looking case, the instrument applies the same shock
combination to the treatment and the outcome, so
$b_h^{\mathrm{E\text{-}HELP}}$ equals $\beta_h^E$ in
equation~\eqref{eq:forward-portable-vector-form} for any weights. Whether $\beta_h^E$ is
the forward-looking partial-equilibrium effect, however, depends on whether the
weights hold the path fixed. In the exactly identified case, the offset is complete, so $\beta_h^E$ is that
effect. In the under-identified case, the weights generally leave part
of the path unabsorbed, and $\beta_h^E$ is then the effect of an experiment that
offsets only the Euclidean projection of the path response on the column space of
$\mathbf R_H^E$, the part that the available date-$t$ shocks can offset. We refer
to $\beta_h^E$ in both cases as the information-constrained effect, because it is
constrained by the news shocks for which proxies are available, and it equals the
forward-looking partial-equilibrium effect when the offset is complete. We report the share of the path
response left unabsorbed,
$\|[\mathbf I_{H+1}-\mathbf R_H^E(\mathbf R_H^E)^+]\mathbf r_H^g\|^2
/\|\mathbf r_H^g\|^2$, as a diagnostic of how far this experiment departs from
the fixed-path experiment.\footnote{The share is zero if and only if
$\mathbf r_H^g$ lies in the column space of $\mathbf R_H^E$, in which case
$\beta_h^E$ is the forward-looking partial-equilibrium effect. The share alone
does not bound the difference between the two effects, because that difference
also depends on the outcome loadings and on the controlled first-stage
denominator.} The following proposition, whose proof is in
Appendix~\ref{app:forward-composite-help-proof}, covers both
cases.

	\begin{proposition}[Forward-looking E-HELP-IV]
		\label{cor:forward-composite-help}
		Suppose \eqref{eq:exposure-dyn}, \eqref{eq:outcome-forward}, and
		\eqref{eq:svma-forward} hold, the proxies satisfy the relevance,
		exogeneity, and normalization conditions above, the exposure-weighted
		proxies satisfy the panel exclusion restrictions, $\Var(\sig)>0$, and the
		weights in
		\eqref{eq:forward-composite-instrument} are \eqref{eq:forward-fixing-mp}.
		If $\breve G_{0,H}^{E}\neq0$, the population coefficient on $\sig G_t$
		in \eqref{eq:forward-composite-help}, with instrument
		$\sig\mathrm z_{g,t}^{E}$, satisfies
		$b_h^{\mathrm{E\text{-}HELP}}=\beta_h^{E}$, where $\beta_h^{E}$ is given by
		\eqref{eq:forward-portable-vector-form}.
	\end{proposition}

	The decomposition method
	of \citet{donaldson2026} has an analogous ex ante form for this environment,
	which also offsets the expected confounder path with date-$t$ shocks but, like its
	backward-looking form, normalizes by the unrestricted treatment response rather
	than the controlled one.

	\subsection{Transformed treatments and outcomes}
	\label{sec:generalized-target}

	Sections~\ref{sec:portable-markov}--\ref{sec:multiple-aggregate-confounders} focus on the identification target $\beta_h^P$, the
	ratio of the response of $Y_{i,t+h}$ to that of $\sig G_t$ under the
	fixed-path experiment. In many applications, however, the treatment of
	interest is a unit-level variable rather than the exposure-weighted
	aggregate $\sig G_t$, and both the outcome and the treatment may be
	transformations that combine multiple dates within the window. The target
	is then the ratio of the response of a transformed outcome
	$\mathcal Y_{i,t+h}$ to that of a transformed unit-level treatment
	$\mathcal G_{i,t+h_{\mathcal G}}$, with $h_{\mathcal G}\leq h$, under the
	same fixed-path experiment, and we denote it by $\mathcal B_h^{P}$. For example, the
	state-level cumulative fiscal multiplier of \citet{donaldson2026}, which applies
	the cumulative definition of \citet{ramey2018} to the design of
	\citet{nakamurasteinsson2014}, is the ratio of state
	output cumulated from $t$ through $t+h$ to state government spending cumulated over the same window.

	The data-generating process retains the outcome equation, the aggregate
	process, and the exposure restriction of Section~\ref{sec:portable-markov}, with a single exposure margin and a single confounder.
	To define $\mathcal B_h^{P}$, we complete it with a transformed outcome and a
	unit-level treatment process. The transformed outcome is a linear
	combination of the outcomes $Y_{i,t},\ldots,Y_{i,t+h}$,
	\begin{align}
		\label{eq:outcome-transformed}
		\mathcal Y_{i,t+h}
		&=
		\sum_{j=0}^{h}\upsilon_{i,j}Y_{i,t+j}+\mathcal Y^{0}_{i,t-1},
	\end{align}
	with weights $\upsilon_{i,j}$ that may differ across units and a
	predetermined term $\mathcal Y^{0}_{i,t-1}$. The weights and the
	predetermined term are measurable with respect to the information
	$\mathcal F_{t-1}$ available at date $t-1$, which contains the history of
	the aggregates, the shocks, and the unit outcomes, together with the
	predetermined unit loadings. The treatment process $\mathcal G_{i,t+h_{\mathcal G}}$ is specified in Assumption~\ref{ass:generalized}, under which it is determined by the shocks dated $t$ through $t+h_{\mathcal G}$, by the state inherited from before the window, and by unit-specific disturbances. The
	horizon $h_{\mathcal G}\leq h$ records when the treatment is measured,
	ranging from zero for a treatment dated $t$ to $h$ for a treatment that
	spans the outcome window.

	\begin{assumption}[The treatment process and the instruments]
		\label{ass:generalized}
		\begin{enumerate}[label=(\roman*),leftmargin=2em]
			\item The transformed treatment satisfies
			\begin{align}
				\label{eq:treatment-linear}
				\mathcal G_{i,t+h_{\mathcal G}}
				&=
				\underbrace{
					\sum_{j=0}^{h_{\mathcal G}}\mathbf d_{i,j}'\boldsymbol\epsilon_{t+j}
				}_{\text{aggregate component}}
				+
				\underbrace{
					\mathcal G^{0}_{i,t-1}
				}_{\text{predetermined component}}
				+
				\underbrace{
					\omega_{i,t+h_{\mathcal G}}
				}_{\text{idiosyncratic component}},
			\end{align}
			where $\mathbf d_{i,j}=(d_{i,j,g},d_{i,j,r})'$ and $\mathcal G^{0}_{i,t-1}$
			are $\mathcal F_{t-1}$-measurable, and $\omega_{i,t+h_{\mathcal G}}$ is
			uncorrelated with $\sig$, with $\boldsymbol\epsilon_{t+j}$ for all
			$j\leq h$, and with $\eta_{g,t+s}$ and $\eta_{r,t+s}$ for
			$0\leq s\leq h$.
			\item For $x,y\in\{g,r\}$ and $0\leq s,j\leq h$,
			\[
			\E[z_{x,t+s}\mid\mathcal F_{t-1},\mathcal H_i]=0,\qquad
			\E[z_{x,t+s}\epsilon_{y,t+j}\mid\mathcal F_{t-1},\mathcal H_i]
			=\lambda_x{\bf1}\{x=y,\ s=j\},
			\]
			where $\mathcal H_i$ collects the predetermined unit loadings and
			$\lambda_x=\E[z_{x,t}\epsilon_{x,t}]\neq0$.
			\item The exposure-weighted proxies satisfy the panel exclusion
			restrictions of Section~\ref{sec:twfe-hedge} with respect to $u_{it}$,
			the transformed disturbance $\sum_{j=0}^{h}\upsilon_{i,j}u_{i,t+j}$, and
			$\omega_{i,t+h_{\mathcal G}}$.
		\end{enumerate}
	\end{assumption}

	With $\boldsymbol\epsilon_{t+j}$ collecting the structural shocks, the
	identification target is the generalized partial-equilibrium effect
	\begin{align}
		\label{eq:generalized-target}
		\mathcal B_h^{P}
		&:=
		\left.
		\frac{
			\partial\mathcal Y_{i,t+h}/\partial\epsilon_{g,t}
		}{
			\partial\mathcal G_{i,t+h_{\mathcal G}}/\partial\epsilon_{g,t}
		}
		\right|_{R_{t+s}\ \text{fixed},\;s=0,\ldots,h}.
	\end{align}
	We assume that this ratio is common across units, a restriction that
	Proposition~\ref{prop:generalized} uses. The definition nests two cases of
	interest. It reduces to the benchmark target, $\mathcal B_h^{P}=\beta_h^P$, when
	$h_{\mathcal G}=0$, $\mathcal G_{i,t}=\sig G_t$, and
	$\mathcal Y_{i,t+h}=Y_{i,t+h}$. It also covers the cumulative fiscal
	multiplier above.\footnote{That multiplier uses
		$\mathcal Y_{i,t+h}=\sum_{j=0}^{h}(Y_{i,t+j}-Y_{i,t-1})/Y_{i,t-1}$ and
		$\mathcal G_{i,t+h}=\sum_{j=0}^{h}(G_{i,t+j}-G_{i,t-1})/Y_{i,t-1}$, so
		$\upsilon_{i,j}=Y_{i,t-1}^{-1}$, $\mathcal Y^{0}_{i,t-1}=-(h+1)$, and
		$h_{\mathcal G}=h$. Projecting each $G_{i,t+j}$ on the shocks dated $t$
		through $t+j$, conditional on $\mathcal F_{t-1}$, splits
		$\mathcal G_{i,t+h}$ into the three components of equation
		\eqref{eq:treatment-linear}. This projection does not establish the linear
		representation or the instrument restrictions of
		Assumption~\ref{ass:generalized}, nor the common controlled ratio in
		equation~\eqref{eq:generalized-target}, which remain maintained
		assumptions.} When the treatment spans the window ($h_{\mathcal G}>0$), the
	denominator of equation~\eqref{eq:generalized-target} is the controlled
	response of $\mathcal G_{i,t+h_{\mathcal G}}$, which generally differs from
	its unrestricted response even when the confounder feeds back into the
	treatment only with a lag. A method that normalizes by the unrestricted
	treatment response, such as the decomposition method of
	Section~\ref{sec:donaldson-connection}, then differs from $\mathcal B_h^P$
	under lagged feedback as well.

	We identify $\mathcal B_h^{P}$ by applying the HELP-IV design of
	Section~\ref{sec:dyn-identification} to the transformed pair,
	\begin{align}
		\label{eq:lpiv-generalized}
		\mathcal Y_{i,t+h}
		&=
		\mathcal B_h^{\mathrm{HELP}}\mathcal G_{i,t+h_{\mathcal G}}
		+
		\sum_{s=0}^{h}
		c_{h,s}\sig R_{t+s}
		+
		\mu_{i,h}
		+
		\lambda_{t,h}
		+
		\xi^{\mathcal G}_{i,t,h},
	\end{align}
	in which the regressor vector and its instruments are
	\[
	\boldsymbol X^{\mathcal G}_{it,h}
	=(\mathcal G_{i,t+h_{\mathcal G}},\sig R_t,\ldots,\sig R_{t+h})',
	\qquad
	\mathbf Z_{it,h}=\sig(z_{g,t},z_{r,t},\ldots,z_{r,t+h})'.
	\]
	The following
	proposition establishes that $\mathcal B_h^{\mathrm{HELP}}$ identifies $\mathcal B_h^{P}$.

	\begin{proposition}[Identification and consistency]
		\label{prop:generalized}
		Fix $h\geq0$. Suppose the model of Section~\ref{sec:portable-markov} and
		Assumption~\ref{ass:generalized} hold, the ratio in
		\eqref{eq:generalized-target} is common across units, $\Var(\sig)>0$, $\E[R_tz_{r,t}]\neq0$, and
		$\mathcal D_h^{\mathcal G}\neq0$, where $\mathcal D_h^{\mathcal G}$ is the
		covariance of $\dot s_{ig}=\sig-\E[\sig]$ with the controlled response of
		$\mathcal G_{i,t+h_{\mathcal G}}$ in the denominator of
		\eqref{eq:generalized-target}.
		\begin{enumerate}[label=(\roman*),leftmargin=2em]
			\item The population moment system
			$\E[\widetilde{\mathbf Z}_{it,h}\xi^{\mathcal G}_{i,t,h}]=\mathbf0$ has a
			unique solution, whose first element satisfies
			\begin{align}
				\label{eq:help-identifies-generalized}
				\mathcal B_h^{\mathrm{HELP}}
				&=
				\mathcal B_h^{P}.
			\end{align}
			\item If, in addition, the aggregate block, the proxies, and the unit
			loadings are jointly stationary and ergodic, the variables in
			\eqref{eq:lpiv-generalized} have finite second moments, and the
			residualized panel moments converge to their population counterparts,
			then $\widehat{\mathcal B}_h^{\mathrm{HELP}}\xrightarrow{p}\mathcal B_h^P$
			as $T\to\infty$.
		\end{enumerate}
	\end{proposition}

	The proof is in Appendix~\ref{app:generalized}. Intuitively, under
	Assumption~\ref{ass:generalized} and the specification of
	$\mathcal Y_{i,t+h}$, each transformation combines realizations at dates
	within the response window, with unit-specific weights and loadings, and adds components determined
	before the window opens. The dated path instruments fix the confounder path
	throughout the window, so the design recovers the controlled response of the
	within-window combination, including a cumulative sum. The weights
	$\upsilon_{i,j}$ enter the numerator and the loadings $\mathbf d_{i,j}$ the
	denominator of each unit's ratio, and the common ratio in
	equation~\eqref{eq:generalized-target} ensures that the ratio does not vary
	across units. The design also removes the components $\mathcal Y^{0}_{i,t-1}$ and
	$\mathcal G^{0}_{i,t-1}$, because they are orthogonal to the instruments, which are unpredictable
	from pre-window information. The
	coefficient on $\mathcal G_{i,t+h_{\mathcal G}}$ is therefore the ratio of the
	controlled responses of $\mathcal Y_{i,t+h}$ and
	$\mathcal G_{i,t+h_{\mathcal G}}$, namely $\mathcal B_h^{P}$. The same design extends to the environments of Sections~\ref{sec:multiple-exposure-margins}--\ref{sec:forward-portable-section} once its instruments are replaced. With multiple exposure margins or confounders, the instruments of Sections~\ref{sec:multiple-exposure-margins} and~\ref{sec:multiple-aggregate-confounders} apply. In the forward-looking environment of Section~\ref{sec:forward-portable-section}, the composite instrument in equation~\eqref{eq:forward-composite-instrument} replaces the dated proxies, as in the fiscal application of Section~\ref{sec:app-donaldson}.

	\subsubsection{Aggregate shifters and omitted-history bias}
	\label{sec:gen-orth}

	Previous applied work estimates targets defined on transformed treatments and
	outcomes with a shift-share IV regression
	\citep{nakamurasteinsson2014,chodorowreich2019,donaldson2026},
	\begin{align}
		\label{eq:lp-twfe-generalized}
		\mathcal Y_{i,t+h}
		&=
		\mathcal B_h\mathcal G_{i,t+h_{\mathcal G}}
		+
		\mu_{i,h}
		+
		\lambda_{t,h}
		+
		v_{i,t,h},
	\end{align}
	in which the transformed treatment is instrumented by the shift-share
	variable $\sig G_t$ and the path controls are omitted. We denote the
	resulting IV coefficient by $\mathcal B_h^{\mathrm{SS}}$. This regression is a
	TWFE-IV local projection, but its treatment instrument is the
	exposure-weighted aggregate treatment itself rather than the
	exposure-weighted proxy of Section~\ref{sec:twfe-hedge}. In what follows, we first
	characterize the bias of $\mathcal B_h^{\mathrm{SS}}$, which in addition to the
	HEDGE bias contains a component that arises because the shifter $G_t$ is
	not an innovation. We then explain why the HELP-IV
	design in regression~\eqref{eq:lpiv-generalized} removes both components and how a
	treatment instrument that is unpredictable from pre-window information can
	be obtained in practice.

	To characterize the bias of $\mathcal B_h^{\mathrm{SS}}$, we compare it with
	$\mathcal B_h^{\mathrm{TWFE}}$, the coefficient in regression~\eqref{eq:lp-twfe-generalized}
	when the treatment is instead instrumented by the proxy $\sig z_{g,t}$ of
	equation~\eqref{eq:g-proxy}, as in the TWFE-IV design of
	Section~\ref{sec:twfe-hedge}. Because the proxy is a date-$t$ innovation,
	the bias of $\mathcal B_h^{\mathrm{TWFE}}$ is the HEDGE bias alone. The comparison
	therefore isolates the additional bias from the shift-share variable, the
	\emph{omitted-history bias}. Write
	$G_t=\sum_{\varsigma\in\{g,r\}}\sum_{s\geq0}\psi^{G}_{\varsigma,s}\epsilon_{\varsigma,t-s}$,
	with $\psi^{G}_{\varsigma,s}=[\Psi_s]_{G,\varsigma}$ from
	equation~\eqref{eq:svma-dyn} and $\sigma_\varsigma^2=\Var(\epsilon_{\varsigma,t})$,
	let
	$\mathcal S^{\mathcal Y}_{\varsigma,\ell}=\E[\dot s_{ig}\,\partial\mathcal Y_{i,t+h}/\partial\epsilon_{\varsigma,t+h-\ell}]$
	and
	$\mathcal S^{\mathcal G}_{\varsigma,\ell}=\E[\dot s_{ig}\,\partial\mathcal G_{i,t+h_{\mathcal G}}/\partial\epsilon_{\varsigma,t+h-\ell}]$
	denote the exposure-weighted responses of the transformed outcome and
	treatment to $\epsilon_{\varsigma,t+h-\ell}$ without holding the confounder
	path fixed. The index $\ell$ is the distance from the shock date to $t+h$, so
	$\ell=h+s$ refers to the shock dated $t-s$. Define
	\begin{align}
		\label{eq:raw-first-stage}
		\Phi_h^G
		&=
		\sum_{\varsigma\in\{g,r\}}
		\sum_{s\geq0}
		\psi^G_{\varsigma,s}
		\sigma_\varsigma^2
		\mathcal S^{\mathcal G}_{\varsigma,h+s}.
	\end{align}

	\begin{proposition}[HEDGE and omitted-history bias]
		\label{lemma:raw-fails}
		Suppose the conditions of Proposition~\ref{prop:generalized} hold, the
		exposure-weighted shifter $\sig G_t$ satisfies the exclusion restrictions
		of Assumption~\ref{ass:generalized}(iii), $\Phi_h^G\neq0$, $\mathcal S^{\mathcal G}_{g,h}\neq0$, and
		$\psi^G_{r,0}=0$. Then
		$\mathcal B_h^{\mathrm{TWFE}}=\mathcal S^{\mathcal Y}_{g,h}/\mathcal S^{\mathcal G}_{g,h}$
		and
		\begin{align}
			\label{eq:raw-fails}
			\mathcal B_h^{\mathrm{SS}}-\mathcal B_h^P
			&=
			\underbrace{
				\mathcal B_h^{\mathrm{TWFE}}-\mathcal B_h^P
			}_{\text{HEDGE bias}}
			\nonumber\\
			&\quad+
			\underbrace{
				(\Phi_h^G)^{-1}
				\sum_{\varsigma\in\{g,r\}}
				\sum_{s\geq1}
				\psi^G_{\varsigma,s}
				\sigma_\varsigma^2
				\left(
				\mathcal S^{\mathcal Y}_{\varsigma,h+s}
				-
				\mathcal B_h^{\mathrm{TWFE}}
				\mathcal S^{\mathcal G}_{\varsigma,h+s}
				\right)
			}_{\text{omitted-history bias}}.
		\end{align}
	\end{proposition}

	The proof is in Appendix~\ref{app:raw-fails}. The condition
	$\psi^G_{r,0}=0$ means that $G_t$ contains no contemporaneous confounder
	shock, which rules out impact feedback from the confounder to the
	treatment. Without it, $\mathcal B_h^{\mathrm{SS}}$ contains a third term,
	$(\Phi_h^G)^{-1}\psi^G_{r,0}\sigma_r^2(\mathcal S^{\mathcal Y}_{r,h}
	-\mathcal B_h^{\mathrm{TWFE}}\mathcal S^{\mathcal G}_{r,h})$, from the
	contemporaneous confounder shock, which HELP-IV also removes. The omitted-history bias
	arises because $G_t$ carries shocks dated before the window opens.
	These shocks reach the transformed treatment through
	$\mathcal G^{0}_{i,t-1}$ and the transformed outcome through
	$\mathcal Y^{0}_{i,t-1}$, the lagged aggregates in
	equation~\eqref{eq:outcome-dyn}, the persistence of $Y_{i,t+j}$, and the
	aggregates dated $t$ through $t+h$, which also carry them, so the
	coefficient mixes their effects with the response to
	the current shock. The bias therefore vanishes generically only if
	$\E[G_t\mid\mathcal F_{t-1}]=0$, that is, only if $G_t$ is itself an
	innovation. The term $\E[G_t\mid\mathcal F_{t-1}]$ is the expected value of
	the shifter given pre-window information, so the omitted-history bias is the
	time-series counterpart of the bias that recentering removes in shift-share
	designs \citep{borusyakhull2023}. This requirement is the condition that the
	shifter be unpredictable, under which
	\citet{manuel2026} shows that shift-share regressions, with controls for
	the rule that assigns exposure, recover the units' direct responses to the
	treatment path. Proposition~\ref{lemma:raw-fails} goes one step further, because it
	quantifies the bias when this condition fails. It also shows that the HEDGE
	bias remains even when $G_t$ is an innovation, so that the bias must be removed
	separately by holding the confounder path fixed.

	HELP-IV removes both biases simultaneously. On the one hand, its treatment
	instrument $\sig z_{g,t}$ is unpredictable from pre-window information,
	which excludes the past shocks behind the omitted-history bias
	$\mathcal B_h^{\mathrm{SS}}-\mathcal B_h^{\mathrm{TWFE}}$. On the other hand, its path
	instruments $\sig z_{r,t},\ldots,\sig z_{r,t+h}$ hold the confounder path
	fixed, which removes the HEDGE bias $\mathcal B_h^{\mathrm{TWFE}}-\mathcal B_h^P$.

	The remaining practical question is how to obtain a treatment instrument with
	this property. When an external proxy satisfying equation~\eqref{eq:g-proxy} is
	available, it serves this role directly. When no external proxy is available,
	a treatment shock identified from an SVAR has the same property if the
	moving-average representation is invertible and the identifying
	restrictions separate the treatment shock from the contemporaneous
	confounder shock \citep{barnichonmesters2020}.
	Appendix~\ref{app:svar-inference} accounts for its estimation.

\medskip
\noindent\textit{Section summary.} The extensions in this section differ in
their environments but rest on one identification principle. As in
Section~\ref{sec:dyn-identification}, the treatment coefficient of each design
equals the coefficient obtained with a single instrument, the exposure-weighted
combined proxy. Its weights render it uncorrelated with every value of the
relevant confounder paths, or with as much of the expected path as the available
news shocks allow, and it is unpredictable from information before the
treatment shock. This instrument therefore moves the treatment while
holding those paths fixed, so that the reduced form and the first stage are
measured under the same fixed-path experiment and their ratio identifies the
partial-equilibrium effect, or the information-constrained effect when the news
shocks are few. Each extension tailors the design to the environment
that a class of empirical applications faces, yet every such adaptation
preserves this principle. An advantage of this shared foundation is that every
design remains an IV regression, so that estimation and inference for all
designs share a common framework, which the next section develops.

	\section{Inference for HELP-IV designs}
	\label{sec:help-inference}

This section develops inference for the HELP-IV designs of
Sections~\ref{sec:portable-markov} and~\ref{sec:confounders}. We focus our discussion on
the benchmark HELP-IV design of Section~\ref{sec:portable-markov}, whose sampling uncertainty arises,
with proxies, only from the IV moment, and composite HELP-IV, whose sampling uncertainty
arises from both the IV moment and the estimated offset weights on the confounder proxies.
Inference for the designs with multiple exposure margins, multiple confounders, the
targeting rule, forward-looking outcomes, and transformed treatments and outcomes is
discussed briefly in this section and developed in detail in Appendix~\ref{app:inference}.
The section closes with shocks estimated from an SVAR.

	\subsection{Benchmark HELP-IV}
	\label{sec:help-estimation}

	The sample analogue of the HELP-IV moment system in
Section~\ref{sec:dyn-identification} delivers the estimator
\begin{equation}
	\widehat{\boldsymbol\vartheta}_h
	=\widehat\Sigma_h^{-1}\frac1{NT}\sum_{t,i}
	\widetilde{\mathbf Z}_{it,h}\widetilde Y_{i,t+h},
	\qquad
	\widehat\Sigma_h=\frac1{NT}\sum_{t,i}\widetilde{\mathbf Z}_{it,h}
	\widetilde{\boldsymbol X}_{it,h}',
	\label{eq:main-estimator}
\end{equation}
where $N$ is the number of units, $T$ is the number of dates, $\mathbf Z_{it,h}$ collects the
exposure-weighted proxies, $\boldsymbol X_{it,h}$ collects the exposure-weighted treatment
and confounder path, and a tilde now denotes sample residualization on the unit and time
effects and on any included controls, such as the lagged outcome. The coefficient on $\sig G_t$, $\widehat b_h^{\mathrm{HELP}}=e_1'\widehat{\boldsymbol\vartheta}_h$,
estimates the partial-equilibrium effect $\beta_h^P$, where $e_1$ is the first unit vector.

To characterize the asymptotic variance of this estimator, we define the population
counterpart $\Sigma_h$ of $\widehat\Sigma_h$ and the long-run variance $\Omega_h$ of the
date-level IV moment $N^{-1}\sum_i\widetilde{\mathbf Z}_{it,h}\widetilde\xi_{i,t,h}$,
\begin{equation}
	\Sigma_h=\E\left[\lim_{N\to\infty}\frac1N\sum_i
	\widetilde{\mathbf Z}_{it,h}\widetilde{\boldsymbol X}_{it,h}'\right],
	\qquad
	\Omega_h=\lim_{N,T\to\infty}\Var\left(\frac1{\sqrt T}\sum_t\frac1N\sum_i
	\widetilde{\mathbf Z}_{it,h}\widetilde\xi_{i,t,h}\right),
	\label{eq:main-moment}
\end{equation}
where $\xi_{i,t,h}$ is the error of the local projection at the population coefficients.
The date-level IV moment provides one observation per date, because all units share the same
aggregate proxies at a given date. The estimator therefore converges at rate $\sqrt T$ rather
than $\sqrt{NT}$ \citep{almuzarasancibrian2025}, and $\Omega_h$ normalizes the sum over dates by
$1/\sqrt T$.

We estimate the long-run variance with a kernel estimator
\begin{equation}
	\widehat\Omega_h=\frac1T\sum_{t=1}^{T}\sum_{\tau=1}^{T}
	k\Bigl(\frac{t-\tau}{b_T}\Bigr)
	\Bigl(\frac1N\sum_i\widetilde{\mathbf Z}_{it,h}\widehat\xi_{i,t,h}\Bigr)
	\Bigl(\frac1N\sum_i\widetilde{\mathbf Z}_{i\tau,h}\widehat\xi_{i,\tau,h}\Bigr)',
	\label{eq:hac-main}
\end{equation}
where $\widehat\xi_{i,t,h}$ denotes the fitted residuals, $k$ is a kernel with
$k(0)=1$, such as the Bartlett kernel $k(x)=(1-|x|)_+$, and $b_T$ is the bandwidth. The
kernel estimator is consistent for $\Omega_h$ when $b_T\to\infty$ with $b_T=o(\sqrt T)$ and the
date-level IV moment has finite fourth moments and summable fourth-order cumulants (see
Proposition~\ref{prop:hac} for the proof). Because the estimator first averages the moment over
units at each date and then weights the dates by the kernel, the resulting standard errors are
robust to any correlation across units at a given date, provided the cross-sectional approximation
error is negligible (Assumption~\ref{ass:regular}), and to serial correlation across dates, which
the common shocks induce.

The following proposition establishes the asymptotic distribution of
$\widehat b_h^{\mathrm{HELP}}$, which rests on a central limit theorem for the date-level IV moment
under regularity conditions on its serial dependence, the rank of the first stage, and the
cross-sectional approximation error (Assumption~\ref{ass:regular} in the appendix).

	\begin{proposition}[Inference for HELP-IV]
		\label{thm:help}
		Under the regularity conditions in Assumption~\ref{ass:regular}, the conditions of
		Proposition~\ref{prop:help-identification}, and
		$\widehat\Omega_h\xrightarrow{p}\Omega_h$,
		\begin{equation}
			\sqrt T\,\widehat V_h^{-1/2}\bigl(\widehat b_h^{\mathrm{HELP}}-\beta_h^P\bigr)
			\xrightarrow{d} N(0,1),
			\label{eq:help-clt}
		\end{equation}
		where $\widehat V_h=e_1'\widehat\Sigma_h^{-1}\widehat\Omega_h
		\widehat\Sigma_h^{-1\prime}e_1$, $\widehat V_h\xrightarrow{p}V_h$, and
		$V_h=e_1'\Sigma_h^{-1}\Omega_h\Sigma_h^{-1\prime}e_1$.
	\end{proposition}

	The proof is in Appendix~\ref{app:theory}. The normal approximation requires a controlled first
stage that is strong relative to its sampling error. With weak proxies, the approximation can be
poor, and Appendix~\ref{app:weak} gives an Anderson--Rubin-type confidence set for
$b_h^{\mathrm{HELP}}$ that remains valid when the controlled first stage is weak.

Proposition~\ref{thm:help} extends to multiple-confounder
HELP-IV and to HELP-IV with multiple exposure margins, because both designs keep the date-level IV
moment and only add regressors and instruments, which enlarges $\Sigma_h$ and $\Omega_h$. The
treatment coefficient remains a scalar with multiple confounders, so Proposition~\ref{thm:help} holds
with these larger matrices under the conditions of Corollary~\ref{cor:vec-conf} in place of those of
Proposition~\ref{prop:help-identification}. The treatment coefficient becomes a vector with multiple
exposure margins, because each regressor is interacted with each margin. This vector, identified under
the conditions of Corollary~\ref{cor:margins}, requires a matrix that selects it in
place of $e_1$, and $V_h$ becomes its covariance matrix, which also delivers inference on contrasts
among these coefficients (Proposition~\ref{prop:limit}).

Proposition~\ref{thm:help} also extends to targeting-rule HELP-IV
(Section~\ref{sec:combined-conf}), but inference must then account for a sample imbalance in its
instruments. This imbalance arises because the residual exposure $\sig^\perp$ is orthogonal to exposure
to the one-way confounders in the population but only up to order $N^{-1/2}$ in a sample. The imbalance
leaves a component of the one-way confounders in the IV moment, and this component does not average out
because the imbalance is fixed over time. The component therefore generally adds a cross-sectional
sampling term to the variance. This term is zero in every sample when exposure to each one-way
confounder lies in the span of the characteristics $\mathbf a_i$, and it is negligible under
population balance when $T/N\to0$ (Proposition~\ref{prop:targeting}). When the exposures to the
one-way confounders and their path coefficients are identified, the term has a sample analogue,
which we add to the time-series variance (Appendix~\ref{app:targeting}). When they are not, the
time-series standard error is complete only in these two cases, and it understates the uncertainty
when balance holds only in the population and $T/N$ does not vanish.

	\subsection{Composite HELP-IV}
	\label{sec:composite-estimation}

	Composite HELP-IV offers an alternative design for multiple confounders and for
	forward-looking partial-equilibrium effects. Its inference, however, differs from that in
	Section~\ref{sec:help-estimation}, because the design first estimates the weights on the confounder proxies
	and then constructs the instrument from these estimates. With the composite instrument, the
	regression in Section~\ref{sec:additional-composite-help} delivers the estimator
	\begin{equation}
		\widehat b_h^{\mathrm{HELP\text{-}C}}
		=\Bigl(\sum_{t,i}\widetilde{s_{ig}\widehat{\mathrm z}_{g,t,h}^C}\,
		\widetilde{s_{ig}G_t}\Bigr)^{-1}
		\sum_{t,i}\widetilde{s_{ig}\widehat{\mathrm z}_{g,t,h}^C}\,
		\widetilde Y_{i,t+h},
		\label{eq:composite-estimator-main}
	\end{equation}
	where $\widehat{\mathrm z}_{g,t,h}^C=z_{g,t}
	+\sum_{s=0}^h\widehat{\boldsymbol w}_s'\boldsymbol z_{r,t+s}$ is the estimated composite
	instrument.

	The estimated weights $\widehat{\boldsymbol w}_\ell$ solve the sample analogue of the recursion
	in Section~\ref{sec:additional-composite-help},
	\begin{equation}
		\widehat{\boldsymbol w}_\ell
		=-\widehat\Sigma_{rz,0}^{-1}
		\Bigl(\widehat{\mathbf v}_\ell
		+\sum_{s=0}^{\ell-1}\widehat\Sigma_{rz,\ell-s}\widehat{\boldsymbol w}_s\Bigr),
		\qquad \ell=0,\ldots,h,
		\label{eq:weights-main}
	\end{equation}
	with the sample cross moments
	\begin{equation}
		\begin{aligned}
			\widehat{\mathbf v}_\ell
			&=\frac1T\sum_t(\mathbf R_{t+\ell}-\bar{\mathbf R}_\ell)(z_{g,t}-\bar z_g),\\
			\widehat\Sigma_{rz,\ell}
			&=\frac1T\sum_t(\mathbf R_{t+\ell}-\bar{\mathbf R}_\ell)
			(\boldsymbol z_{r,t}-\bar{\boldsymbol z}_r)',
		\end{aligned}
		\label{eq:weight-inputs-main}
	\end{equation}
	where bars denote sample means. These sample cross moments suffice for the recursion, because
	covariance stationarity makes the covariance of $\mathbf R_{t+\ell}$ with $\boldsymbol z_{r,t+s}$
	depend only on $\ell-s$ and the causality restrictions set it to zero for $s>\ell$.

	The estimation error in the weights is governed by the weight equations, the stacked form of this
	recursion,
	\begin{equation}
		\widehat{\underline\Sigma}_{rz,h}\widehat{\underline{\boldsymbol w}}_h
		+\widehat{\underline{\mathbf v}}_h=\mathbf0,
		\label{eq:weight-moment-main}
	\end{equation}
	whose population counterpart is
	$\underline\Sigma_{rz,h}\underline{\boldsymbol w}_h+\underline{\mathbf v}_h=\mathbf0$. Here,
	$\underline{\boldsymbol w}_h=(\boldsymbol w_0',\ldots,\boldsymbol w_h')'$ stacks the weights,
	$\underline{\mathbf v}_h=(\mathbf v_0',\ldots,\mathbf v_h')'$ stacks the cross moments of the
	confounders with the treatment proxy, and
	$\underline\Sigma_{rz,h}=(\Sigma_{rz,\ell-s}{\bf1}\{s\leq\ell\})_{\ell,s=0}^h$ is the block
	lower-triangular matrix of their cross moments with the confounder proxies, whose zero blocks
	reflect the causality restrictions.

	To characterize the asymptotic variance of the composite estimator, we define the population
	cross moment $\Sigma_h^C$ of the instrument and the treatment and the derivative
	$D_{\boldsymbol w_s,h}$ of the composite IV moment
	$N^{-1}\sum_i\widetilde{s_{ig}\mathrm z_{g,t,h}^C}\,\widetilde\xi^C_{i,t,h}$ with respect
	to the weights,
	\begin{equation}
		\Sigma_h^C=\E\left[\lim_{N\to\infty}\frac1N\sum_i
		\widetilde{s_{ig}\mathrm z_{g,t,h}^C}\,\widetilde{s_{ig}G_t}\right],
		\qquad
		D_{\boldsymbol w_s,h}=\E\left[\lim_{N\to\infty}\frac1N\sum_i
		\widetilde{s_{ig}\boldsymbol z_{r,t+s}'}\,
		\widetilde\xi^C_{i,t,h}\right],
		\label{eq:composite-moments-main}
	\end{equation}
	where $\xi^C_{i,t,h}$ is the error of the composite regression at the population coefficient,
	and $D_{\boldsymbol w,h}=(D_{\boldsymbol w_0,h},\ldots,D_{\boldsymbol w_h,h})$ collects these
	derivatives. The derivative is generally nonzero, because the composite error
	$\xi^C_{i,t,h}$ contains the omitted path terms $\sum_q\sum_s c_{q,h,s}\sig R_{q,t+s}$, which are
	correlated with the confounder proxies. It vanishes when the confounder path does not affect
	the outcome (Proposition~\ref{prop:exposure-rank}). The asymptotic variance of $\widehat b_h^{\mathrm{HELP\text{-}C}}$ combines the long-run
	variance $\Omega_{gg,h}^C$ of the composite IV moment, the long-run variance $\Omega_{ww,h}^C$
	of the weight equations, and their long-run covariance $\Omega_{gw,h}^C$,
	\begin{equation}
		V_h^C=(\Sigma_h^C)^{-2}\bigl\{\Omega_{gg,h}^C
		-2\,\Omega_{gw,h}^C\underline\Sigma_{rz,h}^{-1\prime}D_{\boldsymbol w,h}'
		+D_{\boldsymbol w,h}\underline\Sigma_{rz,h}^{-1}\Omega_{ww,h}^C
		\underline\Sigma_{rz,h}^{-1\prime}D_{\boldsymbol w,h}'\bigr\}.
		\label{eq:composite-variance-main}
	\end{equation}
	With known weights, the variance reduces to $(\Sigma_h^C)^{-2}\Omega_{gg,h}^C$, the first term
	in braces. Estimating the weights adds the other two terms, because
	$\underline\Sigma_{rz,h}^{-1}$ maps the error in the weight equations into an error in the
	weights, and $D_{\boldsymbol w,h}$ maps that error into the IV moment. The last term,
	$D_{\boldsymbol w,h}\underline\Sigma_{rz,h}^{-1}\Omega_{ww,h}^C
	\underline\Sigma_{rz,h}^{-1\prime}D_{\boldsymbol w,h}'$, is the variance of this transmitted
	error. The middle term, $-2\,\Omega_{gw,h}^C\underline\Sigma_{rz,h}^{-1\prime}D_{\boldsymbol w,h}'$,
	is twice its covariance with the IV moment, which can take either sign, so estimating the weights
	can raise or lower the variance.

	To estimate this variance, we collect its three long-run objects in one matrix, the long-run
	variance of the composite IV moment stacked with the weight equations,
	\begin{equation}
		\Omega_h^C=\lim_{N,T\to\infty}\Var\left(\frac1{\sqrt T}\sum_t
		\begin{pmatrix}
			N^{-1}\sum_i\widetilde{s_{ig}\mathrm z_{g,t,h}^C}\,\widetilde\xi^C_{i,t,h}\\[2pt]
			\mathbf R_tz_{g,t}+\mathbf R_t\boldsymbol z_{r,t}'\boldsymbol w_0\\
			\vdots\\
			\mathbf R_{t+h}z_{g,t}
			+\sum_{s=0}^{h}\mathbf R_{t+h-s}\boldsymbol z_{r,t}'\boldsymbol w_s
		\end{pmatrix}\right).
		\label{eq:composite-meat}
	\end{equation}
	The first row of the vector is the composite IV moment, and the remaining rows are the
	date-$t$ terms of the weight equations. Accordingly, $\Omega_{gg,h}^C$ is the first diagonal
	element of $\Omega_h^C$, $\Omega_{ww,h}^C$ is its lower-right block, and $\Omega_{gw,h}^C$ is
	the remainder of its first row. We estimate $\Omega_h^C$ by applying the kernel estimator in
	equation~\eqref{eq:hac-main} to this vector, evaluated at the estimated instrument, the fitted
	residuals, and the estimated weights. Replacing the blocks of $\Omega_h^C$, and $\Sigma_h^C$,
	$D_{\boldsymbol w,h}$, and $\underline\Sigma_{rz,h}$, in
	equation~\eqref{eq:composite-variance-main} with their sample counterparts then yields
	$\widehat V_h^C$.

	The following proposition establishes the asymptotic distribution of
	$\widehat b_h^{\mathrm{HELP\text{-}C}}$, which rests on a central limit theorem for this vector.

	\begin{proposition}[Inference for composite HELP-IV]
		\label{thm:composite}
		Under Assumption~\ref{ass:regular}, the conditions of
		Proposition~\ref{prop:composite-conf}, and
		$\widehat\Omega_h^C\xrightarrow{p}\Omega_h^C$,
		\begin{equation}
			\sqrt T\,(\widehat V_h^C)^{-1/2}
			\bigl(\widehat b_h^{\mathrm{HELP\text{-}C}}-\beta_h^P\bigr)
			\xrightarrow{d} N(0,1),
			\label{eq:composite-clt}
		\end{equation}
		where $\widehat V_h^C\xrightarrow{p}V_h^C$ and $V_h^C$ is given by
		\eqref{eq:composite-variance-main}.
	\end{proposition}

	The proof is in Appendix~\ref{app:theory}. Proposition~\ref{thm:composite} extends to
	three further environments. First, with multiple exposure margins
	(Section~\ref{sec:multiple-exposure-margins}), composite HELP-IV instruments
	$\mathbf s_{ig}G_t$ with $\mathbf s_{ig}\mathrm z_{g,t,h}^C$ and identifies the margin-specific
	effects (Proposition~\ref{prop:exposure-rank}). The weights depend only on aggregate moments, so a
	single composite instrument serves all margins. With the exposure vector $\mathbf s_{ig}$ in place
	of $s_{ig}$, the composite IV moment, $\Sigma_h^C$, and $D_{\boldsymbol w_s,h}$ become $M\times1$,
	$M\times M$, and $M\times K$, and the matrix counterpart of
	equation~\eqref{eq:composite-variance-main} is the covariance matrix of the margin-specific
	coefficients. Second, with forward-looking outcomes (Section~\ref{sec:forward-portable-section}),
	the weights change, because E-HELP-IV confines the composite instrument to proxies dated $t$. These
	weights are estimated from sample cross moments of the confounder path with the date-$t$ proxies.
	The unit-covariance normalization of Section~\ref{sec:forward-portable-section} is not imposed,
	because rescaling the proxies rescales the weights and leaves the composite instrument unchanged
	(Appendix~\ref{app:weights}). The estimated weights are differentiable in these moments, with
	Jacobian $W^E$, when $\mathbf R_H^E$ has full column rank (Lemma~\ref{lem:news-weights}).
	Accordingly, $W^E$ replaces $-\underline\Sigma_{rz,h}^{-1}$, the derivative with respect to these
	weights replaces $D_{\boldsymbol w,h}$, and the cross moments that determine these weights replace
	the weight equations. Proposition~\ref{thm:composite} then holds with the information-constrained
	effect $\beta_h^E$ in place of $\beta_h^P$, under the conditions of
	Proposition~\ref{cor:forward-composite-help}. Third, with transformed treatments and outcomes
	(Section~\ref{sec:generalized-target}), the regression changes, because $\mathcal Y_{i,t+h}$
	replaces the outcome and $\mathcal G_{i,t+h_{\mathcal G}}$ replaces the treatment regressor, with
	the same instruments and weights. Propositions~\ref{thm:help} and~\ref{thm:composite} then hold with
	the generalized partial-equilibrium effect $\mathcal B_h^P$ of
	equation~\eqref{eq:generalized-target} in place of $\beta_h^P$, under the conditions of
	Proposition~\ref{prop:generalized}, with $\Sigma_h$, $\Omega_h$, $\Sigma_h^C$, $D_{\boldsymbol w,h}$,
	and $\Omega_h^C$ computed from the transformed regression. The extensions combine. For example,
	E-HELP-IV for a cumulative multiplier uses the transformed regression with the forward-looking
	weights.

	Shocks estimated from an SVAR can also serve as instruments in place of the proxies, as in both
	applications \citep{barnichonmesters2020}. This step assumes that the aggregates follow a correctly specified
	SVAR of finite order whose shocks can be recovered from current and past aggregates, an assumption
	that the identification results of Sections~\ref{sec:portable-markov} and~\ref{sec:confounders}
	do not require. The sampling error of the SVAR then
	propagates to the coefficient through the instruments and, for composite HELP-IV, through the
	weights, which the fitted responses now determine. We therefore further generalize
	Propositions~\ref{thm:help} and~\ref{thm:composite} by stacking the SVAR estimating equations with
	the IV moment (Propositions~\ref{thm:help-S} and~\ref{thm:composite-S}). The variance then also
	contains the variance of this sampling error and its covariance with the IV moment, so this
	correction, as with the weights, can raise or lower the standard error. The three extensions in the
	previous paragraph carry over, because in each case the estimated SVAR enters through the instruments
	and the weights (Appendix~\ref{app:svar-extensions}).

	\section{Monte Carlo simulations}
	\label{sec:montecarlo}

This section uses Monte Carlo simulations to assess the identification and
inference performance of the HELP-IV designs. We first consider a single
confounder, for which we evaluate HELP-IV in the backward-looking environment and
E-HELP-IV in the forward-looking environment and benchmark both designs against
TWFE-IV and the decomposition method of \citet{donaldson2026}. We then consider
multiple confounders, for which we compare the performance of multiple-confounder
HELP-IV, composite HELP-IV, and targeting-rule HELP-IV. We also summarize the
evaluation of the other designs and environments studied in this paper.

	\subsection{A single confounder}
	\label{sec:mc-single}

	\subsubsection{Data-generating process}
	\label{sec:mc-single-dgp}

 The outcome equation is
	\begin{align*}
		Y_{it}
		&=\mu_i+\lambda_t+0.37\,Y_{i,t-1}
		+\sig\bigl(0.62\,G_t+0.28\,G_{t-1}-0.13\,G_{t-2}
		+\zeta^g_1\E_tG_{t+1}+\zeta^g_2\E_tG_{t+2}\bigr)\\
		&\quad+\sir\bigl(-0.45\,R_t+0.21\,R_{t-1}+0.09\,R_{t-2}
		+\zeta^r_1\E_tR_{t+1}+\zeta^r_2\E_tR_{t+2}\bigr)
		+u_{it}.
	\end{align*}
	In the backward-looking environment, the coefficients on expectations
	are zero. In the forward-looking environment,
	$(\zeta^g_1,\zeta^g_2,\zeta^r_1,\zeta^r_2)=(0.30,0.12,-0.25,0.10)$, so the outcome
	has a forward-looking horizon of $F=2$ periods. The exposures $\sig$ and $\sir$ are related by
	\[
	\sir=0.8\,\sig+\uisr,
	\]
	so exposure to the confounder rises with exposure to the treatment. The
	aggregate block is an SVAR(2) with two-way general equilibrium,
	\begin{align*}
		G_t &= 0.5\,R_t+0.70\,G_{t-1}+0.03\,R_{t-1}-0.10\,G_{t-2}+\epsilon_{g,t},\\
		R_t &= 0.5\,G_t+0.05\,G_{t-1}+0.10\,R_{t-1}-0.05\,R_{t-2}+\epsilon_{r,t}
		+\textstyle\sum_{\ell=1}^{6}\epsilon^E_{r,t-\ell,\ell},
	\end{align*}
	where $\epsilon_{g,t}$ and $\epsilon_{r,t}$ are the structural shocks to the
	treatment and to the confounder, and the news shock
	$\epsilon^E_{r,t,\ell}$, $\ell=1,\ldots,6$, is realized at date $t$ and shifts
	$R_{t+\ell}$. The news shocks are present only in the forward-looking
	environment. We assume that the instruments are
	proxies that measure the shocks with error, generated by
	$z_{x,t}=\epsilon_{x,t}+\eta_{x,t}$ for $x\in\{g,r\}$ and by
	$z^E_{r,t,\ell}=\epsilon^E_{r,t,\ell}+\eta^E_{r,t,\ell}$ for the news
	shocks. All primitive random variables are mutually independent and
	normal with mean zero. The measurement errors have standard deviation $0.5$,
	and all other primitive random variables are standard normal, including the
	unit and time effects $\mu_i$ and $\lambda_t$, the exposure $\sig$, the
	disturbances $\uisr$ and $u_{it}$, and the shocks.

	\subsubsection{Results}
	\label{sec:mc-single-results}

	We simulate $N=200$ units over $T=200$ periods in each of $2{,}000$
	replications and report the results at horizons $h=0$, $2$, and $4$. We
	consider the backward-looking environment (Panel A of Table~\ref{tab:mc-single})
	and the forward-looking environment (Panel B). In the forward-looking
	environment, we assume that the researcher observes the instruments of only
	the first two of the six news shocks. This assumption corresponds to the case
	of under-identification, which is the common case in
	applications. Appendix~\ref{app:mc-forward} reports the case of exact identification, in which the researcher observes the instruments of all six news shocks, so that E-HELP-IV holds the expected confounder path fixed completely. For each estimator, we report the
	median bias and the root mean squared error (RMSE) of its estimates across
	replications. For HELP-IV, we also show its inference performance by the
	share of replications in which the nominal 90 percent confidence interval
	covers the target and by the median length of that interval.\footnote{Every
		panel regression includes unit and time effects, the lagged outcome
		$Y_{i,t-1}$, and $\sig G_{t-1}$, $\sig G_{t-2}$, $\sig R_{t-1}$, and
		$\sig R_{t-2}$, which capture the lagged aggregates in the outcome equation. The cross-sectional step
		of the decomposition method includes the same controls. When the weights or the instruments are estimated, the standard error of HELP-IV
uses the sandwich covariance of Section~\ref{sec:help-inference} or
Appendix~\ref{app:svar-inference}, which stacks the first-step moments with the IV
moment.}

	\begin{table}[!ht]\centering
		\caption{TWFE-IV, the decomposition method, and HELP-IV with a single
			confounder}
		\label{tab:mc-single}
		\adjustbox{max width=\textwidth}{%
			\begin{tabular}{cc rr rr rrrr}
\toprule
 & & \multicolumn{2}{c}{TWFE-IV} & \multicolumn{2}{c}{Decomposition} & \multicolumn{4}{c}{HELP-IV}\\
\cmidrule(lr){3-4}\cmidrule(lr){5-6}\cmidrule(lr){7-10}
$h$ & Target & Bias & RMSE & Bias & RMSE & Bias & RMSE & Coverage & Length\\
 & & (1) & (2) & (3) & (4) & (5) & (6) & (7) & (8)\\
\midrule
\multicolumn{10}{l}{\textit{Panel A. Backward-looking environment}}\\[1pt]
0 & 0.62 & $-0.18$ & 0.18 & $-0.16$ & 0.16 & $\phantom{-}0.00$ & 0.01 & 0.89 & 0.02\\
2 & 0.66 & $\phantom{-}0.34$ & 0.35 & $-0.16$ & 0.22 & $-0.01$ & 0.12 & 0.88 & 0.40\\
4 & 0.21 & $\phantom{-}0.59$ & 0.61 & $-0.06$ & 0.22 & $-0.02$ & 0.17 & 0.88 & 0.53\\
\addlinespace
\multicolumn{10}{l}{\textit{Panel B. Forward-looking environment}}\\[1pt]
0 & 0.87 & $-0.10$ & 0.12 & $-0.21$ & 0.22 & $-0.01$ & 0.08 & 0.95 & 0.28\\
2 & 0.79 & $\phantom{-}0.60$ & 0.63 & $-0.15$ & 0.23 & $\phantom{-}0.04$ & 0.21 & 0.93 & 0.69\\
4 & 0.29 & $\phantom{-}0.80$ & 0.88 & $-0.05$ & 0.22 & $\phantom{-}0.02$ & 0.29 & 0.92 & 0.88\\
\bottomrule
\end{tabular}
}

		\vspace{0.6em}
		{\footnotesize
			\begin{minipage}{0.84\textwidth}
				\textit{Notes.} $N=T=200$ and $2{,}000$ replications. The target is the
				partial-equilibrium effect $\beta_h^P$ in Panel A and the
				information-constrained effect $\beta_h^E$ in Panel B
				(Sections~\ref{sec:dyn-pe-effect} and~\ref{sec:forward-portable-section}).
				The decomposition method is that of \citet{donaldson2026}, applied
				in Panel B in its ex ante form. The ex ante form offsets the expected
				confounder path with date-$t$ shocks. In Panel B, HELP-IV is composite
				E-HELP-IV.
				Bias is the median bias. Coverage and length refer to the nominal 90
				percent interval, and length is its median length.
		\end{minipage}}
	\end{table}

	Table~\ref{tab:mc-single} shows that the bias of TWFE-IV changes sign across
	horizons in both environments. The bias is negative on impact, at $-0.18$ in
	the backward-looking environment and $-0.10$ in the forward-looking
	environment, and positive thereafter, reaching $0.59$ and $0.80$ at $h=4$.
	This pattern reflects the two components of the HEDGE bias, which
	Proposition~\ref{lemma:hedge-bias} derives for the backward-looking
	environment and which carry opposite signs under our data-generating process.
	The exposure-correlation component is negative, because a treatment shock
	raises the confounder, and the confounder lowers the outcome most in the
	units that are most exposed to the treatment. The aggregate-feedback
	component is positive, because the confounder feeds back into the treatment
	and prolongs its response. The former dominates on impact and the latter from
	$h=2$ on.

	HELP-IV, by contrast, is centered on the target in both panels, with a median
	bias of at most $0.04$ in absolute value and an RMSE of at most $0.29$. In the
	forward-looking environment, its target is the information-constrained
	effect, because the available instruments do not fully offset the expected confounder path.
	The outcome up to $h=4$ depends on this path from $t$ to $t+6$, and the proxies, $z_{r,t}$ and
	the first two news proxies, leave 2.0 percent of the path's response to the treatment shock unabsorbed
	(Section~\ref{sec:forward-portable-section}). At these horizons, the target differs from the
	forward-looking partial-equilibrium effect by at most $0.03$.

	The coverage of the intervals of HELP-IV differs between the two environments. In the backward-looking environment,
	the intervals cover the target in 88--89 percent of replications, close to
	the nominal 90 percent. In the forward-looking environment, they cover it in
	92--95 percent of replications and are therefore mildly
	conservative.\footnote{The over-coverage
		stems from the first step, which estimates the composite weights. The
		estimated weights offset the confounder path in the sample and not only in
		the population, so the estimates are less dispersed than with the population
		weights. With the first-step term, the median length of the intervals is
		close to the range of the estimates, but the $t$-statistic remains less
		dispersed than a standard normal variable
		(Appendix~\ref{app:mc-forward}).}

	The decomposition method does not recover the target in this design, in which the
	confounder moves the treatment on impact, and its median bias is negative at every
	horizon in both panels. It shares the numerator of the target, the outcome response
	with the confounder path held fixed, and therefore removes the exposure-correlation
	component and the part of the aggregate-feedback component that operates through the
	outcome response. It divides, however, by the unrestricted impact response of the
	treatment, as equations~\eqref{eq:don-statistic} and~\eqref{eq:portable-response-form}
	show. Under our data-generating process, the controlled impact response is about
	$0.75$ times the unrestricted one, so in Panel A the decomposition method converges
	to about $0.75$ times the target.

	\subsubsection{Further designs and environments}
	\label{sec:mc-single-extensions}

	Five further exercises are reported in Appendix~\ref{app:mc}. First, when the measurement error of $z_{g,t}$ or of $z_{r,t}$ in the backward-looking environment has a standard deviation of $2$ or $3$ instead of $0.5$, HELP-IV remains centered on the target, with a median bias of at most $0.04$ in absolute value and coverage of 89--94 percent, although its estimates are less precise (Appendix~\ref{app:mc-weak}). Second, when the instruments are shocks estimated from a correctly specified aggregate
	SVAR (Appendix~\ref{app:svar-inference}), HELP-IV and composite HELP-IV still perform well, with
	median biases of at most $0.02$ in absolute value and coverage of 87--90 percent
	(Appendix~\ref{app:mc-internal}). Third, under one-way general equilibrium, in which
	$R$ does not enter the equation for $G_t$, HELP-IV and the decomposition method are both
	centered on the target, with similar RMSEs (Appendix~\ref{app:mc-don}). Fourth, when the
	data-generating process has a second exposure margin, as in
	Section~\ref{sec:multiple-exposure-margins}, HELP-IV with multiple exposure margins recovers both margin-specific
	partial-equilibrium effects, with coverage of 85--89 percent (Appendix~\ref{app:mc-margins}).

	The fifth exercise evaluates E-HELP-IV with transformed treatments and
	outcomes (Section~\ref{sec:generalized-target}) in a setting closer to
	applications. Instead of a stylized data-generating process, we simulate data
	from the forward-looking two-region New Keynesian model of
	\citet{donaldson2026}, which Appendix~\ref{app:mc-model} describes. The
	target is the cumulative fiscal multiplier when the expected path of the real
	interest rate is held fixed. We first consider one-way general equilibrium,
	in which government spending does not respond to the real rate, as in the
	specification of \citet{donaldson2026}. Both E-HELP-IV and her ex ante
	decomposition method are then centered on the target, but the intervals of
	E-HELP-IV are 32--39 percent shorter, although a single outlying estimate raises
	its RMSE at $h=2$ and $h=4$ (Appendix~\ref{app:mc-transformed}). We then relax this restriction and let
	spending respond to the lagged real rate. To keep this feedback empirically
	plausible, we calibrate its strength to the local-projection estimate of the
	response of US defense spending to a monetary policy shock, relative to that of the
	federal funds rate. This calibration is not a precise estimate of that response in US data. It shows the size of the bias of the decomposition method when the feedback from the real rate to spending is of this strength. E-HELP-IV remains centered on the target, and its
	intervals cover the target in 92--93 percent of replications. The
	decomposition method, by contrast, overstates the target, because a higher
	real rate lowers spending and thereby shrinks the unrestricted spending
	response by which the method normalizes. As this reduction accumulates, the
	overstatement widens with the horizon, from 8 percent at $h=4$, one year after
	the shock, to 19 percent in population at $h=8$, two years after the shock
	(Appendix~\ref{app:mc-transformed}). The shift-share design, which uses the
	change in aggregate spending as the shifter, overstates the target by $0.48$ to
	$0.66$ in both settings.

	\subsection{Multiple aggregate confounders}
	\label{sec:mc-multiple}

	\subsubsection{Data-generating process}
	\label{sec:mc-multiple-dgp}

	We now turn to a data-generating process with $K=8$ confounders,
	$R_{1,t},\ldots,R_{8,t}$, of which the first $K_1=2$ are two-way and the other
	$K-K_1=6$ are one-way. The outcome equation is
	\begin{align*}
		Y_{it}
		&=\mu_i+\lambda_t+0.8\,Y_{i,t-1}+\sig G_t
		+\sum_{q=1}^{8}s_{ir_q}R_{q,t}+u_{it}.
	\end{align*}
	The lagged outcome has the coefficient $0.8$, so that the effect of the
	treatment remains sizable at long horizons. Treatment exposure is
	$\sig=0.6\,a_i+\sig^\perp$, where $a_i$ is the targeting characteristic of
	Assumption~\ref{ass:targeting}, and $\sig^\perp$ has
	variance $0.64$. Exposure to confounder $q$ is $s_{ir_q}=0.8\,\sig+u_i^{r_q}$
	for the two-way confounders and $s_{ir_q}=0.8\,a_i+u_i^{r_q}$ for the one-way
	confounders, so $\sig^\perp$ is orthogonal to exposure to every one-way
	confounder, as Assumption~\ref{ass:targeting} requires.

	We assume that the aggregate block follows the SVAR(1)
	\begin{align*}
		G_t&=0.5\,G_{t-1}+0.1\textstyle\sum_{q=1}^{K_1}R_{q,t-1}
		+\epsilon_{g,t}+\varpi\textstyle\sum_{q=1}^{K_1}\epsilon_{r_q,t},\\
		R_{q,t}&=0.5\,R_{q,t-1}+0.1\,G_t+\epsilon_{r_q,t},
		\qquad q=1,\ldots,8,
	\end{align*}
	where $\epsilon_{g,t}$ and $\epsilon_{r_q,t}$ are the structural shocks to the
	treatment and to confounder $q$. Each confounder responds to the treatment
	within the period with the coefficient $0.1$, smaller than the $0.5$ of
	Section~\ref{sec:mc-single-dgp} because eight confounders respond at once. The
	two-way confounders feed back into the treatment with a lag, and when
	$\varpi>0$ their shocks also move the treatment within the period. The
	one-way confounders feed back into neither the treatment nor the two-way
	confounders, as Assumption~\ref{ass:one-way-block} requires. We assume that the
	instruments are proxies that measure the shocks with error,
	$z_{g,t}=\epsilon_{g,t}+\eta_{g,t}$ and
	$z_{r_q,t}=\epsilon_{r_q,t}+\eta_{r_q,t}$, where $\eta_{g,t}$ has standard
	deviation $0.5$ and $\eta_{r_q,t}$ has standard deviation $\sigma_\eta$. All primitive random variables
	are mutually independent and normal with mean zero, and all except the
	measurement errors and $\sig^\perp$ are standard normal.

	We consider two settings of the data-generating process, which vary the
	measurement error of the confounder proxies, $\sigma_\eta$, and the
	within-period response of the treatment, $\varpi$. The first setting has a
	large measurement error and no within-period response,
	$(\sigma_\eta,\varpi)=(1,0)$ (Panel A of Table~\ref{tab:mc-multi}). The second
	setting has a small measurement error and a within-period response,
	$(\sigma_\eta,\varpi)=(0.5,0.5)$ (Panel B). The confounder shock accounts for
	half of the variance of its proxy in the first setting and for 80 percent in
	the second. The two settings serve two purposes. They verify that the three
	HELP-IV designs recover the target under different conditions, and they show
	when multiple-confounder or composite HELP-IV is preferable.
	Appendix~\ref{app:mc-multi} reports the other two combinations of $\sigma_\eta$
	and $\varpi$.

	\subsubsection{Results}
	\label{sec:mc-multiple-results}

	We simulate $N=200$ units over $T=200$ periods in each of $2{,}000$
	replications and report the results at horizons $h=2$, $4$, $6$, and $8$. We
	study horizons longer than those of Section~\ref{sec:mc-single} to show how
	each design performs as the number of path controls of multiple-confounder
	HELP-IV, $8(h+1)$, grows.

	\begin{table}[!ht]\centering
		\caption{TWFE-IV and multiple-confounder,
			composite, and targeting-rule HELP-IV}
		\label{tab:mc-multi}
		\resizebox{\textwidth}{!}{\footnotesize\setlength{\tabcolsep}{3.1pt}%
			\begin{tabular}{cc rr rrrr rrr rrr}
\toprule
 & & \multicolumn{2}{c}{TWFE-IV} & \multicolumn{4}{c}{Multiple-confounder} & \multicolumn{3}{c}{Composite} & \multicolumn{3}{c}{Targeting rule}\\
\cmidrule(lr){3-4}\cmidrule(lr){5-8}\cmidrule(lr){9-11}\cmidrule(lr){12-14}
$h$ & Target & Bias & RMSE & Bias & RMSE & Coverage & Length & RMSE & Coverage & Length & RMSE & Coverage & Length\\
 & & (1) & (2) & (3) & (4) & (5) & (6) & (7) & (8) & (9) & (10) & (11) & (12)\\
\midrule
\multicolumn{14}{l}{\textit{Panel A. Large measurement error, no within-period response}}\\[1pt]
2 & 1.29 & $\phantom{-}1.03$ & 1.11 & $-0.05$ & 0.20 & 0.88 & 0.63 & 0.23 & 0.90 & 0.76 & 0.16 & 0.89 & 0.51\\
4 & 0.99 & $\phantom{-}1.01$ & 1.16 & $-0.09$ & 0.32 & 0.88 & 0.99 & 0.38 & 0.91 & 1.24 & 0.22 & 0.86 & 0.70\\
6 & 0.67 & $\phantom{-}0.77$ & 1.00 & $-0.10$ & 1.38 & 0.91 & 1.44 & 0.49 & 0.92 & 1.66 & 0.26 & 0.86 & 0.81\\
8 & 0.44 & $\phantom{-}0.52$ & 0.84 & $-0.08$ & 960.84 & 0.96 & 2.29 & 0.60 & 0.93 & 2.02 & 0.29 & 0.87 & 0.88\\
\addlinespace
\multicolumn{14}{l}{\textit{Panel B. Small measurement error, within-period response}}\\[1pt]
2 & 1.21 & $\phantom{-}1.12$ & 1.21 & $-0.06$ & 0.18 & 0.85 & 0.54 & 0.25 & 0.88 & 0.74 & 0.16 & 0.88 & 0.53\\
4 & 0.89 & $\phantom{-}1.12$ & 1.31 & $-0.09$ & 0.25 & 0.82 & 0.69 & 0.35 & 0.90 & 1.09 & 0.21 & 0.86 & 0.64\\
6 & 0.60 & $\phantom{-}0.86$ & 1.18 & $-0.10$ & 0.30 & 0.80 & 0.78 & 0.44 & 0.90 & 1.40 & 0.24 & 0.84 & 0.71\\
8 & 0.39 & $\phantom{-}0.58$ & 1.05 & $-0.09$ & 0.35 & 0.81 & 0.88 & 0.50 & 0.92 & 1.64 & 0.25 & 0.84 & 0.74\\
\bottomrule
\end{tabular}
}

		\vspace{0.6em}
		{\footnotesize
			\begin{minipage}{0.94\textwidth}
				\textit{Notes.} $N=T=200$, $K=8$, and $2{,}000$ replications. The
				target is the partial-equilibrium effect $\beta_h^P$. Panel A
				sets $(\sigma_\eta,\varpi)=(1,0)$ and Panel B $(0.5,0.5)$. Bias is the
				median bias, of TWFE-IV in column (1) and of multiple-confounder
				HELP-IV in column (3), and Table~\ref{tab:mc-multiapp} reports it for
				the other two designs. Coverage and length
				refer to the nominal 90 percent interval, and length is its median
				length.
		\end{minipage}}
	\end{table}

	Table~\ref{tab:mc-multi} shows that TWFE-IV is biased, with a median bias of
	$0.52$ to $1.12$, whereas the three HELP-IV designs recover the partial-equilibrium effect $\beta_h^P$ up to a finite-sample bias. The median
	bias of multiple-confounder HELP-IV is $-0.10$ to $-0.05$, and those of
	composite and targeting-rule HELP-IV lie between $-0.09$ and $-0.02$
	(Appendix~\ref{app:mc-multi}). These biases are at most about one-sixth of the
	bias of TWFE-IV, but at $h=8$, where the target is smallest, they reach 15 to 23
	percent of the target. In finite samples, the bias arises from both the
	lagged outcome $Y_{i,t-1}$, whose coefficient is $0.8$, and the large number
	of parameters to estimate with eight confounders. The bias vanishes as $T$ grows.

The three HELP-IV designs differ, however, in their precision, stability, and
interval coverage. Targeting-rule HELP-IV is the most precise design. At every
horizon and in both panels, the design has the smallest RMSE, $0.16$ to
$0.29$, and the shortest intervals. Because the characteristic $a_i$ absorbs
the six one-way confounders through a moment across units, the design needs
to hold fixed only the paths of the two two-way confounders, whereas
multiple-confounder HELP-IV holds fixed the paths of all eight confounders.

	When no targeting rule is available, the choice between multiple-confounder and composite
	HELP-IV depends on the setting. In Panel A, where the confounder proxies are noisy and
	the treatment does not respond to the confounders within the period, composite
	HELP-IV has the advantage at long horizons. The intervals of composite HELP-IV
	are 15 to 25 percent longer up to $h=6$ and 12 percent shorter at $h=8$, but the
	RMSEs diverge far more at long horizons. The RMSE of multiple-confounder HELP-IV rises to $1.4$ at $h=6$
	and exceeds $900$ at $h=8$, whereas that of composite HELP-IV stays at or below
	$0.6$. This large RMSE is due to extreme estimates in a few
	replications. For example, $8$ of the $2{,}000$ estimates of
	multiple-confounder HELP-IV at $h=6$ and $61$ at $h=8$ deviate from the target
	by more than eight times the target, whereas no estimate of composite HELP-IV
	does. In Panel B, where the proxies are precise and the treatment
	responds within the period, multiple-confounder HELP-IV has the advantage,
	and the RMSE of composite HELP-IV is $1.4$ to $1.5$ times larger. The intervals
	of composite HELP-IV, however, cover the target in 88--92 percent of
	replications, whereas those of multiple-confounder HELP-IV cover it in only
	80--85 percent.

	The two designs hold the confounder paths fixed in different ways, as
	Section~\ref{sec:additional-composite-help} explains, and this difference
	explains the ranking in each panel. In Panel A, the noisy proxies carry little
	information about the $8(h+1)$ path values, so the cross-moment matrix that
	multiple-confounder HELP-IV inverts is nearly singular in some replications,
	and these replications produce the extreme estimates. Composite HELP-IV avoids
	this problem, because it keeps a single endogenous regressor and inverts only
	the $8\times8$ impact covariance $\Sigma_{rz,0}$ in
	recursion~\eqref{eq:composite-weights}. In Panel B, the precise proxies keep
	the cross-moment matrix well conditioned, so multiple-confounder HELP-IV
	produces no extreme estimates. The comparison is then dominated by the
	finite-sample error of composite HELP-IV, which arises because
	recursion~\eqref{eq:composite-weights} imposes the causality restrictions only
	in population.\footnote{Varying the measurement error and the within-period response
		one at a time (Appendix~\ref{app:mc-multi}) shows that the measurement error
		alone determines whether multiple-confounder HELP-IV produces extreme
		estimates. The within-period response widens the gap in precision. With a
		small measurement error, the RMSE of composite HELP-IV is $1.1$ to $1.2$
		times that of multiple-confounder HELP-IV without the response and $1.4$ to
		$1.5$ times with it.} In sum, targeting-rule HELP-IV is the most precise
	design when a targeting rule is available. Between the other two designs,
	composite HELP-IV is the more reliable choice when the proxies are noisy and
	the horizon is long, whereas multiple-confounder HELP-IV is more precise when
	the proxies are accurate.

	\section{Empirical applications}
	\label{sec:applications}

	We consider two applications in which we use the HELP-IV framework developed in this paper to re-estimate partial-equilibrium effects reported in the existing literature. The first application (Section~\ref{sec:app-donaldson}) revisits the state-level fiscal multiplier that \citet{nakamurasteinsson2014} and \citet{duporguerrero2017} estimate with a shift-share design. 	\citet{donaldson2026} and \citet{manuel2026} use this multiplier as a leading example and correct it for general-equilibrium bias from one-way confounders. We re-estimate the same multiplier but treat the interest rate
	as a two-way confounder. This application therefore permits a direct
	comparison with the decomposition method of \citet{donaldson2026}. We also
	combine HELP-IV with the targeting rule of \citet{manuel2026} to address the
	one-way confounders for which no proxy is available.

	The second application (Section~\ref{sec:app-cns}) revisits the regional
	stock-wealth design of \citet[henceforth CNS]{chodorowreichnenovsimsek2021},
	which features multiple two-way confounders. CNS control for four confounders
	with a control-function method. We instead hold their paths fixed with
	composite HELP-IV and carry the estimate through the CNS model to its
	implications for the marginal propensity to consume out of stock wealth and for
	the aggregate labor market.

	\subsection{The cross-sectional fiscal multiplier}
	\label{sec:app-donaldson}

	The object of interest in this application is the partial-equilibrium
	multiplier, the partial-equilibrium effect of state defense spending
	on state output. \citet{nakamurasteinsson2014} estimate the cross-sectional
	multiplier with state and time fixed effects, arguing that the time fixed
	effects absorb the effects of aggregate confounders such as monetary policy.
	Following \citet{donaldson2026}, we estimate its cumulative version,
	\begin{align}
		\label{eq:app-don-mult}
		\mathcal Y_{i,t+h}=\mathcal B_h^{\mathrm{SS}}\,\mathcal G_{i,t+h}+\iota_h\Delta y_{i,t-1}
		+\mu_{i,h}+\lambda_{t,h}+v_{i,t,h},
	\end{align}
	where
	$\mathcal Y_{i,t+h}=\sum_{j=0}^{h}(Y_{i,t+j}-Y_{i,t-1})/Y_{i,t-1}$
	and
	$\mathcal G_{i,t+h}=\sum_{j=0}^{h}(G_{i,t+j}-G_{i,t-1})/Y_{i,t-1}$
	are the cumulative output and spending of Section~\ref{sec:generalized-target},
	as in \citet{ramey2018}, with $Y_{i,t}$
	real per capita gross state product and $G_{i,t}$ real per capita
	defense-contract spending. Both variables are therefore measured relative to
	the same pre-shock level of state output. The control $\Delta y_{i,t-1}$ is
	lagged output growth, $\mu_{i,h}$ and $\lambda_{t,h}$ are horizon-specific
	state and time fixed effects, and $\mathcal B_h^{\mathrm{SS}}$ is the shift-share IV multiplier.

	We use the panel of \citet{donaldson2026}, which contains real
	defense-contract spending and gross state product for $46$ states at annual
	frequency over $1969$--$2007$.\footnote{\citet{donaldson2026} excludes North
		Dakota, South Dakota, Wyoming, and Alaska because their gross state product
		changes by more than twenty percent in multiple years. She reports that
		retaining these small states leaves the point estimates similar but
		increases their variance. We use her sample so that our estimates and hers
		are computed on the same states.} We
	instrument $\mathcal G_{i,t+h}$ with the shift-share variable $\sig G_t$ of
	Section~\ref{sec:gen-orth}, where
	$G_t=(G^{\mathrm{US}}_t-G^{\mathrm{US}}_{t-1})/Y^{\mathrm{US}}_{t-1}$ is the
	one-year change in real per capita national defense spending relative to real
	per capita national output, and $\sig$ is the average of
	$(G_{i,t}/Y_{i,t})/(G^{\mathrm{US}}_t/Y^{\mathrm{US}}_t)$ over the five
	years before the estimation sample, 1964--1968, following
	\citet{donaldson2026}.

	The first row of Table~\ref{tab:app-don} reports the shift-share IV multiplier. It
	rises from $1.06$ in year 1 to $2.35$ in year 4, close to the estimates in
	\citet{donaldson2026}. If more exposed states are less sensitive to the
	interest rate, the monetary tightening that follows a defense-spending
	expansion depresses their output less, and the shift-share design attributes
	this smaller decline to defense spending, which overstates the multiplier. We
	therefore treat the interest rate as a two-way confounder. We first hold its path
	fixed with backward-looking HELP-IV and E-HELP-IV and then add one-way
	confounders through the targeting rule.

	The single-confounder rows of Table~\ref{tab:app-don} treat the interest rate
	as the only confounder. Backward-looking HELP-IV
	(Section~\ref{sec:portable-markov}), applied to the cumulative outcome and
	treatment as in Section~\ref{sec:generalized-target}, uses instruments
	constructed from a bivariate proxy-SVAR in national defense spending and the
	federal funds rate. In this proxy-SVAR, the \citet{romerromer2004} series
	identifies the monetary policy shock, and the defense-spending shock is
	orthogonalized against it. The resulting multiplier rises from $0.66$ in
	year 1 to $1.69$ in year 4, which indicates that the output response builds up
	more slowly than the spending response. In every year, the multiplier lies
	$0.40$ to $0.66$ below the shift-share IV multiplier.

	To facilitate comparison, we also estimate
	the decomposition method of \citet{donaldson2026}, implemented as in her paper (Appendix~\ref{app:se-don-spec2}), which yields a multiplier
	rising from $0.97$ to $1.85$. The decomposition lies
	above HELP-IV, by between $0.16$ and $0.31$. Part of this gap arises because the
	decomposition divides by the unrestricted spending response
	(Section~\ref{sec:donaldson-connection}). To measure this part, we apply the
	decomposition to the proxy-SVAR shocks that HELP-IV uses and replace the
	unrestricted spending response by the controlled one. This replacement, which
	removes the feedback of monetary policy shocks into defense spending, lowers
	the multiplier by $0.05$ in year 1 and by between $0.12$ and $0.15$ in years 2
	to 4, about half of the gap in years 2 and 3 and most of it in year 4. The
	rest reflects the different shock measures of the two procedures.

	\begin{table}[!ht]\centering
		\caption{Cumulative fiscal multipliers over the first four years}
		\label{tab:app-don}
		\begin{tabular}{lcccc}
			\toprule
			& \multicolumn{4}{c}{Year}\\
			\cmidrule(lr){2-5}
			& 1 & 2 & 3 & 4\\
			\midrule
			Shift-share IV, $\mathcal B_h^{\mathrm{SS}}$      & 1.06 & 1.45 & 1.90 & 2.35\\
			& (0.53) & (0.80) & (0.93) & (1.11)\\
			\addlinespace
			\emph{Single confounder} \\
			\quad HELP-IV, $\mathcal B_h^{\mathrm{HELP}}$      & 0.66 & 1.00 & 1.32 & 1.69\\
			& (0.38) & (0.40) & (0.36) & (0.53)\\
			\quad Decomposition, $\mathcal B_h^{\mathrm{Don}}$ & 0.97 & 1.25 & 1.59 & 1.85\\
			& (0.50) & (0.59) & (0.68) & (0.82)\\
			\quad E-HELP-IV, $\mathcal B_h^{\mathrm{E\text{-}HELP}}$ & 0.60 & 0.90 & 1.21 & 1.65\\
			& (0.41) & (0.59) & (0.75) & (1.03)\\
			\addlinespace
			\emph{Multiple confounders} \\
			\quad Targeting-rule HELP-IV, $\mathcal B_h^{\mathrm{HELP\text{-}T}}$ & 0.21 & 0.35 & 0.50 & 0.64\\
			& (0.31) & (0.26) & (0.24) & (0.32)\\
			\bottomrule
		\end{tabular}

		\vspace{0.6em}
		{\footnotesize
			\begin{minipage}{0.92\textwidth}
				\textit{Notes.} Cumulative fiscal multipliers on the $46$-state
				defense-spending panel, annual, $1969$--$2007$. $\mathcal B_h^{\mathrm{SS}}$ is the
				shift-share IV multiplier instrumented by $\sig G_t$. The single-confounder
				rows hold the interest-rate path fixed. $\mathcal B_h^{\mathrm{HELP}}$ is
				backward-looking HELP-IV, $\mathcal B_h^{\mathrm{Don}}$ is the decomposition of
				\citet{donaldson2026}, and $\mathcal B_h^{\mathrm{E\text{-}HELP}}$ is E-HELP-IV
				with two monetary proxies. $\mathcal B_h^{\mathrm{HELP\text{-}T}}$
				is targeting-rule HELP-IV with the state MPC and income-tax share
				as targeting characteristics. Standard errors are in parentheses. For
				$\mathcal B_h^{\mathrm{SS}}$, the standard errors come from a block bootstrap
				(Appendix~\ref{app:boot}). For $\mathcal B_h^{\mathrm{Don}}$, they use the
				baseline covariance estimator of \citet[Appendix~A.7]{donaldson2026}. For
				the HELP-IV rows, they are analytic, and a block bootstrap
				gives similar values (Appendix~\ref{app:se-donaldson},
				Table~\ref{tab:app-linboot}). Targeting-rule standard errors contain only
				the time-series component, which gives the complete standard error under
				exact sample balance (Appendix~\ref{app:se-fiscal}).
		\end{minipage}}
	\end{table}

	We also estimate the fiscal multiplier with
	E-HELP-IV, introduced in Section~\ref{sec:forward-portable-section}, which allows
	outcomes to respond to the anticipated interest-rate path. We use the
	monetary proxies of \citet{romerromer2004} and
	\citet{gertlerkaradi2015} in a four-variable SVAR that also contains national
	defense spending, ordered first, and the federal funds rate. The SVAR delivers
	the identified shocks and the interest-rate responses from
	which the weights of the composite instrument are constructed. Combining the
	two monetary policy shocks with the defense-spending shock holds the
	interest-rate path almost fixed, leaving $2.9$ percent of the response of the
	rate through $t+5$ unabsorbed (the diagnostic of
	Section~\ref{sec:forward-portable-section}). The first-stage $F$ statistic
	lies between $16.9$ and $33.1$ across years. The resulting multiplier rises from $0.60$ in year 1 to $1.65$ in year 4 and
		differs from backward-looking HELP-IV by at most $0.11$ in every year.
		Because the two monetary proxies do not offset the rate path completely, this
		estimate measures the information-constrained effect rather than the
		forward-looking partial-equilibrium effect.

	Finally, we estimate the fiscal multiplier allowing for unspecified one-way
	confounders in addition to the interest rate, using the targeting-rule
	HELP-IV of Section~\ref{sec:combined-conf}. Following
	\citet{manuel2026}, we project defense exposure on predetermined state
	characteristics that govern its allocation and use the residual from that
	projection in the instruments, which is equivalent to controlling for the
	interactions of the characteristics with the estimated shocks
	(Appendix~\ref{app:targeting}). The first characteristic is the state-level
	marginal propensity to consume $\mathrm{MPC}_i$ of \citet{bellifemine2023},
	which, due to data availability, we measure in $2011$ and assume to be a
	persistent state characteristic. The second is the state income-tax
	share $\mathrm{Tax}_i$, tax revenue relative to personal income over
	$1964$--$1968$, the years over which exposure is measured. We select it from a
	set of economic, fiscal, and demographic characteristics in the Correlates of
	State Policy database as the characteristic, observed for every state, that
	adds most to the explanatory power of the MPC.\footnote{\citet{manuel2026} builds
		his rule from the same MPC measure together with trade openness and
		per capita income. Neither of these two characteristics predicts defense
		exposure in our sample once the MPC is included. The estimates
		reported below are similar when we target on the MPC alone, which carries
		most of the predictive power of the rule.} The estimated targeting rule is
	\begin{align}
		\label{eq:app-don-targeting-rule}
		\sig
		&=
		\underset{(1.25)}{4.69}
		-\underset{(3.16)}{9.93}\,\mathrm{MPC}_i
		-\underset{(0.068)}{0.123}\,\mathrm{Tax}_i,
		\qquad
		R^2=0.23,
		\quad
		N=46,
	\end{align}
	with heteroskedasticity-robust standard errors in parentheses and $p$-values
	of $0.003$ for the MPC and $0.077$ for the tax share.

	Table~\ref{tab:app-don} reports the estimates. The multiplier rises from
	$0.21$ in year 1 to $0.64$ in year 4, magnitudes comparable to those
	\citet{manuel2026} obtains once he controls for his targeting rule. It stays
	below one in every year, and it lies below each of the single-confounder
	estimates, which suggests that aggregate channels beyond the interest rate
	contribute to the conventional multiplier.  These point estimates are closer to those of the Neoclassical model of \citet{nakamurasteinsson2014}, whose relative multiplier lies below $0.5$, than to those of their New Keynesian model with consumption--labor complementarity. The relative multiplier of the latter model, about $1.5$, matches the
	conventional estimate. The single-confounder estimates in year 2, $1.00$ for
	HELP-IV and $0.90$ for E-HELP-IV, lie instead close to the relative multiplier
	of $0.83$ in their baseline New Keynesian model with separable preferences.

	\subsection{Regional effects of stock-wealth shocks}
	\label{sec:app-cns}

	The second application revisits CNS, a design with multiple two-way confounders. The object of interest is the response of county employment and payroll to a
	stock-market gain when the paths of the other aggregates that move with the
	market are held fixed.   To identify the target, CNS exploit variation in lagged county stock wealth relative to local labor
	income and estimate the following equation:

	\begin{align}
		\label{eq:cns-twfe}
		\Delta_{a,t-1,t+h}y
		&=
		\mathcal B_h^{\mathrm{CNS}}
		\left(
		S_{a,t-1}R_{a,t-1,t}
		\right)
		+
		\Gamma_hX_{a,t-1}
		+
		\epsilon_{a,t-1,t+h}.
	\end{align}

	\noindent Here, $\Delta_{a,t-1,t+h}y$ is the change in log county employment or payroll
	between $t-1$ and $t+h$, $S_{a,t-1}$ is stock wealth relative to local labor
	income, and $R_{a,t-1,t}$ is the county portfolio return. The outcome and the
	treatment are thus a transformed outcome and a unit-level treatment as in
	Section~\ref{sec:generalized-target}, so the target is the partial-equilibrium
	effect $\mathcal B_h^{P}$ defined there. $X_{a,t-1}$ includes county fixed effects, state-by-quarter fixed effects,
	eight lags of the wealth-return interaction, a Bartik industry control, and
	interactions of $S_{a,t-1}$ with four aggregate confounders.

	These four aggregate confounders are the holding return on a five-year
	zero-coupon Treasury bond, national house-price growth, and the present-value
	growth of aggregate labor income and of non-corporate business income. CNS
	include them because they regard differential responses to aggregates that
	co-move with the market as the main threat to identification, and the
	interactions are intended to absorb those responses through a control
	function. As Section~\ref{sec:why-instrument} explains, however, such controls
	restrict realized values of the confounders and do not hold their paths fixed,
	so we re-estimate the specification with composite HELP-IV.\footnote{We adopt composite HELP-IV rather than multiple-confounder HELP-IV
		because the multiple-confounder first stage is large relative to the sample, which is
		the case in which composite HELP-IV is preferable, as
		Section~\ref{sec:additional-composite-help} explains. At the two-year
		horizon, multiple-confounder HELP-IV would instrument the stock return together with
		each of the four confounders over eight quarters. Its first stage would
		therefore estimate $33$ coefficients from $96$ quarters of aggregate
		variation.}

	We estimate the model on the same sample as CNS, $1991$Q$3$ through
	$2015$Q$2$.\footnote{The county panel and the aggregate series are taken from
		the replication package of \citet{chodorowreichnenovsimsek2021}, available at
		\url{https://www.openicpsr.org/openicpsr/project/123521/version/V1/view}.}
	Composite HELP-IV holds the paths of the four confounders fixed through its
	instrument rather than through controls. We estimate the shocks from a
	five-variable national SVAR in the aggregate market return and the four
	confounders, identified by a Cholesky decomposition with the market return
	ordered first. The composite instrument is then formed as in
	equation~\eqref{eq:joint-composite-instrument}, with the market-return shock in
	place of the treatment proxy and the four confounder shocks in place of the
	confounder proxies. Its offset weights solve recursion~\eqref{eq:composite-weights},
	with the SVAR impulse responses in place of the proxy cross moments. The
	composite instrument therefore moves the market return while leaving the four
	confounder paths through $t+h$ unchanged.\footnote{Ordering the
		market return last lets it respond on impact to the four confounder shocks,
		so that the market-return shock no longer moves the confounders on impact.
		This ordering leaves the estimates little changed. The ordering
		within the confounder block is irrelevant, because the estimator depends only
		on the space spanned by those shocks.}
	Accordingly, we estimate equation~\eqref{eq:cns-twfe} without the four
	interactions and instrument $S_{a,t-1}R_{a,t-1,t}$ with the product of the
	composite instrument and the county's stock-wealth exposure, averaged over
	$1990$--$1993$.\footnote{This window overlaps the first two years of the
		sample. Measuring exposure over the four quarters before the sample begins
		changes the nontradable payroll coefficient from $0.83$ to $0.91$, with a
		standard error of $0.77$.} Inference follows
	Appendix~\ref{app:svar-composite} and accounts for the estimation of the SVAR.

	Table~\ref{tab:app-cns-main} reports the estimated labor market responses of
	employment and payroll at $h=7$, two years after the shock, as in CNS, who focus on this horizon because the labor market response starts to stabilize there.
	The row \emph{CNS benchmark} reproduces Table~1 of CNS. To isolate the role
	of the confounder controls, the row \emph{No confounder controls} drops the
	four interactions while keeping the remaining controls, and the row
	\emph{Composite HELP-IV} applies the estimator described above. \emph{No
		confounder controls} is the common reference point for both corrections:
	comparing it with \emph{CNS benchmark} isolates the contribution of the
	control function, and comparing it with \emph{Composite HELP-IV} shows the
	effect of holding the four confounder paths fixed under the maintained
	identification assumptions. Composite HELP-IV also uses shocks estimated from an
	SVAR and an exposure measure averaged over a fixed period, so the second
	comparison reflects these design choices as well as holding the confounder paths
	fixed.

	We highlight two main findings. First, measured from the CNS benchmark,
	composite HELP-IV moves every estimate by more than dropping the confounder
	controls does. Consider the
	nontradable payroll coefficient in Column~(4), which in the CNS model measures
	the local income that the additional spending generates. It equals $3.24$
	basis points per one percent of local annual labor income in the benchmark
	specification and $2.62$ once the confounder controls are omitted, but only
	$0.83$ once the confounder paths are held fixed. Similarly, except for tradable employment, the composite HELP-IV estimates lie
	between $31$ and $81$ percent below the CNS benchmark for the remaining
	coefficients. Under the maintained assumptions, the partial-equilibrium
	effect that CNS seek to measure is therefore considerably smaller than their benchmark
	estimate. On a common sample, the two nontradable payroll estimates differ by
	$2.28$ basis points, with a standard error of $0.85$.

	Second, the difference between the nontradable and the tradable responses is
	much smaller under composite HELP-IV than in the CNS benchmark. In the CNS
	benchmark the nontradable industries respond strongly while the tradable
	industries do not respond, with employment coefficients of $2.02$ and $-0.11$
	and payroll coefficients of $3.24$ and $0.71$. CNS interpret this difference
	as evidence that the response operates through local consumption demand,
	because news about productivity or a change in the cost of capital would raise
	activity in both sectors. Under composite HELP-IV, the two sectors respond by similar amounts, with employment coefficients of $0.39$ and $0.43$ and payroll coefficients of $0.83$ and $0.49$. These estimates therefore give weaker support to the consumption wealth effect on which the remainder of the CNS analysis rests.

	\begin{table}[!ht]
		\centering
		\caption{Stock-wealth effects on county employment and payroll at the two-year horizon}
		\label{tab:app-cns-main}
		\adjustbox{max width=\textwidth}{%
			\begin{tabular}{lcccccc}
				\toprule
				& \multicolumn{2}{c}{All industries}
				& \multicolumn{2}{c}{Nontradable}
				& \multicolumn{2}{c}{Tradable}\\
				\cmidrule(lr){2-3}\cmidrule(lr){4-5}\cmidrule(lr){6-7}
				& Employment & Payroll & Employment & Payroll & Employment & Payroll\\
				& (1) & (2) & (3) & (4) & (5) & (6)\\
				\midrule
				CNS benchmark & 0.77 & 2.18 & 2.02 & 3.24 & $-0.11$ & 0.71\\
				& (0.36) & (0.63) & (0.80) & (1.00) & (0.57) & (0.69)\\
				\addlinespace
				No confounder controls & 0.65 & 2.09 & 1.17 & 2.62 & $-0.23$ & 0.84\\
				& (0.32) & (0.57) & (0.97) & (1.02) & (0.53) & (0.60)\\
				\addlinespace
				Composite HELP-IV       & 0.45 & 1.02 & 0.39 & 0.83 & 0.43 & 0.49\\
				& (0.43) & (0.54) & (0.63) & (0.74) & (0.69) & (0.93)\\
				\bottomrule
		\end{tabular}}

		\vspace{0.6em}
		{\footnotesize
		\begin{minipage}{0.96\textwidth}
			\textit{Notes.} Horizon $h=7$ response of county employment and payroll
			to a stock-wealth increase equal to one percent of local annual labor
			income. The sample and variables follow
			\citet{chodorowreichnenovsimsek2021}. Coefficients are reported in basis
			points. The CNS benchmark enters the four aggregate confounders as
			interactions with county stock exposure, the second row drops those
			interactions and keeps the remaining controls, and composite HELP-IV replaces
			them with the composite instrument. Standard errors are in parentheses. For the first two rows they
			are clustered by county and quarter. For composite HELP-IV they are analytic
			standard errors that include estimation of the national SVAR
			(Appendix~\ref{app:se-donaldson}). A block bootstrap gives similar
			values (Table~\ref{tab:app-linboot}).
		\end{minipage}}
	\end{table}

	\begin{table}[!ht]
		\centering
		\caption{Implied economic magnitudes}
		\label{tab:app-cns-implied}
		\begin{tabular}{lccc}
			\toprule
			& Formula & CNS & Composite HELP-IV\\
			\midrule
			MPC out of stock wealth (cents per \$)
			& $b_{\mathrm{NT}}/[(1-\alpha)M]$
			& 3.2
			& 0.8\\
			Aggregate labor bill (\%)
			& $\lambda_{L}b_{\mathrm{NT}}$
			& 1.7
			& 0.44\\
			Aggregate hours (\%)
			& $\lambda_{H}b_{\mathrm{NT}}$
			& 0.77
			& 0.27\\
			\bottomrule
		\end{tabular}

		\vspace{0.6em}
		{\footnotesize
		\begin{minipage}{0.95\textwidth}
			\textit{Notes.} The coefficient $b_{\mathrm{NT}}$ is the nontradable payroll
			estimate of Table~\ref{tab:app-cns-main}. For the MPC, $(1-\alpha)=2/3$ is the
			labor share and $M=1.5$ is the local multiplier. For the aggregate labor bill,
			$\lambda_L=S^A\times0.20$, where $S^A=2.67$ is the aggregate
			wealth-to-labor-bill ratio and $0.20$ is the stock-valuation shock. For
			aggregate hours, $\lambda_H=\lambda_L/(1+\kappa^A)$, where the wage-adjustment
			parameter $\kappa^{A}$ is recomputed in each column from the all-industry
			payroll and employment coefficients of the same specification. This gives
			$1.26$ in the CNS column, which \citet{chodorowreichnenovsimsek2021} round to
			$1.3$, and $0.62$ under composite HELP-IV. The two aggregate rows hold
			monetary policy unchanged, which is a counterfactual distinct from the
			fixed-path restriction behind $b_{\mathrm{NT}}$, and
			\citet{chodorowreichnenovsimsek2021} report them as lower bounds.
		\end{minipage}}
	\end{table}

CNS carry the estimated nontradable payroll coefficient, $b_{\mathrm{NT}}$,
into their model of consumption wealth effects with geographic heterogeneity,
an exercise that rests on the portability of the estimated effect. We take this
model, and hence its consumption channel, as given and ask how its implications
change when $b_{\mathrm{NT}}$ holds the confounder paths fixed. In that
model, $b_{\mathrm{NT}}$ is the product of the household marginal propensity
to consume (MPC) out of stock wealth, the labor share $1-\alpha$, and the
local spending multiplier $M$, so the MPC equals
$b_{\mathrm{NT}}/[(1-\alpha)M]$. The same coefficient also scales the
aggregate responses to a twenty percent increase in stock valuations that
monetary policy does not counter. The aggregate labor bill rises by
$\lambda_{L}b_{\mathrm{NT}}$, where $\lambda_{L}$ scales the valuation shock
by the aggregate ratio of stock wealth to the labor bill. This mapping holds
because the model equates the proportional response of the aggregate labor
bill with that of local nontradable payroll. Aggregate hours rise by
$\lambda_{H}b_{\mathrm{NT}}$, with $\lambda_{H}=\lambda_{L}/(1+\kappa^{A})$,
because the aggregate degree of wage adjustment $\kappa^{A}$ divides that
response between wages and hours. We recompute $\kappa^{A}$ in each
specification from its all-industry payroll and employment coefficients.

	Table~\ref{tab:app-cns-implied} compares the three economic magnitudes implied
	by composite HELP-IV with those implied by the CNS benchmark. All three are
	substantially smaller under composite HELP-IV. The implied MPC declines from
	$3.2$ to $0.8$ cents per dollar per year. CNS note that their estimate of
	$3.2$ cents is close to the annual discount rates commonly assumed in the
	literature and is therefore consistent with the permanent-income benchmark.
	The composite HELP-IV estimate falls well below that benchmark and is closer to
	the small stock-wealth effects estimated from regional variation by
	\citet{casequigleyshiller2005}.

	The implied responses of the aggregate labor bill and aggregate hours to a
	twenty percent increase in stock valuations likewise decline, from $1.7$ to
	$0.44$ percent and from $0.77$ to $0.27$ percent, respectively. Although CNS
	report these responses as lower bounds, the decline implies a smaller
	demand-stabilization role for policy after a stock-valuation shock.
	Holding the confounder paths fixed thus changes not only the estimated
	coefficient but also the model-based implications that CNS draw from it.

	\section{Conclusion}
	\label{sec:conclusion}

TWFE-IV regressions in exposure designs are widely used to estimate dynamic
partial-equilibrium effects, on the assumption that their time effects absorb
aggregate confounders. We show that time effects generally suffice to recover
these effects only under two conditions, namely that exposures to the treatment
and to the confounders are uncorrelated and that the confounders do not feed
back into the treatment. When either condition fails, the TWFE-IV estimate
suffers from the HEDGE bias. We then develop the HELP-IV framework, which
corrects for the HEDGE bias by augmenting the panel IV regression with
path controls and proxy instruments, and adapt it to four more
general environments (multiple exposure margins, multiple confounders,
forward-looking outcomes, and transformed treatments and outcomes). Our
simulations show that, in finite samples, HELP-IV is centered on the target
effect and the inference developed in this paper delivers coverage generally
close to the nominal level. Our applications show that HELP-IV can deliver
estimates of the state-level fiscal multiplier and of stock-wealth effects on
the labor market that differ from previous TWFE-IV estimates, with distinct
implications for structural models.

This paper opens avenues for future research in the following directions. First, the
core idea of HELP-IV is that the moment conditions of a panel IV regression can
implicitly deliver the policy counterfactuals of \citet{simszha2006md} and
\citet{mckaywolf2023}. This idea can be extended to models with heterogeneous
treatment effects, spatial or production linkages across units, and
state-dependent dynamics. Second, HELP-IV is a simple yet general
regression-based design. As granular panel data and exposure measures become
increasingly available, including those constructed from firm and transaction
records with machine-learning and text-based methods, it can be used to study
causal relationships with exposure designs in macro-finance
\citep{ottonellowinberry2020,jeenas2026}, international economics
\citep{amitiitskhokikonings2022}, household finance
\citep{miansufi2014,guren2021}, and labor economics \citep{autordornhanson2013},
as well as in related business research, such as pricing and marketing.

	\bibliographystyle{plainnat}
	\bibliography{references}

	\clearpage
	\appendix
	\numberwithin{equation}{section}
	\numberwithin{table}{section}
	\numberwithin{figure}{section}
	\numberwithin{proposition}{section}
	\numberwithin{lemma}{section}
	\numberwithin{corollary}{section}
	\numberwithin{assumption}{section}
	\numberwithin{remark}{section}
	\titleformat{\section}
	{\normalfont\Large\scshape\filcenter}{Appendix \thesection.}{0.7em}{}
	\section{Identification results}
	\label{app:proofs}

	Throughout this appendix, $\dot s_{ig}=\sig-\E[\sig]$,
	$\sigma_s^2=\Var(\sig)>0$, and aggregate variables are centered.
	Tildes on panel variables denote population residualization on unit and date
	effects. The controlled treatment path $\breve G_j$ is defined in
	Section~\ref{sec:dyn-pe-effect}. Set
	$\sigma_x^2=\Var(\epsilon_{x,t})$ and
	$\lambda_x=\E[z_{x,t}\epsilon_{x,t}]=\kappa_x\sigma_x^2$,
	$x\in\{g,r\}$; the proxy scale $\lambda_x$ is distinct from the time effect
	$\lambda_t$. With $K$ confounders,
	$\boldsymbol\Lambda_r=\E[\boldsymbol z_{r,t}\boldsymbol\epsilon_{r,t}']$ is
	diagonal and nonsingular. With additional controls, the arguments require the
	same exclusion and rank conditions for the residualized moments; the product
	factorizations below refer to unit and date effects alone.
	With a single confounder, the composite instrument is
	\begin{align}
		\mathrm z_{g,t,h}^C
		&=
		z_{g,t}+\sum_{s=0}^{h}w_s z_{r,t+s},
		\label{eq:path-neutral-instrument}
	\end{align}
	the scalar case of \eqref{eq:joint-composite-instrument}, whose offset
	weights $w_0,\ldots,w_h$ leave the confounder path unchanged,
	\begin{align}
		\E\bigl[R_{t+\ell}\mathrm z_{g,t,h}^C\bigr]
		&=
		0,
		\qquad \ell=0,\ldots,h,
		\label{eq:path-neutrality}
	\end{align}
	a property that we call path neutrality. In structural units, the offsetting confounder shocks $e_0,\ldots,e_h$ solve
	\begin{align}
		R_\ell^g+\sum_{s=0}^{\ell}e_sR_{\ell-s}^r
		&=
		0,
		\qquad \ell=0,\ldots,h,
		\label{eq:help-offset-experiment}
	\end{align}
	and $w_s=e_s\lambda_g/\lambda_r$ converts them into proxy units.

	\subsection{Proof of Proposition~\ref{lemma:hedge-bias}}
	\label{proof:hedge-bias}
	\begin{proof}
		Exposure is independent of the centered aggregate variables, so
		residualization multiplies each of them by $\dot s_{ig}$, for example
		$\widetilde{\sig G_t}=\dot s_{ig}G_t$, and cross moments factor, for example
		\[
		\E[\widetilde{\sig z_{g,t}}\widetilde{\sig G_t}]
		=\sigma_s^2\E[z_{g,t}G_t].
		\]
		In addition, $\E[\dot s_{ig}\uisr]=0$.
		Iterating \eqref{eq:outcome-dyn} and \eqref{eq:pe-outcome-controlled},
		and using $\E[\dot s_{ig}\sir]=\phi\sigma_s^2$ from \eqref{eq:exposure-dyn},
		panel exclusion, and causality gives
		\begin{align*}
			\E[\widetilde{\sig z_{g,t}}\widetilde Y_{i,t+h}]
			&=\sigma_s^2\lambda_g
			\left\{\sum_{j=0}^h\varphi_jG_{h-j}^g
			+\phi\sum_{j=0}^h\bar\gamma_jR_{h-j}^g\right\},\\
			\E[\widetilde{\sig z_{g,t}}\widetilde{\sig G_t}]
			&=\sigma_s^2\lambda_gG_0^g,\\
			\breve Y_{i,h}&=\sig\sum_{j=0}^h\varphi_j\breve G_{h-j}.
		\end{align*}
		The TWFE-IV coefficient is the ratio of the first two moments, and
		$\beta_h^P=\breve Y_{i,h}/(\sig\breve G_0)$, which gives
		\eqref{eq:hedge-bias-lemma}. If $G_j^r=0$ for all $j\geq0$, then
		$\breve G_j=G_j^g$ for all $j\leq h$, and the first difference is zero.
	\end{proof}

	\subsection{Proof of Proposition~\ref{prop:help-identification}}
	\label{app:proof-help-identification}
	\begin{proof}
		Let $\underline R_h^g=(R_0^g,\ldots,R_h^g)'$,
		$\underline e_h=(e_0,\ldots,e_h)'$, and
		\[
		T_h^r=(R_{\ell-s}^r{\bf1}\{s\leq\ell\})_{\ell,s=0}^h,
		\qquad d_h=(G_0^r,0,\ldots,0)',
		\]
		so that \eqref{eq:help-offset-experiment} is
		$\underline e_h=-(T_h^r)^{-1}\underline R_h^g$.
		Write $\Sigma_h=\E[\widetilde{\mathbf Z}_{it,h}
		\widetilde{\boldsymbol X}_{it,h}']$. Dividing each proxy row by its
		nonzero scale gives
		\[
		\{\sigma_s^2\operatorname{diag}(\lambda_g,
		\lambda_r I_{h+1})\}^{-1}\Sigma_h
		=\begin{pmatrix}G_0^g&\underline R_h^{g\prime}\\d_h&T_h^{r\prime}\end{pmatrix}.
		\]
		The row vector $(1,\underline e_h')$ therefore satisfies
		\[
		(1,\underline e_h')\begin{pmatrix}G_0^g&\underline R_h^{g\prime}\\d_h&T_h^{r\prime}\end{pmatrix}
		=\bigl(G_0^g+\underline e_h'd_h,\,0'\bigr)
		=\bigl(\breve G_0,\,0'\bigr).
		\]
		Let $y_h^g$ and $\underline y_h^r=(y_{h,0}^r,\ldots,y_{h,h}^r)'$ denote the
		reduced-form moments $\E[\widetilde{\mathbf Z}_{it,h}\widetilde Y_{i,t+h}]$
		divided by the same row scales, so that $y_{h,s}^r$ is the exposure-weighted
		response of $Y_{i,t+h}$ to $\epsilon_{r,t+s}$. As in the proof of
		Proposition~\ref{lemma:hedge-bias},
		\[
		y_h^g=\sum_{j=0}^h\varphi_jG_{h-j}^g+\phi\sum_{j=0}^h\bar\gamma_jR_{h-j}^g,
		\qquad
		y_{h,s}^r=\sum_{j=0}^{h-s}\varphi_jG_{h-s-j}^r
		+\phi\sum_{j=0}^{h-s}\bar\gamma_jR_{h-s-j}^r;
		\]
		the terms in $\uisr$ and $u_{it}$ vanish by independence and exclusion. The
		offsets $e_0,\ldots,e_h$ hold the path fixed, so
		$R_\ell^g+\sum_{s\leq\ell}e_sR_{\ell-s}^r=0$ and
		$G_\ell^g+\sum_{s\leq\ell}e_sG_{\ell-s}^r=\breve G_\ell$ for $\ell=0,\ldots,h$.
		Hence $y_h^g+\underline e_h'\underline y_h^r=\sum_{j=0}^h\varphi_j\breve G_{h-j}
		=\breve G_0\beta_h^P$. Applying $(1,\underline e_h')$ to the normalized moment system
		\eqref{eq:help-population-system} gives
		$\breve G_0b_h^{\mathrm{HELP}}=\breve G_0\beta_h^P$.
		By the Schur complement with respect to the block $T_h^{r\prime}$, whose
		determinant is $(R_0^r)^{h+1}$,
		\[
		\det\Sigma_h=(\sigma_s^2)^{h+2}\lambda_g\lambda_r^{h+1}
		\breve G_0(R_0^r)^{h+1}\ne0,
		\qquad \E[R_tz_{r,t}]=\lambda_rR_0^r\ne0.
		\]
		The solution is unique and $b_h^{\mathrm{HELP}}=\beta_h^P$.
	\end{proof}

	\subsection{Proof of Corollary~\ref{cor:margins}}
	\label{app:margins}
	\label{app:margins-model}
	\label{app:margins-iv}
	\begin{proof}
		Let $f_{t,h}=(G_t,R_t,\ldots,R_{t+h})'$,
		$p_{t,h}=(z_{g,t},z_{r,t},\ldots,z_{r,t+h})'$, and
		\begin{equation}
			\mathbf S=\Var(\mathbf s_{ig}),\qquad
			\mathbf A_h=\E[p_{t,h}f_{t,h}'].
			\label{eq:margin-rank}
		\end{equation}
		Use the exposure-first ordering
		$\boldsymbol X_{it,h}=\mathbf s_{ig}\otimes f_{t,h}$ and
		$\mathbf Z_{it,h}=\mathbf s_{ig}\otimes p_{t,h}$. Independence gives
		\begin{equation}
			\boldsymbol\Sigma_h=\mathbf S\otimes\mathbf A_h,\qquad
			\boldsymbol\Sigma_h^{-1}=\mathbf S^{-1}\otimes\mathbf A_h^{-1}.
			\label{eq:margins-kronecker-first-stage}
		\end{equation}
		For every $m$ and every $j,s\in\{0,\ldots,h\}$,
		\[
		\E[(s_{ig,m}-\E s_{ig,m})\uisr z_{r,t+j}R_{t+s}]
		=\E[(s_{ig,m}-\E s_{ig,m})\uisr]\E[z_{r,t+j}R_{t+s}]=0,
		\]
		and the corresponding treatment-proxy and idiosyncratic-error moments are
		zero. Because $\E[(\mathbf s_{ig}-\E\mathbf s_{ig})\sir]=\mathbf S\boldsymbol\phi$
		with $\boldsymbol\phi=(\phi_1,\ldots,\phi_M)'$, the reduced-form moment
		factors as $\E[\widetilde{\mathbf Z}_{it,h}\widetilde Y_{i,t+h}]
		=(\mathbf S\otimes I)\mathbf a_h$, where the block $\mathbf a_{h,m}$ is the
		reduced-form vector of a single-margin model with outcome loadings
		$\beta_{m,\ell}$ and exposure-correlation coefficient $\phi_m$, divided by its exposure
		variance. Hence
		$\boldsymbol\vartheta_h=(\mathbf S^{-1}\otimes\mathbf A_h^{-1})
		(\mathbf S\otimes I)\mathbf a_h=(I_M\otimes\mathbf A_h^{-1})\mathbf a_h$,
		and each block $\mathbf A_h^{-1}\mathbf a_{h,m}$ solves a single-margin
		system. The row transformation in the preceding proof gives its first element,
		so
		\[
		(I_M\otimes e_1')\boldsymbol\vartheta_h
		=(\beta_{1,h}^P,\ldots,\beta_{M,h}^P)'=\boldsymbol\beta_h^P.
		\]
		Nonsingularity of $\mathbf S$ and $\mathbf A_h$ gives uniqueness.
		Under Assumption~\ref{ass:regular}, convergence of the sample moments and
		the continuous mapping theorem give consistency.
	\end{proof}

	\subsubsection{Network exposure}
	\label{app:margins-network}
	For a predetermined network $(a_{ij})$, the choice
	\begin{equation}
		s_{ig,1}=\sig,\qquad s_{ig,2}=\sum_j a_{ij}s_{jg}
		\label{eq:network-margin}
	\end{equation}
	is covered by the same proof; $s_{ig,2}$ is a known exposure mapping in the
	sense of \citet{aronow2017}. Identification requires $\det\mathbf S>0$, so a
	network margin proportional to own exposure supplies no additional
	coefficient, as in the reflection problem of \citet{manski1993} and the
	identification condition for peer effects of \citet{bramoulle2009}.

	\subsection{Proof of Corollary~\ref{cor:vec-conf}}
	\label{app:confounders}
	\begin{proof}
		Let $\underline{\mathbf v}_h$ and $\underline\Sigma_{rz,h}$ be as in
		Section~\ref{sec:composite-estimation}, and let
		$p_{r,t,h}=(\boldsymbol z_{r,t}',\ldots,\boldsymbol z_{r,t+h}')'$.
		The confounder block is
		\begin{equation}
			\E[p_{r,t,h}\mathbf R_{t,h}']=\underline\Sigma_{rz,h}',
			\label{eq:multiple-confounder-first-stage}
		\end{equation}
		which is nonsingular because $\det\underline\Sigma_{rz,h}=(\det\Sigma_{rz,0})^{h+1}\ne0$.
		Let $\underline{\boldsymbol w}_h=-\underline\Sigma_{rz,h}^{-1}\underline{\mathbf v}_h$. Then
		\[
		(1,\underline{\boldsymbol w}_h')\E[(z_{g,t},p_{r,t,h}')'
		(G_t,\mathbf R_{t,h}')]
		=\bigl(\E[G_t\mathrm z_{g,t,h}^C],0'\bigr),
		\]
		where $\mathrm z_{g,t,h}^C$ is the composite \eqref{eq:joint-composite-instrument}.
		Let $\mathbf R_\ell^r$ denote the $K\times K$ matrix of responses of
		$\mathbf R_{t+\ell}$ to $\boldsymbol\epsilon_{r,t}$, $\mathbf R_\ell^g$ the response of
		$\mathbf R_{t+\ell}$ to $\epsilon_{g,t}$, and $\mathbf G_\ell^r$ the $K$-vector of
		responses of $G_{t+\ell}$ to $\boldsymbol\epsilon_{r,t}$. Then
		$\Sigma_{rz,\ell}=\mathbf R_\ell^r\boldsymbol\Lambda_r$ and
		$\mathbf v_\ell=\lambda_g\mathbf R_\ell^g$, so the recursion gives
		$\boldsymbol w_\ell=\lambda_g\boldsymbol\Lambda_r^{-1}\mathbf e_\ell$, where
		$\{\mathbf e_\ell\}$ is the sequence of offsetting confounder shocks that holds the
		$K$ paths fixed. The composite therefore moves $G_{t+\ell}$ by
		$\lambda_g\{G_\ell^g+\sum_{s\leq\ell}(\mathbf G_{\ell-s}^{r})'\mathbf e_s\}
		=\lambda_g\breve G_\ell$ and leaves every confounder path unchanged. In
		particular, $\E[G_t\mathrm z_{g,t,h}^C]=\lambda_g\breve G_0\ne0$, and the
		same row operation on the reduced form gives
		$\lambda_g\sum_{j=0}^h\varphi_j\breve G_{h-j}
		=\lambda_g\breve G_0\beta_h^P$. The exposure factor $\sigma_s^2$
		cancels, the complete moment matrix is nonsingular by the Schur complement,
		and $b_h^{\mathrm{HELP}}=\beta_h^P$.
	\end{proof}

	\subsection{Proof of Proposition~\ref{prop:composite-conf}}
	\label{app:composite-conf}
	\begin{proof}
		By causality, \eqref{eq:joint-composite-orthogonality} is the block-triangular system
		$\underline\Sigma_{rz,h}\underline{\boldsymbol w}_h=-\underline{\mathbf v}_h$,
		whose rows give \eqref{eq:composite-weights}. Since $\det\underline\Sigma_{rz,h}=(\det\Sigma_{rz,0})^{h+1}\ne0$, the weights exist uniquely. With these weights,
		the composite moves $G_{t+\ell}$ by $\lambda_g\breve G_\ell$ and leaves
		every confounder path unchanged (proof of Corollary~\ref{cor:vec-conf}), so
		\begin{align}
			\E[\widetilde{\sig\mathrm z_{g,t,h}^C}\widetilde{\sig G_t}]
			&=\sigma_s^2\lambda_g\breve G_0,\nonumber\\
			\E[\widetilde{\sig\mathrm z_{g,t,h}^C}\widetilde Y_{i,t+h}]
			&=\sigma_s^2\lambda_g\breve G_0\beta_h^P.
			\label{eq:composite-proof-moments}
		\end{align}
		The denominator is nonzero. Their ratio is
		$b_h^{\mathrm{HELP\text{-}C}}=\beta_h^P$.
	\end{proof}

	\subsection{Proof of Corollary~\ref{cor:combined-conf}}
	\label{app:target-identification}
	\begin{proof}
		Let $p^{(1)}_{t,h}$ collect the treatment proxy and the proxies for the
		retained two-way block, and let $\sigma_\perp^2=\E[(\sig^\perp)^2]>0$.
		Projection orthogonality and Assumption~\ref{ass:targeting} imply
		\[
		\E[\sig^\perp\sig]=\sigma_\perp^2,\qquad
		\E[\sig^\perp s_{ir_q}p^{(1)}_{t,h}R_{q,t+j}]
		=\E[\sig^\perp s_{ir_q}]\E[p^{(1)}_{t,h}R_{q,t+j}]=0,
		\quad q>K_1.
		\]
		For a retained two-way confounder, set
		$\tilde\phi_q=\E[\sig^\perp s_{ir_q}]/\sigma_\perp^2$, which can differ from
		$\phi_q$ because $\E[\sig^\perp u_i^{r_q}]$ can be nonzero. Write
		$s_{ir_q}R_{q,t+j}=\tilde\phi_q\sig R_{q,t+j}
		+(s_{ir_q}-\tilde\phi_q\sig)R_{q,t+j}$. The second term is orthogonal to
		$\sig^\perp$ times any aggregate variable, and the unrestricted path
		coefficients $c_{q,h,s}$ absorb the first. Thus the retained moment system is
		the system of Corollary~\ref{cor:vec-conf}, with exposure factor
		$\sigma_\perp^2$ and exposure-correlation coefficients $\tilde\phi_q$.
		By Assumption~\ref{ass:one-way-block}, shocks in the omitted block
		leave both $G$ and the retained confounder paths unchanged, and their
		offsets exist because the impact matrix of the block is nonsingular. Let
		$\breve G_j^{(1)}$ and $\breve G_j^{(1,2)}$ denote the treatment paths
		when the retained paths alone and when all confounder paths are held fixed.
		The offsets of the omitted block therefore preserve the already controlled
		treatment path:
		\[
		\breve G_j^{(1)}=\breve G_j^{(1,2)},\qquad j=0,\ldots,h.
		\]
		The retained system consequently identifies the full fixed-path target,
		$b_h^{\mathrm{HELP\text{-}T}}=\beta_h^P$.
	\end{proof}

	\subsection{Proof of Proposition~\ref{cor:forward-composite-help}}
	\label{app:forward-composite-help-proof}
	\begin{proof}
		For the normalized date-$t$ proxies, exclusion gives
		\begin{align*}
			\E[\widetilde{\sig\mathrm z_{g,t}^{E}}\widetilde{\sig G_t}]
			&=\sigma_s^2\{1+(\mathbf g_0^E)'\mathbf e_H^E\}
			=\sigma_s^2\breve G_{0,H}^E,\\
			\E[\widetilde{\sig\mathrm z_{g,t}^{E}}\widetilde Y_{i,t+h}]
			&=\sigma_s^2\{b_h+(\mathbf e_H^E)'\mathbf c_h^E\}.
		\end{align*}
		The denominator is nonzero, so the IV ratio is $\beta_h^E$ in
		\eqref{eq:forward-portable-vector-form}. Further,
		\[
		\mathbf r_H^g+\mathbf R_H^E\mathbf e_H^E
		=\{I-\mathbf R_H^E(\mathbf R_H^E)^+\}\mathbf r_H^g.
		\]
		This vector is zero if and only if
		$\mathbf r_H^g\in\col(\mathbf R_H^E)$. Under exact spanning, the date-$t$
		shock combination leaves $R_t,\ldots,R_{t+H}$ unchanged. In the linear model,
		the revision of $\E_{t+k}R_{t+k+f}$ caused by date-$t$ shocks equals the
		response of $R_{t+k+f}$, so with $H\geq\bar h+F$ the expectations in
		\eqref{eq:outcome-forward} formed at dates $t,\ldots,t+\bar h$ are also
		unchanged, and $\beta_h^E$ is the forward-looking partial-equilibrium effect.
		Otherwise the
		IV ratio is the information-constrained effect. The ratio identity
		holds in both cases.
	\end{proof}

	\subsection{Proof of Proposition~\ref{prop:generalized}}
	\label{app:generalized}
	\begin{proof}
		Let $e_0,\ldots,e_h$ be the offsetting confounder shocks in
		\eqref{eq:help-offset-experiment}, and define
		\[
		\zeta_{t,h}=z_{g,t}/\lambda_g+
		\sum_{s=0}^h e_sz_{r,t+s}/\lambda_r.
		\]
		The conditional restrictions in part (ii) of Assumption~\ref{ass:generalized} imply,
		for every integrable predetermined loading $a_i$,
		\[
		\E[a_i\zeta_{t,h}]=0,\qquad
		\E[a_i\zeta_{t,h}\epsilon_{g,t}]=\E[a_i],\qquad
		\E[a_i\zeta_{t,h}\epsilon_{r,t+s}]=e_s\E[a_i].
		\]
		Let $\breve{\mathcal G}_i$ and $\breve{\mathcal Y}_i$ denote the unit-specific
		controlled responses in \eqref{eq:generalized-target}, so that
		$\breve{\mathcal Y}_i=\mathcal B_h^P\breve{\mathcal G}_i$, and write
		$\mathcal D_h^{\mathcal G}=\E[\dot s_{ig}\breve{\mathcal G}_i]$.
		Part (i) of the assumption and the exclusion restrictions in parts (i)
		and (iii) give
		\begin{align*}
			\E[\dot s_{ig}\zeta_{t,h}\mathcal G_{i,t+h_{\mathcal G}}]
			&=\mathcal D_h^{\mathcal G},\\
			\E[\dot s_{ig}\zeta_{t,h}\mathcal Y_{i,t+h}]
			&=\mathcal B_h^P\mathcal D_h^{\mathcal G},\\
			\E[\dot s_{ig}\zeta_{t,h}\sig R_{t+s}]&=0,
			\qquad s=0,\ldots,h.
		\end{align*}
		The row operation defined by $\zeta_{t,h}$ eliminates the path block.
		The remaining equation is
		$\mathcal D_h^{\mathcal G}\mathcal B_h^{\mathrm{HELP}}
		=\mathcal D_h^{\mathcal G}\mathcal B_h^P$. The moment matrix has determinant
		\[
		\det\E[\widetilde{\mathbf Z}_{it,h}
		\widetilde{\boldsymbol X}_{it,h}^{\mathcal G\prime}]
		=\lambda_g\lambda_r^{h+1}\mathcal D_h^{\mathcal G}
		(\sigma_s^2R_0^r)^{h+1}\ne0.
		\]
		Uniqueness follows. Under the stated convergence of panel moments,
		\[
		\widehat{\boldsymbol\vartheta}_h^{\mathcal G}
		=\widehat\Sigma_h^{-1}\widehat r_h
		\xrightarrow{p}\Sigma_h^{-1}r_h,
		\qquad \widehat{\mathcal B}_h^{\mathrm{HELP}}\xrightarrow{p}\mathcal B_h^P.
		\]
	\end{proof}

	\subsection{Proof of Proposition~\ref{lemma:raw-fails}}
	\label{app:raw-fails}
	\begin{proof}
		The exclusion restrictions for the proxy and the shifter and the response
		definitions give
		\[
		\mathcal B_h^{\mathrm{TWFE}}=\frac{\mathcal S^{\mathcal Y}_{g,h}}
		{\mathcal S^{\mathcal G}_{g,h}},\qquad
		\mathcal B_h^{\mathrm{SS}}=
		(\Phi_h^G)^{-1}\sum_{\varsigma\in\{g,r\}}\sum_{s\geq0}
		\psi^G_{\varsigma,s}\sigma_\varsigma^2
		\mathcal S^{\mathcal Y}_{\varsigma,h+s}.
		\]
		Subtracting $\mathcal B_h^{\mathrm{TWFE}}$ gives
		\begin{align*}
			\mathcal B_h^{\mathrm{SS}}-\mathcal B_h^{\mathrm{TWFE}}
			&=(\Phi_h^G)^{-1}\sum_{\varsigma\in\{g,r\}}\sum_{s\geq0}
			\psi^G_{\varsigma,s}\sigma_\varsigma^2
			\{\mathcal S^{\mathcal Y}_{\varsigma,h+s}
			-\mathcal B_h^{\mathrm{TWFE}}\mathcal S^{\mathcal G}_{\varsigma,h+s}\}\\
			&=\frac1{\Phi_h^G}\sum_{\varsigma\in\{g,r\}}\sum_{s\geq1}
			\psi^G_{\varsigma,s}\sigma_\varsigma^2
			\{\mathcal S^{\mathcal Y}_{\varsigma,h+s}
			-\mathcal B_h^{\mathrm{TWFE}}\mathcal S^{\mathcal G}_{\varsigma,h+s}\},
		\end{align*}
		because $\mathcal S^{\mathcal Y}_{g,h}
		-\mathcal B_h^{\mathrm{TWFE}}\mathcal S^{\mathcal G}_{g,h}=0$ and
		$\psi^G_{r,0}=0$. Adding $\mathcal B_h^{\mathrm{TWFE}}-\mathcal B_h^P$ proves
		\eqref{eq:raw-fails}.
	\end{proof}

	\section{Estimation and inference with shocks estimated from an SVAR}
	\label{app:svar-inference}

	This appendix extends the results of Section~\ref{sec:help-inference} to
	instruments that are shocks estimated from a structural vector autoregression
	(SVAR), as in both applications. In population, each structural shock is a
	proxy in the sense of \eqref{eq:g-proxy} or \eqref{eq:r-proxy}, with
	$\kappa=1$ and no measurement error, so the identification results of
	Sections~\ref{sec:portable-markov} and~\ref{sec:confounders} apply. The
	estimated shocks, however, depend on the SVAR estimates, whose sampling error
	enters the coefficient at first order through the instruments and, for
	composite HELP-IV, also through the offset weights. The covariance is
	therefore the sandwich of the moment conditions that stack the panel IV
	moment with the SVAR estimating equations.

	\subsection{The SVAR first step}
	\label{app:svar-setup}

	For estimation in this appendix, we assume that the aggregate process admits
	a correctly specified, stable SVAR of fixed order $p_A$,
	\begin{equation}
		\mathbf W_t=\sum_{j=1}^{p_A}A_j\mathbf W_{t-j}
		+\Psi_0\boldsymbol\epsilon_t,
		\label{eq:main-var}
	\end{equation}
	where $\mathbf W_t=(G_t,\mathbf R_t')'$ and
	$\boldsymbol\epsilon_t=(\epsilon_{g,t},\boldsymbol\epsilon_{r,t}')'$ with $K$
	confounders. This finite-order specification is an additional estimation
	assumption; the identification results of Sections~\ref{sec:portable-markov}
	and~\ref{sec:confounders} do not require it. In this appendix, we normalize
	the shocks so that each moves its own variable one for one on impact; their
	variances can differ from one. The estimated shocks are
	$\widehat{\boldsymbol\epsilon}_t=\widehat\Psi_0^{-1}
	(\mathbf W_t-\sum_{j=1}^{p_A}\widehat A_j\mathbf W_{t-j})$.

	Let $\theta$ collect the VAR coefficients, the impact matrix, and any proxy
	moments used for identification, in a smooth, locally identified
	parameterization. The estimator $\widehat\theta$ solves exactly identified
	estimating equations $T^{-1}\sum_tk_t(\widehat\theta)=\mathbf0$, which stack
	the VAR normal equations, the equations for the residual covariance matrix,
	the proxy moments, and the identifying restrictions, as written out in
	\eqref{eq:main-var-primitive}; the restrictions do not
	vary over time. In a proxy-SVAR, a proxy identifies a shock through its
	covariance with the reduced-form residuals; the proxy then enters only
	through $\theta$, and the panel IV uses the estimated shock. Let
	$A_\theta=\E[\partial k_t/\partial\theta']$ be the Jacobian of these
	equations and $\Omega^k=\lim_{T\to\infty}\Var(T^{-1/2}\sum_tk_t)$ their
	long-run variance at the population parameters; Appendix~\ref{app:svar} gives
	both. Lemma~\ref{lem:svar} shows that
	\begin{equation}
		\sqrt T(\widehat\theta-\theta)
		\xrightarrow{d} N\bigl(\mathbf0,\ A_\theta^{-1}\Omega^kA_\theta^{-1\prime}\bigr).
		\label{eq:main-svar-clt}
	\end{equation}

	\subsection{Benchmark HELP-IV}
	\label{app:svar-help}

	For the single-confounder baseline of Section~\ref{sec:help-estimation}, the
	population and sample instrument vectors, marked by the superscript
	$\mathrm S$ for SVAR shocks, are
	\begin{equation}
		\mathbf Z^{\mathrm S}_{it,h}
		=s_{ig}(\epsilon_{g,t},\epsilon_{r,t},\ldots,\epsilon_{r,t+h})',
		\qquad
		\widehat{\mathbf Z}^{\mathrm S}_{it,h}
		=s_{ig}(\widehat\epsilon_{g,t},\widehat\epsilon_{r,t},\ldots,
		\widehat\epsilon_{r,t+h})',
		\label{eq:main-shock-instruments}
	\end{equation}
	and $\widehat b_h^{\mathrm{HELP},\mathrm S}$ is the estimator
	\eqref{eq:main-estimator} with $\widehat{\mathbf Z}^{\mathrm S}_{it,h}$ in
	place of $\mathbf Z_{it,h}$. Let $\Sigma_h$ be \eqref{eq:main-moment}
	evaluated at $\mathbf Z^{\mathrm S}_{it,h}$. The derivative of the IV moment
	with respect to the SVAR parameters is
	\begin{equation}
		D_h=\E\left[\lim_{N\to\infty}\frac1N\sum_i
		\frac{\partial\widetilde{\mathbf Z}^{\mathrm S}_{it,h}}{\partial\theta'}
		\widetilde\xi_{i,t,h}\right],
		\label{eq:main-QD}
	\end{equation}
	evaluated at the population coefficient, holding the observed data fixed.
	The derivative of each estimated shock is given by \eqref{eq:own-shock-diff}.
	The multiple-confounder case stacks the shock blocks as in
	Section~\ref{sec:help-estimation}.

	The coefficients and the SVAR parameters jointly solve the exactly identified
	moment conditions that stack the IV moment with $k_t$. Their Jacobian with
	respect to $(\boldsymbol\vartheta_h',\theta')$ is block triangular,
	\begin{equation}
		\mathbf J_h^{\mathrm S}=\begin{pmatrix}-\Sigma_h&D_h\\ \mathbf0&A_\theta
		\end{pmatrix},
		\label{eq:svar-help-jacobian}
	\end{equation}
	and their long-run variance is
	\begin{equation}
		\Omega_h^{\mathrm S}=\lim_{N,T\to\infty}\Var\left(\frac1{\sqrt T}\sum_t
		\begin{pmatrix}N^{-1}\sum_i\widetilde{\mathbf Z}^{\mathrm S}_{it,h}
			\widetilde\xi_{i,t,h}\\ k_t\end{pmatrix}\right).
		\label{eq:svar-help-meat}
	\end{equation}
	Let $\widehat{\mathbf J}_h^{\mathrm S}$ denote \eqref{eq:svar-help-jacobian}
	with the population moments replaced by their sample analogues at the
	estimates.

	\begin{proposition}[Inference for HELP-IV with SVAR shocks]
		\label{thm:help-S}
		Under Assumption~\ref{ass:regular}, the conditions of
		Proposition~\ref{prop:help-identification}, and
		$\widehat\Omega_h^{\mathrm S}\xrightarrow{p}\Omega_h^{\mathrm S}$,
		\begin{equation}
			\sqrt T\,(\widehat V_h^{\mathrm S})^{-1/2}
			\bigl(\widehat b_h^{\mathrm{HELP},\mathrm S}-\beta_h^P\bigr)
			\xrightarrow{d} N(0,1),
			\label{eq:help-two-contributions}
		\end{equation}
		where $\widehat V_h^{\mathrm S}=e_1'(\widehat{\mathbf J}_h^{\mathrm S})^{-1}
		\widehat\Omega_h^{\mathrm S}(\widehat{\mathbf J}_h^{\mathrm S})^{-1\prime}e_1$,
		$\widehat V_h^{\mathrm S}\xrightarrow{p}V_h^{\mathrm S}$, and
		$V_h^{\mathrm S}=e_1'(\mathbf J_h^{\mathrm S})^{-1}\Omega_h^{\mathrm S}
		(\mathbf J_h^{\mathrm S})^{-1\prime}e_1$.
	\end{proposition}

	Partitioning $\Omega_h^{\mathrm S}$ into the IV-moment block $\Omega_{gg}$,
	the cross block $\Omega_{gk}$, and the SVAR block $\Omega_{kk}$, the
	covariance $V_h^{\mathrm S}$ is the panel term
	$e_1'\Sigma_h^{-1}\Omega_{gg}\Sigma_h^{-1\prime}e_1$, plus a first-step term
	from $\Omega_{kk}$, plus a cross term from $\Omega_{gk}$
	(Appendix~\ref{app:cross-covariance}). The cross term can be nonzero,
	because the regression residual can contain omitted aggregate history, on
	which the estimated VAR coefficients also depend. Treating the estimated
	shocks as known sets $D_h=\mathbf0$.

	\subsection{Composite HELP-IV}
	\label{app:svar-composite}

	With SVAR shocks, the primitive instruments of the composite are
	$\epsilon_{g,t}$ and $\boldsymbol\epsilon_{r,t}$. Under the own-impact
	normalization, the inputs of the recursion \eqref{eq:composite-weights} are
	\begin{equation}
		\mathbf v_\ell^{\mathrm S}=\E[\mathbf R_{t+\ell}\epsilon_{g,t}]
		=\mathbf R_\ell^g\sigma_g^2,
		\qquad
		\Sigma_{rz,\ell}^{\mathrm S}=\E[\mathbf R_{t+\ell}\boldsymbol\epsilon_{r,t}']
		=\mathbf R_\ell^r\boldsymbol\Sigma_r,
		\label{eq:svar-covariance-blocks}
	\end{equation}
	where $\mathbf R_\ell^g$ and $\mathbf R_\ell^r$ are the responses of
	$\mathbf R_{t+\ell}$ to $\epsilon_{g,t}$ and to $\boldsymbol\epsilon_{r,t}$,
	$\sigma_g^2=\Var(\epsilon_{g,t})$, and
	$\boldsymbol\Sigma_r=\Var(\boldsymbol\epsilon_{r,t})$ is diagonal. The
	population weights $\boldsymbol w_s^{\mathrm S}$ solve
	\eqref{eq:composite-weights} with these inputs. They differ from the weights
	of Section~\ref{sec:composite-estimation}, because a proxy that equals a
	multiple of a shock plus noise scales the corresponding covariance blocks by
	that multiple, but both composites satisfy
	\eqref{eq:joint-composite-orthogonality} and identify $\beta_h^P$. The
	population and estimated composites are
	\begin{equation}
		\mathrm z_{g,t,h}^{C,\mathrm S}
		=\epsilon_{g,t}
		+\sum_{s=0}^h(\boldsymbol w_s^{\mathrm S})'\boldsymbol\epsilon_{r,t+s},
		\qquad
		\widehat{\mathrm z}_{g,t,h}^{C,\mathrm S}
		=\widehat\epsilon_{g,t}
		+\sum_{s=0}^h(\widehat{\boldsymbol w}_s^{\mathrm S})'
		\widehat{\boldsymbol\epsilon}_{r,t+s}.
		\label{eq:main-composites}
	\end{equation}

	The estimated weights use the fitted SVAR rather than sample cross moments.
	We calculate the responses recursively,
	$\widehat\Psi_\ell=\sum_{j=1}^{\min(p_A,\ell)}\widehat A_j\widehat\Psi_{\ell-j}$ for
	$\ell\geq1$, with $\widehat\Psi_0$ the estimated impact matrix, and substitute
	the responses $\widehat{\mathbf R}_\ell^g$ and $\widehat{\mathbf R}_\ell^r$ of
	$\mathbf R_{t+\ell}$ and the estimated shock variances into the recursion,
	\begin{equation}
		\widehat{\boldsymbol w}_\ell^{\mathrm S}
		=-\bigl(\widehat{\mathbf R}_0^r\widehat{\boldsymbol\Sigma}_r\bigr)^{-1}
		\Bigl(\widehat{\mathbf R}_\ell^g\widehat\sigma_g^2
		+\sum_{s=0}^{\ell-1}\widehat{\mathbf R}_{\ell-s}^r\widehat{\boldsymbol\Sigma}_r
		\widehat{\boldsymbol w}_s^{\mathrm S}\Bigr),
		\qquad \ell=0,\ldots,h.
		\label{eq:internal-weights-main}
	\end{equation}
	The variance factors arise from the own-impact normalization; under a
	unit-variance normalization they are identity matrices, and
	Lemma~\ref{lem:normalization-invariance} maps the two sets of weights into
	each other. Because the inputs are the responses and variances implied by the
	fitted SVAR, $\widehat{\boldsymbol w}_s^{\mathrm S}$ can differ from the result
	of applying \eqref{eq:weights-main} to
	$\widehat{\boldsymbol\epsilon}_t$. The estimator $\widehat b_h^{\mathrm{HELP\text{-}C},\mathrm S}$
	is \eqref{eq:composite-estimator-main} with
	$\widehat{\mathrm z}_{g,t,h}^{C,\mathrm S}$ in place of
	$\widehat{\mathrm z}_{g,t,h}^C$.

	Let $\Sigma_h^C$ and $D_{\boldsymbol w_s,h}$ be
	\eqref{eq:composite-moments-main} with the structural shocks in place of the
	proxies and with the weights $\boldsymbol w_s^{\mathrm S}$.
	The weights and the shocks are both smooth functions of $\theta$. Let
	\begin{equation}
		D_h^w=\sum_{s=0}^h D_{\boldsymbol w_s,h}
		\frac{\partial\boldsymbol w_s^{\mathrm S}}{\partial\theta'},
		\label{eq:composite-weight-derivative-main}
	\end{equation}
	which varies the weights through $\theta$ and holds the shocks fixed, and let
	$D_h^\epsilon$ vary the shocks through $\theta$ and hold the weights fixed;
	\eqref{eq:composite-shock-derivative-main} in Appendix~\ref{app:svar} gives it
	explicitly. The total derivative of the composite IV moment with respect to
	$\theta$ is $D_h^{\mathrm S}=D_h^w+D_h^\epsilon$. The coefficient and the SVAR
	parameters jointly solve the moment conditions that stack the composite IV
	moment with $k_t$, with Jacobian
	\begin{equation}
		\mathbf J_h^{C,\mathrm S}=\begin{pmatrix}-\Sigma_h^C&D_h^{\mathrm S}\\
			\mathbf0&A_\theta\end{pmatrix}
		\label{eq:svar-composite-jacobian}
	\end{equation}
	with respect to $(b,\theta')$, and with long-run variance
	\begin{equation}
		\Omega_h^{C,\mathrm S}=\lim_{N,T\to\infty}\Var\left(\frac1{\sqrt T}\sum_t
		\begin{pmatrix}N^{-1}\sum_i\widetilde{s_{ig}\mathrm z^{C,\mathrm S}_{g,t,h}}
			\widetilde\xi^C_{i,t,h}\\ k_t\end{pmatrix}\right).
		\label{eq:svar-composite-meat}
	\end{equation}
	Let $\widehat{\mathbf J}_h^{C,\mathrm S}$ denote
	\eqref{eq:svar-composite-jacobian} with the population moments replaced by
	their sample analogues at the estimates.

	\begin{proposition}[Inference for composite HELP-IV with SVAR shocks]
		\label{thm:composite-S}
		Under Assumption~\ref{ass:regular}, the conditions of
		Proposition~\ref{prop:composite-conf}, and
		$\widehat\Omega_h^{C,\mathrm S}\xrightarrow{p}\Omega_h^{C,\mathrm S}$,
		\begin{equation}
			\sqrt T\,(\widehat V_h^{C,\mathrm S})^{-1/2}
			\bigl(\widehat b_h^{\mathrm{HELP\text{-}C},\mathrm S}-\beta_h^P\bigr)
			\xrightarrow{d} N(0,1),
			\label{eq:main-phi-ic}
		\end{equation}
		where $\widehat V_h^{C,\mathrm S}=e_1'(\widehat{\mathbf J}_h^{C,\mathrm S})^{-1}
		\widehat\Omega_h^{C,\mathrm S}(\widehat{\mathbf J}_h^{C,\mathrm S})^{-1\prime}e_1$,
		$\widehat V_h^{C,\mathrm S}\xrightarrow{p}V_h^{C,\mathrm S}$, and
		$V_h^{C,\mathrm S}=e_1'(\mathbf J_h^{C,\mathrm S})^{-1}\Omega_h^{C,\mathrm S}
		(\mathbf J_h^{C,\mathrm S})^{-1\prime}e_1$.
	\end{proposition}

	Relative to Proposition~\ref{thm:composite}, the SVAR estimating equations
	replace the weight equations, and $D_h^{\mathrm S}=D_h^w+D_h^\epsilon$. Both
	parts are functions of the same SVAR estimates, so the sandwich keeps their
	joint covariance and their covariance with the panel moment. The split
	between $D_h^w$ and $D_h^\epsilon$ depends on the shock normalization; their
	sum and $V_h^{C,\mathrm S}$ do not (Lemma~\ref{lem:normalization-invariance}).

	\subsection{Extensions}
	\label{app:svar-extensions}

	\begin{remark}[Extensions with SVAR shocks]
		\label{rem:svar-extensions}
		The extensions of Section~\ref{sec:help-inference} carry over with SVAR shocks
		in place of the proxies. Each instrument component determined by the SVAR is
		a smooth function of $\theta$, so it enters the sandwich through its
		derivative with respect to $\theta$ in the off-diagonal block of the
		Jacobian; the cross-sectional balance conditions for targeting-rule HELP-IV
		remain those of Appendix~\ref{app:targeting}. For HELP-IV with multiple exposure margins and
		multiple-confounder HELP-IV, the instruments are the exposures times the
		estimated shocks, and $D_h$ stacks the shock derivatives over the instrument
		dates, as in \eqref{eq:main-internal-D}. For composite HELP-IV with multiple
		exposure margins, the composite is common to all margins, and $D_h^w$ and
		$D_h^\epsilon$ have one row per margin.

		For E-HELP-IV, the estimated shocks are first put in the unit-covariance
		normalization of Section~\ref{sec:forward-portable-section}: the treatment
		instrument is $\epsilon_{g,t}/\sigma_g^2$, and the confounder instruments are
		$\boldsymbol\Sigma_E^{-1}\boldsymbol\epsilon_{r,t}^{E}$, where
		$\boldsymbol\Sigma_E$ is the diagonal covariance matrix of the date-$t$ shocks
		in $\boldsymbol\epsilon_{r,t}^{E}$. In these coordinates, $\mathbf r_H^g$ and
		$\mathbf R_H^E$ are the responses of $(R_t,\ldots,R_{t+H})'$ implied by the
		fitted SVAR, and the weights are \eqref{eq:forward-fixing-mp}. The derivatives include the
		estimated variance factors; \eqref{eq:explicit-news-W} and
		\eqref{eq:explicit-news-K} give them in the equivalent unit-variance
		coordinates, including the residual-path term under incomplete spanning.

		For targeting-rule HELP-IV implemented through the characteristic--instrument
		controls of Appendix~\ref{app:targeting}, these controls contain the
		estimated shocks, so the derivative is that of the complete concentrated
		moment, \eqref{eq:D-total} with the projection derivative
		\eqref{eq:target-projector-derivative}. For transformed treatments and
		outcomes, the derivatives are formed from the residuals of the transformed
		regression.
	\end{remark}

	\section{Estimation and inference: derivations and proofs}
	\label{sec:appendix}
	\label{app:inference}
	This appendix proves the results of Section~\ref{sec:help-inference} and
	Appendix~\ref{app:svar-inference}. Each estimator solves exactly identified
	estimating equations that stack the panel IV moment with the moments of its
	first step, if any, and its asymptotic covariance is the corresponding
	sandwich, whose middle term keeps all covariances between the panel and
	first-step moments.

	\subsection{Common notation and identification}
	\label{app:design}

	Let $\mathbf s_{ig}=(s_{ig,1},\ldots,s_{ig,M})'$ collect $M$ time-invariant treatment
	exposures, with $M$ fixed. The single-exposure model of Section~\ref{sec:portable-markov} has $M=1$ and
	$s_{ig,1}=s_{ig}$. Let $\mathbf R_t\in\R^K$ collect the confounders and set
	\begin{equation}
		f_{t,h}=(G_t,\mathbf R_t',\ldots,\mathbf R_{t+h}')',\qquad
		p_{t,h}=(z_{g,t},\boldsymbol z_{r,t}',\ldots,\boldsymbol z_{r,t+h}')'.
		\label{eq:blocks}
	\end{equation}
	Designs without a superscript use the instruments of
	Section~\ref{sec:help-inference}. The superscript $\mathrm S$ labels designs
	whose instruments are shocks estimated from the SVAR of
	Appendix~\ref{app:svar}; in these, $z_{g,t}$ and $\boldsymbol z_{r,t}$ are
	replaced by the selected estimated shocks. Instrument validity includes the maintained lead--lag exclusion
	restrictions.

	Write $\boldsymbol\beta_h^P=(\beta_{1,h}^P,\ldots,\beta_{M,h}^P)'$.
	For stationary vector series, use
	$\Omega_{xy}=\sum_{\ell\in\mathbb Z}\Cov(x_t,y_{t-\ell})$ and
	$\LRV(x_t)=\Omega_{xx}$; $\Omega_{yx}=\Omega_{xy}'$.
	In the scalar case, write $\Omega(x,y)=\sum_{\ell\in\mathbb Z}\Cov(x_t,y_{t-\ell})$.

	\subsubsection{Full-path design with multiple exposures}
	We refer to HELP-IV with instrumented path controls
	(Sections~\ref{sec:dyn-identification}, \ref{sec:multiple-exposure-margins},
	and~\ref{sec:vec-conf}) as the full-path design, to distinguish it from
	composite HELP-IV. The common-exposure-basis specification is
	\begin{equation}
		Y_{i,t+h}=(\mathbf s_{ig}\otimes f_{t,h})'\vartheta_h
		+\mu_{i,h}+\lambda_{t,h}+\xi_{i,t,h},
		\label{eq:full}
	\end{equation}
	with regressors $X^F_{it,h}=\mathbf s_{ig}\otimes f_{t,h}$ and instruments
	$Z^F_{it,h}=\mathbf s_{ig}\otimes p_{t,h}$.
	The selector $L_h=I_M\otimes e_1'$ picks the $M$ treatment coefficients. The
	projections of the confounder exposures on $\mathbf s_{ig}$ enter the path
	coefficients, and their orthogonal remainders enter $\xi_{i,t,h}$.

	With only two-way fixed effects, independence of the exposure vector and the
	aggregate process gives
	\begin{equation}
		\Sigma_h^F=S\otimes \mathsf A_h,
		\label{eq:multiple-rank}
	\end{equation}
	where
	$S=\E[(\mathbf s_{ig}-\E\mathbf s_{ig})(\mathbf s_{ig}-\E\mathbf s_{ig})']$ is
	the exposure covariance matrix, $\mathsf A_h=\E[p_{t,h}f_{t,h}']$, and
	aggregate variables are centered. Rank therefore requires both $S$ and
	$\mathsf A_h$ to be nonsingular. With additional controls, the residualized
	$\Sigma_h^F$ gives the rank condition.

	\subsubsection{Composite instruments}
	For a scalar aggregate composite $\mathrm z_{g,t,h}^C$, use
	\begin{equation}
		X^C_{it,h}=\mathbf s_{ig}G_t,\qquad Z^C_{it,h}=\mathbf s_{ig} \mathrm z_{g,t,h}^C.
		\label{eq:composite}
	\end{equation}
	If $\E[\mathbf R_{t+\ell}\mathrm z_{g,t,h}^C]=0$ for $\ell=0,\ldots,h$, the omitted path
	terms in \eqref{eq:full} vanish from the population moment, provided the
	exposures are independent of the aggregate process and the structural
	exclusion holds. With only two-way effects,
	$\Sigma_h^C=S\E[\mathrm z_{g,t,h}^CG_t]$, which path neutrality alone does not
	bound away from zero.

	For multiple exposures the derivative block for $\boldsymbol w_s$ is the
	$M\times K$ matrix $D_{\boldsymbol w_s,h}$ of
	\eqref{eq:composite-moments-main} with $\mathbf s_{ig}$ in place of $s_{ig}$,
	where the residual $u^C$ includes the omitted confounder paths and is
	therefore correlated with the confounder proxies.

	\subsubsection{Residualization and effective sample}
	Joint inference uses a common effective date set of size $T$; fixed leads and
	lags change sums by finitely many endpoint terms. Let $M_C$ denote the sample
	residual-maker for all included controls and fixed effects, applied jointly to
	$Y,X,Z$; all panel moments use that operator. With nonseparable controls, path
	neutrality is imposed on the residualized panel cross moments.

	\begin{proposition}[Path neutrality and exposure rank]
		\label{prop:exposure-rank}
		Suppose only unit and date effects are included, exposure moments factor
		from the aggregate process, the primitive instruments satisfy exclusion for
		$\xi_{i,t,h}$, and, with $\mathsf C_{h,\ell}\in\R^{M\times K}$,
		\[
		Y_{i,t+h}=\mathbf s_{ig}'\boldsymbol\beta_h^PG_t
		+\sum_{\ell=0}^h\mathbf s_{ig}'\mathsf C_{h,\ell}\mathbf R_{t+\ell}
		+\mu_{i,h}+\lambda_{t,h}+\xi_{i,t,h}.
		\]
		\begin{enumerate}[label=(\roman*),leftmargin=2em]
			\item $\Sigma_h^F$ is given by \eqref{eq:multiple-rank}, and
			$\Sigma_h^C=S\,\E[\mathrm z_{g,t,h}^CG_t]$.
			\item If $\E[\mathbf R_{t+\ell}\mathrm z_{g,t,h}^C]=0$ for $\ell=0,\ldots,h$ and
			$S\,\E[\mathrm z_{g,t,h}^CG_t]$ is nonsingular, the composite moment
			identifies $\boldsymbol\beta_h^P$.
			\item The weight derivative is
			\begin{equation}
				D_{\boldsymbol w_s,h}
				=S\sum_{\ell=0}^h\mathsf C_{h,\ell}
				\E[\mathbf R_{t+\ell}\boldsymbol z_{r,t+s}'].
				\label{eq:exposure-explicit-D}
			\end{equation}
		\end{enumerate}
	\end{proposition}
	\begin{proof}
		Put $\dot{\mathbf s}_{ig}=\mathbf s_{ig}-\E\mathbf s_{ig}$.
		Two-way demeaning gives
		$\widetilde{\mathbf s_{ig}\otimes f_{t,h}}
		=(\mathbf s_{ig}-\overline{\mathbf s}_g)\otimes(f_{t,h}-\bar f_h)$, so
		independence implies \eqref{eq:multiple-rank} and, in the same way,
		$\Sigma_h^C=S\E[\mathrm z_{g,t,h}^CG_t]$. Thus $\rank(\Sigma_h^F)=\rank(S)\rank(\E[p_{t,h}f_{t,h}'])$.
		At the target,
		\begin{equation}
			\E[\widetilde Z^C_{it,h}u^C_{i,t,h}]
			=S\sum_{\ell=0}^h\mathsf C_{h,\ell}\E[\mathbf R_{t+\ell}\mathrm z_{g,t,h}^C]
			+\E[\widetilde Z^C_{it,h}\widetilde\xi_{i,t,h}]=0.
			\label{eq:exposure-path-bridge}
		\end{equation}
		Nonsingularity of $\Sigma_h^C$ gives uniqueness. Finally,
		$\partial Z^C_{it,h}/\partial \boldsymbol w_s'=\mathbf s_{ig}\boldsymbol z_{r,t+s}'$.
		Substituting the full-path equation into the residual in
		\eqref{eq:composite-moments-main} and using
		primitive-proxy exclusion to eliminate the $\xi$ term gives
		\eqref{eq:exposure-explicit-D}.
	\end{proof}

	\subsubsection{Sample moments for each design}
	Let $d\in\{\mathrm F,\mathrm C\}$ index the full-path or composite
	regression; a further superscript $\mathrm S$, as in $u^{\mathrm{S,C}}$,
	marks designs with SVAR shocks as instruments. With $\eta$ collecting its first-step parameters, define
	\begin{align}
		u^d_{i,t,h}(\vartheta)
		&=\widetilde Y_{i,t+h}-\widetilde X_{it,h}^{d\prime}\vartheta,\nonumber\\
		g^d_{N,t,h}(\vartheta,\eta)
		&=\frac1N\sum_i\widetilde Z^d_{it,h}(\eta)u^d_{i,t,h}(\vartheta),\nonumber\\
		\Sigma_h^d
		&=\E\left[\lim_{N\to\infty}\frac1N\sum_i
		\widetilde Z^d_{it,h}\widetilde X_{it,h}^{d\prime}\right].
		\label{eq:appendix-design-moments}
	\end{align}
	Here $u^F_{i,t,h}=\widetilde\xi_{i,t,h}$ and
	$u^C_{i,t,h}=\widetilde\xi^C_{i,t,h}$ at the population coefficients.
	For the full-path design $L_h=I_M\otimes e_1'$; for the composite design
	$\vartheta_h=\boldsymbol\beta_h^P$ and $L_h=I_M$.
	Design superscripts are omitted when there is no ambiguity.

	\subsection{Stacked moments and the sandwich covariance}
	\label{app:unified}

	Use $g_{t,h}(\vartheta,\eta)$ for a stationary population version of the
	oracle moment in \eqref{eq:appendix-design-moments}, where $\eta$ collects the
	first-step parameters that enter the instruments, outcomes, regressors, or
	controls. The approximation from exact sample residualization and finite $N$
	to this process is stated below. The Jacobian with respect to $\vartheta$
	is $\Sigma_h=-\E[\partial_{\vartheta'}g_{t,h}]$, which for the designs of
	Section~\ref{sec:help-inference} is \eqref{eq:main-moment}, and the
	derivative with respect to the first-step parameters is
	\begin{equation}
		D_h=\left.\partial_{\eta'}\E[g_{t,h}(\vartheta_h,\eta)]
		\right|_{\eta=\eta_0}.
		\label{eq:QD}
	\end{equation}
	If only $Z$ depends on $\eta$ and the residualization operator is fixed,
	$D_h$ takes the form \eqref{eq:main-QD}, with $\eta$ in place of $\theta$ and
	the residual $\widetilde u_{i,t,h}$ in place of $\widetilde\xi_{i,t,h}$.
	If $Y$, $X$, or the controls are generated, differentiate the entire
	concentrated moment instead. In particular,
	\begin{equation}
		d g_{N,t,h}=\frac1N\sum_i\left[
		(d\widetilde Z_{it,h})\widetilde u_{i,t,h}
		+\widetilde Z_{it,h}\{d\widetilde Y_{i,t+h}
		-(d\widetilde X_{it,h})'\vartheta_h\}\right].
		\label{eq:D-total}
	\end{equation}
	The coefficient vector is held fixed in this derivative.

	Let the first step solve exactly identified estimating equations
	$T^{-1}\sum_tm_{\eta,t}(\widehat\eta)=0$, with nonsingular Jacobian
	$A_\eta=\E[\partial_{\eta'}m_{\eta,t}(\eta_0)]$. Stack the panel and
	first-step moments,
	\begin{equation}
		\mathbf m_{t,h}=\bigl(g_{t,h}(\vartheta_h,\eta_0)',\
		m_{\eta,t}(\eta_0)'\bigr)',
		\label{eq:stacked}
	\end{equation}
	and let $\boldsymbol\Omega_h=\lim_{T\to\infty}
	\Var(T^{-1/2}\sum_t\mathbf m_{t,h})$ be their long-run variance. Under exact
	identification, the Jacobian of the stacked moments with respect to
	$(\vartheta_h',\eta')$ is block triangular,
	\begin{equation}
		\mathbf J_h=\begin{pmatrix}-\Sigma_h&D_h\\ \mathbf0&A_\eta\end{pmatrix},
		\label{eq:stacked-jacobian}
	\end{equation}
	and the asymptotic covariance of $\sqrt T(\widehat\vartheta_h-\vartheta_h)$
	is $V_{\vartheta,h}=S_\vartheta\mathbf J_h^{-1}\boldsymbol\Omega_h
	\mathbf J_h^{-1\prime}S_\vartheta'$, where $S_\vartheta=(I,\mathbf0)$ selects
	the coefficients. With the blocks
	$\Omega_{gg}$, $\Omega_{g\eta}=\Omega_{\eta g}'$, and $\Omega_{\eta\eta}$ of
	$\boldsymbol\Omega_h$,
	\begin{equation}
		\Sigma_h^{-1}\bigl\{\Omega_{gg}-D_hA_\eta^{-1}\Omega_{\eta g}
		-\Omega_{g\eta}A_\eta^{-1\prime}D_h'
		+D_hA_\eta^{-1}\Omega_{\eta\eta}A_\eta^{-1\prime}D_h'\bigr\}
		\Sigma_h^{-1\prime},
		\label{eq:sandwich-general}
	\end{equation}
	and $V_h=L_hV_{\vartheta,h}L_h'$ for the treatment coefficients. For
	overidentified panel moments with weighting matrix $\mathcal W_h$, replace
	$\mathbf J_h^{-1}$ by $(\mathbf J_h'\mathcal W\mathbf J_h)^{-1}\mathbf J_h'\mathcal W$
	with $\mathcal W=\operatorname{diag}(\mathcal W_h,I)$. Without a first step,
	$\mathbf J_h=-\Sigma_h$ and the sandwich is
	$\Sigma_h^{-1}\Omega_{gg}\Sigma_h^{-1\prime}$. The four estimators of
	Section~\ref{sec:help-inference} and Appendix~\ref{app:svar-inference} are
	special cases: HELP-IV has no first step; composite HELP-IV has
	$\eta=\underline{\boldsymbol w}_h$, $m_{\eta,t}$ the weight block of the
	vector in \eqref{eq:composite-meat}, $A_\eta=\underline\Sigma_{rz,h}$, and
	$D_h=D_{\boldsymbol w,h}$
	(Appendix~\ref{app:weights}); and the designs with
	SVAR shocks have $\eta=\theta$, $m_{\eta,t}=k_t$, and $A_\eta=A_\theta$
	(Appendix~\ref{app:svar}).

	Weights that are smooth functions of the SVAR parameters enter through the
	chain rule in $D_h$, not as an independent first step.

	\subsection{Estimated offset weights: dated and news offsets}
	\label{app:weights}

	\subsubsection{Dated offsets}
	The population weights solve the recursion \eqref{eq:composite-weights},
	with $\mathbf v_\ell=\E[\mathbf R_{t+\ell}z_{g,t}]$,
	$\Sigma_{rz,\ell}=\E[\mathbf R_{t+\ell}\boldsymbol z_{r,t}']$, $\Sigma_{rz,0}$
	nonsingular, and $\Sigma_{rz,\ell}=0$ for $\ell<0$. The population weights
	are $\underline{\boldsymbol w}_h=-\underline\Sigma_{rz,h}^{-1}\underline{\mathbf v}_h$,
	and the sample recursion \eqref{eq:weights-main} solves
	\eqref{eq:weight-moment-main}. In the notation of \eqref{eq:stacked}, $\eta=\underline{\boldsymbol w}_h$,
	$m_{\eta,t}$ is the weight block of the vector in \eqref{eq:composite-meat},
	and $A_\eta=\underline\Sigma_{rz,h}$.

	\begin{lemma}[Asymptotic normality of the dated offset weights]
		\label{lem:weight-estimation}
		Suppose $\Sigma_{rz,0}$ is nonsingular and the weight block of the vector in
		\eqref{eq:composite-meat} satisfies a central limit theorem (CLT) with long-run variance
		$\Omega_{ww,h}^C$. For fixed $h$,
		\begin{equation}
			\sqrt T(\widehat{\underline{\boldsymbol w}}_h-\underline{\boldsymbol w}_h)
			\xrightarrow{d} N\bigl(\mathbf0,\ \underline\Sigma_{rz,h}^{-1}
			\Omega_{ww,h}^C\underline\Sigma_{rz,h}^{-1\prime}\bigr).
			\label{eq:weight-result}
		\end{equation}
	\end{lemma}
	\begin{proof}
		The population and sample equations imply
		\begin{equation}
			\widehat{\underline\Sigma}_{rz,h}(\widehat{\underline{\boldsymbol w}}_h-\underline{\boldsymbol w}_h)
			=-(\widehat{\underline{\mathbf v}}_h-\underline{\mathbf v}_h)
			-(\widehat{\underline\Sigma}_{rz,h}-\underline\Sigma_{rz,h})\underline{\boldsymbol w}_h.
			\label{eq:weight-exact-identity}
		\end{equation}
		By the joint law of large numbers (LLN), $\widehat{\underline\Sigma}_{rz,h}\xrightarrow{p}\underline\Sigma_{rz,h}$, which is nonsingular
		because $\det\underline\Sigma_{rz,h}=(\det\Sigma_{rz,0})^{h+1}\ne0$. Because
		$\underline{\mathbf v}_h+\underline\Sigma_{rz,h}\underline{\boldsymbol w}_h=\mathbf0$, the right-hand side of
		\eqref{eq:weight-exact-identity} equals
		$-(\widehat{\underline\Sigma}_{rz,h}\underline{\boldsymbol w}_h
		+\widehat{\underline{\mathbf v}}_h)$, whose block $\ell$ is
		$-(\widehat{\mathbf v}_\ell+\sum_{s\leq\ell}\widehat\Sigma_{rz,\ell-s}\boldsymbol w_s)$. Each sample cross moment
		differs from the sample mean of the corresponding product by the product of
		two sample means of mean-zero variables, which is $O_p(T^{-1})$. Hence block
		$\ell$ equals minus the sample mean of
		$\mathbf R_{t+\ell}z_{g,t}+\sum_{s\leq\ell}\mathbf R_{t+\ell-s}\boldsymbol z_{r,t}'\boldsymbol w_s$
		up to $O_p(T^{-1})$, and
		$\sqrt T$ times the right-hand side converges to
		$N(\mathbf0,\Omega_{ww,h}^C)$. Multiplication by
		$\widehat{\underline\Sigma}_{rz,h}^{-1}$ and Slutsky's theorem prove
		\eqref{eq:weight-result}.
	\end{proof}

	The coefficient and the weights jointly solve the exactly identified moment
	conditions formed by the composite IV moment and \eqref{eq:weight-moment-main}.
	Their long-run variance is \eqref{eq:composite-meat}, and their Jacobian with
	respect to $(b,\underline{\boldsymbol w}_h')$ is
	\begin{equation}
		\mathbf J_h^C=\begin{pmatrix}-\Sigma_h^C&D_{\boldsymbol w,h}\\
			\mathbf0&\underline\Sigma_{rz,h}\end{pmatrix}.
		\label{eq:composite-jacobian}
	\end{equation}
	The first row of $(\mathbf J_h^C)^{-1}$ is
	$(\Sigma_h^C)^{-1}(-1,D_{\boldsymbol w,h}\underline\Sigma_{rz,h}^{-1})$, so
	$e_1'(\mathbf J_h^C)^{-1}\Omega_h^C(\mathbf J_h^C)^{-1\prime}e_1$ equals
	\eqref{eq:composite-variance-main}. Every entry of the vector in
	\eqref{eq:composite-meat} is linear in products whose sample means are the
	aggregate cross moments in \eqref{eq:weight-inputs-main} and the panel cross
	moments of $s_{ig}(z_{g,t},\boldsymbol z_{r,t}',\ldots,\boldsymbol z_{r,t+h}')'$
	with $\widetilde Y_{i,t+h}$ and $\widetilde{s_{ig}G_t}$. Hence $V_h^C$ is also
	the delta-method variance of $\widehat b_h^{\mathrm{HELP\text{-}C}}$ as a
	function of these cross moments, with the weights entering through the
	recursion.

	\subsubsection{Forward-looking offsets and incomplete spanning}
	Fix a common path horizon $H$ at least as large as every reported response
	horizon, and let $\mathcal R_{t,H}=(\mathbf R_t',\ldots,\mathbf R_{t+H}')'$.
	Let $\boldsymbol z^E_{r,t}\in\R^J$, $J\leq K(H+1)$, collect the available
	contemporaneous and news confounder proxies, all dated $t$. For a fixed
	positive definite path metric $P$, define
	\begin{align}
		\mathbf r_H^g&=\E[\mathcal R_{t,H}z_{g,t}],\qquad
		\mathbf R_H^E=\E[\mathcal R_{t,H}\boldsymbol z_{r,t}^{E\prime}],\qquad
		\Upsilon=\mathbf R_H^{E\prime}P\mathbf R_H^E,\nonumber\\
		\mathbf e_H^E&=-\Upsilon^{-1}\mathbf R_H^{E\prime}P\mathbf r_H^g,
		\label{eq:news}
	\end{align}
	the residual path $\boldsymbol\varrho_H=\mathbf r_H^g+\mathbf R_H^E\mathbf e_H^E$,
	and the composite $\mathrm z^E_{g,t}=z_{g,t}+(\mathbf e_H^E)'\boldsymbol z^E_{r,t}$,
	where $\mathbf R_H^E$ has full column rank locally. With $P=I$ and the
	normalization of Section~\ref{sec:forward-portable-section},
	\eqref{eq:news} is \eqref{eq:forward-fixing-mp}. The residual path is zero
	if and only if $\mathbf r_H^g\in\operatorname{col}(\mathbf R_H^E)$. In the
	single-exposure model with two-way effects and omitted path coefficient
	vector $\mathbf c_h$ padded with zeros through $H$,
	\begin{equation}
		\beta_h^{E}=\beta_h^P+
		\frac{\mathbf c_h'\boldsymbol\varrho_H}{\E[\mathrm z^E_{g,t}G_t]},
		\label{eq:projection-target}
	\end{equation}
	where $\beta_h^P$ here denotes the forward-looking partial-equilibrium effect
	of Section~\ref{sec:forward-portable-section}, so under incomplete spanning
	the estimand depends on $\mathbf c_h$ and on $P$.

	The differential of \eqref{eq:news} is
	\begin{equation}
		d\mathbf e_H^E=-\Upsilon^{-1}\bigl\{(d\mathbf R_H^E)'P\boldsymbol\varrho_H
		+\mathbf R_H^{E\prime}P\bigl(d\mathbf r_H^g+(d\mathbf R_H^E)\mathbf e_H^E\bigr)\bigr\}.
		\label{eq:news-diff}
	\end{equation}
	An estimated metric adds $\mathbf R_H^{E\prime}(dP)\boldsymbol\varrho_H$ inside
	the braces. For fixed $P$, the Jacobian of the map
	$(\mathbf r_H^g,\ve(\mathbf R_H^E))\mapsto\mathbf e_H^E$ is
	\begin{equation}
		W^E=-\Upsilon^{-1}\Bigl(\mathbf R_H^{E\prime}P,\ \
		(\boldsymbol\varrho_H'P\otimes I_J)\mathcal K_E
		+(\mathbf e_H^E)'\otimes\mathbf R_H^{E\prime}P\Bigr),
		\label{eq:news-jacobian}
	\end{equation}
	where $\mathcal K_E\ve(\mathbf R_H^E)=\ve(\mathbf R_H^{E\prime})$. The term
	in $\boldsymbol\varrho_H$ vanishes under exact spanning. In the notation of
	\eqref{eq:stacked}, the first step estimates
	$(\mathbf r_H^g,\ve(\mathbf R_H^E))$ by sample means, with
	$m_{\eta,t}=((\mathcal R_{t,H}z_{g,t}-\mathbf r_H^g)',
	\ve(\mathcal R_{t,H}\boldsymbol z_{r,t}^{E\prime}-\mathbf R_H^E)')'$ and
	$A_\eta=-I$, so $D_h=D_{\mathbf e,h}W^E$. If aggregate variables and proxies
	are first projected on common predetermined controls, the same formulas apply
	to products of their population residuals.

	\begin{lemma}[Asymptotic normality of a forward-looking offset]
		\label{lem:news-weights}
		Suppose $P$ is fixed and positive definite, $\mathbf R_H^E$ has locally full
		column rank, and $(\widehat{\mathbf r}_H^g,\ve(\widehat{\mathbf R}_H^E))$
		satisfies a CLT jointly with the other stacked moments. Then, whether or not
		$\boldsymbol\varrho_H=\mathbf0$,
		\[
		\sqrt T(\widehat{\mathbf e}_H^E-\mathbf e_H^E)\xrightarrow{d} N\bigl(\mathbf0,\
		W^E\Omega^EW^{E\prime}\bigr),
		\]
		where
		$\Omega^E=\lim_{T\to\infty}\Var\bigl(T^{-1/2}\sum_t((\mathcal R_{t,H}z_{g,t})',
		\ve(\mathcal R_{t,H}\boldsymbol z_{r,t}^{E\prime})')'\bigr)$ and $W^E$ is
		\eqref{eq:news-jacobian}.
	\end{lemma}
	\begin{proof}
		Write $\Delta\mathbf r=\widehat{\mathbf r}_H^g-\mathbf r_H^g$,
		$\Delta\mathbf R=\widehat{\mathbf R}_H^E-\mathbf R_H^E$, and
		$\Delta\mathbf e=\widehat{\mathbf e}_H^E-\mathbf e_H^E$. Since $\Upsilon$ is
		positive definite, \eqref{eq:news} is smooth, so
		$(\Delta\mathbf r,\Delta\mathbf R,\Delta\mathbf e)=O_p(T^{-1/2})$. The normal
		equations give
		\begin{equation}
			0=\Upsilon\Delta\mathbf e+(\Delta\mathbf R)'P\boldsymbol\varrho_H
			+\mathbf R_H^{E\prime}P(\Delta\mathbf r+\Delta\mathbf R\,\mathbf e_H^E)
			+O_p(T^{-1}),
			\label{eq:news-normal-expansion}
		\end{equation}
		so $\widehat{\mathbf e}_H^E$ is a smooth function of the sample cross moments
		with Jacobian $W^E$, and the delta method gives the limit. For projected
		variables $x_t^\perp=x_t-\alpha_x'c_t$ and $z_t^\perp=z_t-\alpha_z'c_t$,
		\[
		d\E[x_t^\perp z_t^{\perp\prime}]
		=-(d\alpha_x)'\E[c_tz_t^{\perp\prime}]
		-\E[x_t^\perp c_t']d\alpha_z=0,
		\]
		so root-$T$ projection errors enter these moments only at second order. An
		estimated $P$ adds $-\Upsilon^{-1}\mathbf R_H^{E\prime}(dP)\boldsymbol\varrho_H$.
	\end{proof}

	\subsubsection{Normalization and regularity}
	If $\boldsymbol z_{r,t}^{E*}=U\boldsymbol z^E_{r,t}$ for an invertible $U$,
	then $\mathbf R_H^{E*}=\mathbf R_H^EU'$,
	$\mathbf e_H^{E*}=U^{-\prime}\mathbf e_H^E$, and
	$(\mathbf e_H^{E*})'\boldsymbol z_{r,t}^{E*}=(\mathbf e_H^E)'\boldsymbol z^E_{r,t}$,
	also when $\boldsymbol\varrho_H\neq\mathbf0$. Rescaling path coordinates
	without adjusting $P$ can change the target. Rank changes and overcomplete
	instrument spaces are excluded.

	\subsection{SVAR and proxy-SVAR first steps}
	\label{app:svar}

	For the SVAR derivatives, use unit-variance computational coordinates
	$\boldsymbol\varepsilon_t$ and an impact matrix $B$. These are related
	to the own-impact shocks $\boldsymbol\epsilon_t$ and $\Psi_0$ of
	Appendix~\ref{app:svar-inference} by \eqref{eq:own-impact-map}; they are not
	an additional shock process. Write \eqref{eq:main-var} as
	$\mathbf W_t=\mathcal A\underline{\mathbf W}_{t-1}+u_t$, where
	$\underline{\mathbf W}_{t-1}=(\mathbf W_{t-1}',\ldots,\mathbf W_{t-p_A}')'$,
	$\mathcal A=(A_1,\ldots,A_{p_A})$, and $u_t=B\boldsymbol\varepsilon_t$ with
	$\E[\boldsymbol\varepsilon_t\boldsymbol\varepsilon_t']=I$.
	Deterministic terms may be included in $\underline{\mathbf W}_{t-1}$ and estimated jointly. Let
	$\Sigma_u=BB'$ and $\Gamma_W=\E[\underline{\mathbf W}_{t-1}\underline{\mathbf W}_{t-1}']$, with $\E[u_t\underline{\mathbf W}_{t-1}']=0$.

	\subsubsection{Structural identification and estimated proxy moments}
	Suppose $r_B(B,\mu,\mathcal A)=0$ supplies the additional restrictions
	selecting a locally unique, nonsingular $B$, with fixed sign conventions.
	Here $\mu=\E[u_tz_t]$ is the moment of a scalar identifying proxy $z_t$, if
	one is used, estimated with OLS residuals. Set
	$\theta=(\ve(\mathcal A)',\ve(B)',\mu')'$; with deterministic restrictions,
	$\mu$ is absent. The SVAR estimates solve the exactly identified estimating
	equations $T^{-1}\sum_tk_t(\widehat\theta)=\mathbf0$, whose date-level
	version stacks the VAR normal equations, the residual-covariance equations,
	the proxy moment, and the restrictions,
	\begin{equation}
		k_t(\theta)=\bigl(\ve(u_t\underline{\mathbf W}_{t-1}')',\ \vh(u_tu_t'-BB')',\
		(u_tz_t-\mu)',\ r_B(B,\mu,\mathcal A)'\bigr)'.
		\label{eq:main-var-primitive}
	\end{equation}
	The restrictions do not vary over time. The covariance restriction can be
	differentiated using
	\begin{equation}
		d\ve(BB')=\{B\otimes I+(I\otimes B)\mathcal K\}\,d\ve(B),
		\label{eq:B-cov}
	\end{equation}
	where $\mathcal K\ve(B)=\ve(B')$, followed by selection of the nonredundant
	lower-triangular entries with the elimination matrix $E_n$, which satisfies
	$E_n\ve(H)=\vh(H)$. With $r_{B,\mathcal A}$, $r_{B,B}$, and $r_{B,\mu}$ the
	derivatives of $r_B$ with respect to $\ve(\mathcal A)$, $\ve(B)$, and $\mu$,
	and $\mathcal L_{Wz}=\E[\underline{\mathbf W}_{t-1}z_t]'\otimes I_n$, the Jacobian
	$A_\theta=\E[\partial k_t/\partial\theta']$ is
	\begin{equation}
		A_\theta=\begin{pmatrix}
			-\Gamma_W\otimes I_n&0&0\\
			0&-E_n\{B\otimes I_n+(I_n\otimes B)\mathcal K\}&0\\
			-\mathcal L_{Wz}&0&-I\\
			r_{B,\mathcal A}&r_{B,B}&r_{B,\mu}
		\end{pmatrix}.
		\label{eq:main-structural-map}
	\end{equation}
	The expected derivative of the covariance equations with respect to
	$\mathcal A$ is zero because $\E[u_t\underline{\mathbf W}_{t-1}']=0$. Ordering the parameters as
	$(\mathcal A,\mu,B)$ makes $A_\theta$ block lower triangular, so it is
	nonsingular when $\Gamma_W$ and the restriction Jacobian
	\[
	\mathcal J_B=\begin{pmatrix}
		E_n\{B\otimes I_n+(I_n\otimes B)\mathcal K\}\\ r_{B,B}
	\end{pmatrix}
	\]
	are nonsingular. With multiple proxies, stack these blocks. The maintained
	restrictions must identify every shock used by HELP-IV; sign restrictions
	alone do not provide the required local point identification.

	\begin{lemma}[Asymptotic normality of the identified SVAR]
		\label{lem:svar}
		Suppose $\Gamma_W$ is nonsingular, the VAR is stable,
		$\E[u_t\underline{\mathbf W}_{t-1}']=0$, and $k_t$ satisfies a CLT jointly with the other stacked
		moments. Suppose the covariance and structural restrictions locally identify
		$B$ with nonsingular Jacobian $\mathcal J_B$. Then
		$\sqrt T(\widehat\theta-\theta_0)\xrightarrow{d}
		N(\mathbf0,A_\theta^{-1}\Omega^kA_\theta^{-1\prime})$, as stated in
		\eqref{eq:main-svar-clt}, with $\Omega^k=\lim_{T\to\infty}\Var(T^{-1/2}\sum_tk_t)$ and $k_t$
		given by \eqref{eq:main-var-primitive}. Write
		$V_\theta=A_\theta^{-1}\Omega^kA_\theta^{-1\prime}$.
	\end{lemma}
	\begin{proof}
		The sample versions of the equations hold exactly at $\widehat\theta$: the
		OLS normal equations, the definition of $\widehat\Sigma_u$ from the OLS
		residuals, the sample proxy moment, and the restrictions. The difference
		between the residual-based and the population-residual sample moments is of
		smaller order: for $\widehat u_t=u_t-(\widehat{\mathcal A}-\mathcal A)\underline{\mathbf W}_{t-1}$,
		\begin{equation}
			T^{-1}\sum_t\widehat u_t\widehat u_t'
			=T^{-1}\sum_tu_tu_t'
			-(\widehat{\mathcal A}-\mathcal A)\widehat\Gamma_W
			(\widehat{\mathcal A}-\mathcal A)',
			\label{eq:covariance-exact-identity}
		\end{equation}
		with $\widehat\Gamma_W=T^{-1}\sum_t\underline{\mathbf W}_{t-1}\underline{\mathbf W}_{t-1}'$, and the subtracted term
		is $O_p(T^{-1})$; the effect of $\widehat{\mathcal A}$ on the sample proxy
		moment is captured by the block $-\mathcal L_{Wz}$ of $A_\theta$. The
		equations are smooth in $\theta$, and the sample Jacobian converges to
		$A_\theta$, which is nonsingular on the locally identified branch. A
		mean-value expansion of the exactly identified equations and the CLT for
		$k_t$ therefore give \eqref{eq:main-svar-clt}, with the restriction rows
		contributing no sampling variation.
	\end{proof}

	\subsubsection{Own-impact normalization of Appendix~\ref{app:svar-inference}}
	Let $e_g$ select the treatment shock, $J_r$ the confounder shocks, and
	$J_R$ the confounders. Define $\Phi_0=I_n$ and
	$\Phi_\ell=\sum_{j=1}^{\min(p_A,\ell)}A_j\Phi_{\ell-j}$.
	The following computational formulas use unit-variance coordinates
	$\boldsymbol\varepsilon_t$, whereas Appendix~\ref{app:svar-inference} uses the own-impact
	shocks $\boldsymbol\epsilon_t$. In the bivariate
	case let $\mathcal S=\operatorname{diag}(B_{gg},B_{rr})$, with nonzero diagonal
	entries and fixed sign conventions. Then
	$\boldsymbol\epsilon_t=\mathcal S\boldsymbol\varepsilon_t$, so that
	$\Var(\boldsymbol\epsilon_t)=\mathcal S\mathcal S'$, and the responses to the
	own-impact shocks are
	\begin{equation}
		\Psi_\ell=\Phi_\ell B\mathcal S^{-1},
		\label{eq:own-impact-map}
	\end{equation}
	which satisfy $\operatorname{diag}(\Psi_0)=(1,1)'$.
	For additional shocks, choose a nonsingular diagonal scale matrix according
	to the maintained normalization. Holding the aggregate observations fixed,
	\begin{align}
		d\boldsymbol\epsilon_t
		&=(d\mathcal S)\boldsymbol\varepsilon_t
		-\mathcal S B^{-1}\{(d\mathcal A)\underline{\mathbf W}_{t-1}+(dB)\boldsymbol\varepsilon_t\},
		\label{eq:own-shock-diff}\\
		d\Psi_\ell
		&=\{(d\Phi_\ell)B+\Phi_\ell(dB)\}\mathcal S^{-1}
		-\Phi_\ell B\mathcal S^{-1}(d\mathcal S)\mathcal S^{-1}.
		\label{eq:own-response-diff}
	\end{align}
	Consequently the covariance blocks for the own-impact instruments are
	\begin{align}
		\mathbf v_\ell&=J_R\Phi_\ell B\mathcal S'e_g,
		&d\mathbf v_\ell&=J_R\{d(\Phi_\ell B)\mathcal S'+\Phi_\ell B(d\mathcal S)'\}e_g,\nonumber\\
		\Sigma_{rz,\ell}&=J_R\Phi_\ell B\mathcal S'J_r',
		&d\Sigma_{rz,\ell}&=J_R\{d(\Phi_\ell B)\mathcal S'+\Phi_\ell B(d\mathcal S)'\}J_r'.
		\label{eq:own-covariance-blocks}
	\end{align}
	These are the blocks $\mathbf R_\ell^g\sigma_g^2$ and
	$\mathbf R_\ell^r\boldsymbol\Sigma_r$ of \eqref{eq:svar-covariance-blocks}.
	Their weight derivatives satisfy
	\begin{equation}
		\Sigma_{rz,0}\,d\boldsymbol w_\ell
		=-d\mathbf v_\ell-(d\Sigma_{rz,0})\boldsymbol w_\ell
		-\sum_{s<\ell}\{(d\Sigma_{rz,\ell-s})\boldsymbol w_s
		+\Sigma_{rz,\ell-s}d\boldsymbol w_s\}.
		\label{eq:own-weight-diff}
	\end{equation}
	Equations \eqref{eq:own-shock-diff} and \eqref{eq:own-weight-diff} give the
	derivatives $D_h^w$ and $D_h^\epsilon$ in Proposition~\ref{thm:composite-S}. They
	include estimation of shock scales. In
	particular, the derivative of the composite moment through the estimated
	shocks, holding the weights fixed, is
	\begin{equation}
		D_h^\epsilon=\E\left[\lim_{N\to\infty}\frac1N\sum_i
		\widetilde{s_{ig}\left(\frac{\partial\epsilon_{g,t}}{\partial\theta'}
			+\sum_{s=0}^h \boldsymbol w_s'
			\frac{\partial\boldsymbol\epsilon_{r,t+s}}{\partial\theta'}\right)}
		\widetilde\xi^C_{i,t,h}\right],
		\label{eq:composite-shock-derivative-main}
	\end{equation}
	with the shock derivatives given by \eqref{eq:own-shock-diff}.

	Let $s_g=e_g'\mathcal S e_g$ and $\mathcal S_r=J_r\mathcal S J_r'$, and denote
	by $\boldsymbol w_\ell^{\mathrm{std}}$ and $\boldsymbol w_\ell^{\mathrm{own}}$ the
	weights computed with unit-variance and with own-impact shocks.

	\begin{lemma}[Normalization invariance]
		\label{lem:normalization-invariance}
		For the dated triangular construction,
		\begin{equation}
			\boldsymbol w_\ell^{\mathrm{own}}
			=s_g \mathcal S_r^{-\prime}\boldsymbol w_\ell^{\mathrm{std}},\qquad
			\mathrm z_{g,t,h}^{C,\mathrm{own}}
			=s_g\mathrm z_{g,t,h}^{C,\mathrm{std}}.
			\label{eq:normalization-weights}
		\end{equation}
		The full-path and composite coefficient estimates and their sandwich
		covariances are the same under both normalizations, including when
		$\mathcal S$ is estimated.
	\end{lemma}
	\begin{proof}
		The covariance blocks obey
		$\mathbf v_\ell^{\mathrm{own}}=s_g\mathbf v_\ell^{\mathrm{std}}$ and
		$\Sigma_{rz,\ell}^{\mathrm{own}}=\Sigma_{rz,\ell}^{\mathrm{std}}\mathcal S_r'$.
		Substitution in the triangular system gives
		\[
		\Sigma_{rz,0}^{\mathrm{std}}\mathcal S_r'
		(s_g\mathcal S_r^{-\prime}\boldsymbol w_\ell^{\mathrm{std}})
		=-s_g\mathbf v_\ell^{\mathrm{std}}
		-\sum_{s<\ell}\Sigma_{rz,\ell-s}^{\mathrm{std}}\mathcal S_r'
		(s_g\mathcal S_r^{-\prime}\boldsymbol w_s^{\mathrm{std}}).
		\]
		Nonsingularity and induction establish the first equality in
		\eqref{eq:normalization-weights}. Since
		$J_r\boldsymbol\epsilon_t=\mathcal S_rJ_r\boldsymbol\varepsilon_t$,
		\[
		\epsilon_{g,t}+\sum_{s=0}^h
		(\boldsymbol w_s^{\mathrm{own}})'J_r\boldsymbol\epsilon_{t+s}
		=s_g\left\{e_g'\boldsymbol\varepsilon_t+
		\sum_{s=0}^h(\boldsymbol w_s^{\mathrm{std}})'
		J_r\boldsymbol\varepsilon_{t+s}\right\}.
		\]
		Thus both sample moments defining the composite IV ratio are multiplied
		by the same nonzero $\widehat s_g$. For the full-path estimator let
		$C_{\mathcal S}=\operatorname{diag}(s_g,\mathcal S_r,\ldots,\mathcal S_r)$; its instrument vector is
		multiplied by $C_{\mathcal S}$. With any fixed residual-maker and any consistent
		sample scale matrix, put
		$\widehat m_h^{\mathrm{std}}=(NT)^{-1}\sum_{t,i}
		\widetilde{\widehat Z}_{it,h}^{\mathrm{std}}\widetilde Y_{i,t+h}$. Then
		\[
		\{\widehat C_{\mathcal S}\widehat\Sigma_h^{\mathrm{std}}\}^{-1}
		\widehat C_{\mathcal S}\widehat m_h^{\mathrm{std}}
		=(\widehat\Sigma_h^{\mathrm{std}})^{-1}\widehat m_h^{\mathrm{std}}.
		\]
		For the population moment, $g^{\mathrm{own}}=C_{\mathcal S}g^{\mathrm{std}}$ and
		\[
		d\E[g^{\mathrm{own}}]
		=(dC_{\mathcal S})\E[g^{\mathrm{std}}]+C_{\mathcal S}d\E[g^{\mathrm{std}}]
		=C_{\mathcal S}d\E[g^{\mathrm{std}}].
		\]
		Premultiplication by $(\Sigma_h^{\mathrm{own}})^{-1}$ cancels
		$C_{\mathcal S}$, so the coefficient block of the sandwich is unchanged. The scalar composite follows with $C_{\mathcal S}=s_g$.
		Separate weight and shock derivatives transform by the product rule; their
		sum is the derivative above. Therefore the sandwich covariances agree.
	\end{proof}

	\subsubsection{Shock and response derivatives}
	Holding the observed aggregate data fixed,
	\begin{align}
		d\boldsymbol\varepsilon_t&=-B^{-1}(d\mathcal A)\underline{\mathbf W}_{t-1}
		-B^{-1}(dB)\boldsymbol\varepsilon_t,\label{eq:shock-diff}\\
		\frac{\partial\boldsymbol\varepsilon_t}{\partial\ve(A_j)'}
		&=-B^{-1}(\mathbf W_{t-j}'\otimes I),&
		\frac{\partial\boldsymbol\varepsilon_t}{\partial\ve(B)'}
		&=-B^{-1}(\boldsymbol\varepsilon_t'\otimes I).\label{eq:shock-jac}
	\end{align}
	Let $\Phi_0=I$ and
	$\Phi_\ell=\sum_{j=1}^{\min(p_A,\ell)}A_j\Phi_{\ell-j}$. Then
	\begin{equation}
		d\Phi_\ell=\sum_{j=1}^{\min(p_A,\ell)}
		\{(dA_j)\Phi_{\ell-j}+A_jd\Phi_{\ell-j}\},\qquad
		d(\Phi_\ell B)=(d\Phi_\ell)B+\Phi_\ell dB.
		\label{eq:irf-diff}
	\end{equation}
	These recursive differentials avoid differentiating a companion-matrix power.

	\subsubsection{Estimated shocks: full-path and composite instruments}
	Let $e_g$ select the treatment shock, $J_r$ the confounder shocks, and $J_R$
	the confounders. For the unit-variance implementation of \eqref{eq:blocks}, use
	$z_{g,t}=e_g'\boldsymbol\varepsilon_t$ and $\boldsymbol z_{r,t}=J_r\boldsymbol\varepsilon_t$.
	The full-path derivative stacks \eqref{eq:shock-jac} over the instrument
	dates. For a dated composite use \eqref{eq:composite-weights} with
	\begin{equation}
		\mathbf v_\ell=J_R\Phi_\ell Be_g,\qquad
		\Sigma_{rz,\ell}=J_R\Phi_\ell BJ_r'.
		\label{eq:internal-blocks}
	\end{equation}
	The shock sequence is serially orthogonal, so future shocks have zero
	covariance with earlier aggregates under the causal VAR representation.
	For each direction $d\theta$, the weights satisfy \eqref{eq:own-weight-diff},
	and
	\begin{equation}
		d\mathrm z_{g,t,h}^C=e_g'd\boldsymbol\varepsilon_t+
		\sum_{s=0}^h\{(d\boldsymbol w_s)'J_r\boldsymbol\varepsilon_{t+s}
		+\boldsymbol w_s'J_r d\boldsymbol\varepsilon_{t+s}\}.
		\label{eq:internal-vdiff}
	\end{equation}
	For a news composite use
	$\mathbf r_H^g=\operatorname{col}_{\ell=0}^H(J_R\Phi_\ell Be_g)$ and
	$\mathbf R_H^E=\operatorname{col}_{\ell=0}^H(J_R\Phi_\ell BJ_E')$, where $J_E$ selects
	the available date-$t$ shock block, apply \eqref{eq:news-diff}, and use
	$d\mathrm z^E_{g,t}=e_g'd\boldsymbol\varepsilon_t
	+(d\mathbf e_H^E)'J_E\boldsymbol\varepsilon_t
	+(\mathbf e_H^E)'J_Ed\boldsymbol\varepsilon_t$. Identification of the SVAR
	does not imply $\boldsymbol\varrho_H=\mathbf0$.

	\subsubsection{Response, weight, and shock matrices}
	Let $\mathbb A$ be the $np_A\times np_A$ companion matrix and
	$J=(I_n,0,\ldots,0)$. Define
	\begin{align}
		\Xi_0&=0,\qquad
		\Xi_\ell=\sum_{j=0}^{\ell-1}\{J(\mathbb A')^{\ell-1-j}\}\otimes\Phi_j,
		\quad \ell\ge1,\nonumber\\
		\Xi^B_\ell&=\big((B'\otimes I_n)\Xi_\ell,\ I_n\otimes\Phi_\ell\big).
		\label{eq:main-response-matrices}
	\end{align}
	Here $\Xi_\ell$ is $n^2\times n^2p_A$ and $\Xi^B_\ell$ is
	$n^2\times(n^2p_A+n^2)$; they map parameter changes into
	$d\ve(\Phi_\ell)$ and $d\ve(\Phi_\ell B)$, respectively. This is the
	matrix form used for VAR response inference by \citet{lutkepohl1990}.
	The HELP-IV weight Jacobian is computed, without numerical differentiation,
	from
	\begin{align}
		J_{\mathbf v_\ell}&=(e_g'\otimes J_R)\Xi^B_\ell,\qquad
		J_{\Sigma_{rz,0}}=(J_r\otimes J_R)\Xi^B_0,\qquad
		J_{\Sigma_{rz,\ell-s}}=(J_r\otimes J_R)\Xi^B_{\ell-s},\nonumber\\
		W_\ell&=-\Sigma_{rz,0}^{-1}\left[J_{\mathbf v_\ell}+(\boldsymbol w_\ell'\otimes I_K)J_{\Sigma_{rz,0}}
		+\sum_{s<\ell}\{(\boldsymbol w_s'\otimes I_K)J_{\Sigma_{rz,\ell-s}}+\Sigma_{rz,\ell-s}W_s\}\right].
		\label{eq:main-weight-matrices}
	\end{align}
	The recursion starts at $\ell=0$ with an empty sum; $W_\ell$ is $K\times\dim\theta$.
	The shock Jacobian is
	\begin{equation}
		\mathcal J_t=-B^{-1}\big(\underline{\mathbf W}_{t-1}'\otimes I_n,\ \boldsymbol\varepsilon_t'\otimes I_n\big),
		\label{eq:main-shock-matrix}
	\end{equation}
	so that $d\boldsymbol\varepsilon_t=\mathcal J_t\,d\theta$, with zero columns
	for $\mu$ when it is part of $\theta$.
	Consequently define
	\begin{align}
		\mathcal P_{t,h}&=\operatorname{col}
		(e_g'\mathcal J_t,J_r\mathcal J_t,\ldots,J_r\mathcal J_{t+h}),\nonumber\\
		\mathcal K^{\epsilon}_{t,h}
		&=e_g'\mathcal J_t+\sum_{s=0}^h \boldsymbol w_s'J_r\mathcal J_{t+s},\nonumber\\
		\mathcal K_{t,h}
		&=\mathcal K^{\epsilon}_{t,h}
		+\sum_{s=0}^h(J_r\boldsymbol\varepsilon_{t+s})'W_s.
		\label{eq:main-internal-J}
	\end{align}
	The matrices multiplying the aggregate estimation error are the residual
	cross-moments
	\begin{align}
		D_h^{\mathrm{S,F}}&=\E\left[\lim_{N\to\infty} N^{-1}\sum_i
		\widetilde{\mathbf s_{ig}\otimes\mathcal P_{t,h}}
		u^{\mathrm{S,F}}_{i,t,h}\right],\nonumber\\
		D_{\epsilon_s,h}&=\E\left[\lim_{N\to\infty} N^{-1}\sum_i
		\widetilde{\mathbf s_{ig}(J_r\boldsymbol\varepsilon_{t+s})'}
		u^{\mathrm{S,C}}_{i,t,h}\right],\nonumber\\
		B_h^w&=\sum_{s=0}^hD_{\epsilon_s,h}W_s,\qquad
		B_h^{\epsilon}=\E\left[\lim_{N\to\infty} N^{-1}\sum_i
		\widetilde{\mathbf s_{ig}\mathcal K^{\epsilon}_{t,h}}
		u^{\mathrm{S,C}}_{i,t,h}\right],\nonumber\\
		D_h^{\mathrm{S,C}}&=B_h^w+B_h^{\epsilon}.
		\label{eq:main-internal-D}
	\end{align}
	Residualization is applied column by column. Each expectation is estimated
	by its sample average with the fitted panel residual. In particular,
	$B_h^w$ differentiates weights holding shocks fixed, and $B_h^{\epsilon}$
	differentiates shocks holding weights fixed. Their sum is the total
	instrument derivative. If included controls are generated, use the total
	concentrated-moment derivative in \eqref{eq:D-total}, including the
	projection derivative in \eqref{eq:target-projector-derivative}.

	\subsubsection{Covariance of the aggregate estimates}
	By Lemma~\ref{lem:svar}, the covariance of the SVAR estimates is
	$V_\theta=A_\theta^{-1}\Omega^kA_\theta^{-1\prime}$; write
	$\Omega_{kk}=\Omega^k$. Its covariance with the panel moment enters the
	sandwich through the cross block $\Omega_{gk}$ of the long-run variance of
	the panel moment stacked with $k_t$. Both blocks are obtained from a single
	joint covariance estimate of the observed residual products. This includes
	uncertainty in any estimated proxy-identification moments.

	For recursive identification, $B$ is the lower-triangular Cholesky factor,
	there is no proxy moment, and $\theta=(\ve(\mathcal A)',\vh(B)')'$. The
	Jacobian is then block diagonal with blocks
	\[
	-\Gamma_W\otimes I_n,\qquad -E_n(I_{n^2}+K_{nn})(B\otimes I_n)E_n',
	\]
	where $K_{nn}=\mathcal K$. Mapped back to $\ve(B)$, the second block gives
	\begin{equation}
		H_B=E_n'\left[E_n(I_{n^2}+K_{nn})
		(B\otimes I_n)E_n'\right]^{-1}.
		\label{eq:main-cholesky}
	\end{equation}
	Under the additional assumption of iid Gaussian innovations,
	\begin{equation}
		V_\theta=\operatorname{diag}\left(
		\Gamma_W^{-1}\otimes\Sigma_u,\quad
		H_B\{2D_n^+(\Sigma_u\otimes\Sigma_u)D_n^{+\prime}\}H_B'\right),
		\label{eq:main-gaussian-cov}
	\end{equation}
	where $D_n$ is the duplication matrix and $D_n^+=(D_n'D_n)^{-1}D_n'$.

	\subsubsection{Equivalent recursive Jacobians}
	\label{app:explicit-jacobians}

	The preceding differentials can be written as matrices in the coordinates
	\[
	\theta=(\ve(A_1)',\ldots,\ve(A_{p_A})',\ve(B)')'.
	\]
	Let $P_{\ell,j}=\partial\ve(\Phi_\ell)/\partial\ve(A_j)'$,
	with $P_{0,j}=0$. The recursion in \eqref{eq:irf-diff} gives
	\begin{equation}
		P_{\ell,j}=
		{\bf1}\{j\leq\min(p_A,\ell)\}
		(\Phi_{\ell-j}'\otimes I)
		+\sum_{k=1}^{\min(p_A,\ell)}(I\otimes A_k)P_{\ell-k,j}.
		\label{eq:explicit-P}
	\end{equation}
	The indicator term is interpreted as zero when $\ell-j<0$.
	Since $\Xi_\ell=(P_{\ell,1},\ldots,P_{\ell,p_A})$, \eqref{eq:explicit-P} gives $\Xi^B_\ell$ in
	\eqref{eq:main-response-matrices}, and \eqref{eq:main-weight-matrices} then
	specifies every column of the weight Jacobian, including the estimation of
	the impact response $\Sigma_{rz,0}=J_RBJ_r'$ and of all earlier weights. The
	shock Jacobian is \eqref{eq:main-shock-matrix} with
	$\underline{\mathbf W}_{t-1}=(\mathbf W_{t-1}',\ldots,\mathbf W_{t-p_A}')'$. Stacking selected rows
	of $\mathcal J_t,\ldots,\mathcal J_{t+h}$ gives $\mathcal P_{t,h}$ in
	\eqref{eq:main-internal-J}, and inserting \eqref{eq:main-weight-matrices}
	into its last line gives $\mathcal K_{t,h}$. Therefore
	\eqref{eq:main-internal-D} is an explicit expectation of known Jacobian
	matrices and the population panel residual.

	For a forward-looking offset built from estimated SVAR shocks, form
	$\mathbf r_H^g$ and $\mathbf R_H^E$ as above. For each coordinate $\theta_j$,
	the derivatives $\partial_j\mathbf r_H^g$ and $\partial_j\mathbf R_H^E$ follow
	from \eqref{eq:main-response-matrices}, and the offset Jacobian has columns
	\begin{equation}
		W_j^E=-\Upsilon^{-1}
		\bigl\{(\partial_j\mathbf R_H^E)'P\boldsymbol\varrho_H
		+\mathbf R_H^{E\prime}P(\partial_j\mathbf r_H^g
		+\partial_j\mathbf R_H^E\,\mathbf e_H^E)\bigr\}.
		\label{eq:explicit-news-W}
	\end{equation}
	The composite-shock Jacobian is
	\begin{equation}
		\mathcal K_{t,H}^E
		=e_g'\mathcal J_t+(J_E\boldsymbol\varepsilon_t)'W^E
		+(\mathbf e_H^E)'J_E\mathcal J_t.
		\label{eq:explicit-news-K}
	\end{equation}

	\begin{lemma}[Closed-form matrices and primitive covariance blocks]
		\label{lem:closed-matrices}
		For fixed $h$ and $p_A$:
		\begin{enumerate}[label=(\roman*),leftmargin=2em]
			\item \eqref{eq:main-response-matrices} equals \eqref{eq:explicit-P};
			\item \eqref{eq:main-weight-matrices} and \eqref{eq:main-shock-matrix} are
			the Jacobians of the weights and of the shocks with respect to $\theta$;
			\item under the conditions of Lemma~\ref{lem:svar}, the structural block of
			$V_\theta$ is \eqref{eq:main-cholesky} under Cholesky identification, and
			$V_\theta$ is \eqref{eq:main-gaussian-cov} under iid Gaussian innovations.
		\end{enumerate}
	\end{lemma}
	\begin{proof}
		Since $\Phi_\ell=J\mathbb A^\ell J'$ and $d\mathbb A=J'\,d\mathcal A$,
		\begin{align*}
			d\Phi_\ell
			&=\sum_{j=0}^{\ell-1}\Phi_j\,d\mathcal A\,
			\mathbb A^{\ell-1-j}J',\\
			d\ve(\Phi_\ell)
			&=\sum_{j=0}^{\ell-1}\{J(\mathbb A')^{\ell-1-j}\otimes\Phi_j\}
			d\ve(\mathcal A)
			=\Xi_\ell\,d\ve(\mathcal A).
		\end{align*}
		Differentiating $\Phi_\ell=\sum_kA_k\Phi_{\ell-k}$ with respect to each
		$A_j$ gives the blocks $P_{\ell,j}$ in \eqref{eq:explicit-P}, so that
		$\Xi_\ell=(P_{\ell,1},\ldots,P_{\ell,p_A})$. The identity
		\[
		d\ve(\Phi_\ell B)=(B'\otimes I_n)d\ve(\Phi_\ell)
		+(I_n\otimes\Phi_\ell)d\ve(B)
		\]
		gives $\Xi^B_\ell$ and, after selection, the blocks $J_{\mathbf v_\ell}$,
		$J_{\Sigma_{rz,0}}$, and $J_{\Sigma_{rz,\ell-s}}$ in
		\eqref{eq:main-weight-matrices}. Vectorizing \eqref{eq:own-weight-diff} with
		$d(\Sigma_{rz,\ell-s}\boldsymbol w_s)=(\boldsymbol w_s'\otimes I_K)d\ve(\Sigma_{rz,\ell-s})
		+\Sigma_{rz,\ell-s}d\boldsymbol w_s$, together with \eqref{eq:shock-diff},
		implies \eqref{eq:main-weight-matrices} and
		\eqref{eq:main-shock-matrix}. Lemma~\ref{lem:svar} gives
		$V_\theta=A_\theta^{-1}\Omega_{kk}A_\theta^{-1\prime}$.

		For lower-triangular $B$ with positive diagonal, put
		$b_\triangle=\vh(B)$. Then
		\[
		d\ve(B)=E_n'\,db_\triangle,\qquad
		d\vh(BB')=E_n(I_{n^2}+K_{nn})(B\otimes I_n)E_n'\,db_\triangle.
		\]
		The Cholesky map is smooth and invertible, which gives
		\eqref{eq:main-cholesky}. Under iid Gaussian innovations,
		\begin{align*}
			\LRV(\ve(u_t\underline{\mathbf W}_{t-1}'))&=\Gamma_W\otimes\Sigma_u,\\
			\LRV(\vh(u_tu_t'-\Sigma_u))&=2D_n^+(\Sigma_u\otimes\Sigma_u)D_n^{+\prime},\\
			\E[\ve(u_t\underline{\mathbf W}_{t-1}')\vh(u_{t-j}u_{t-j}'-\Sigma_u)']&=0\qquad(j\in\mathbb Z).
		\end{align*}
		The second equality follows from Isserlis' identity; the third follows
		from vanishing Gaussian third moments at $j=0$ and conditional
		expectations at $j\ne0$. The sandwich with the block-diagonal Jacobian
		proves \eqref{eq:main-gaussian-cov}.
	\end{proof}

	\begin{lemma}[Expansion of a generated instrument]
		\label{lem:generated-instrument}
		Suppose the identified maps above are twice continuously differentiable
		on a neighborhood of $\theta_0$, with integrable local derivative
		envelopes for the relevant sample moments, and
		$\widehat\theta-\theta_0=O_p(T^{-1/2})$.
		For a fixed horizon, the full-path and composite sample moments satisfy
		\begin{equation}
			\overline g_h(\vartheta_h,\widehat\theta)
			=\overline g_h(\vartheta_h,\theta_0)
			+D_h(\widehat\theta-\theta_0)+o_p(T^{-1/2}),
			\label{eq:generated-moment-expansion}
		\end{equation}
		with $D_h$ given by \eqref{eq:main-internal-D}.
	\end{lemma}
	\begin{proof}
		Let $\Delta\theta=\widehat\theta-\theta_0$.
		Smoothness of $B^{-1}$ and the response and offset maps gives
		\[
		\widehat Z_{it,h}-Z_{it,h}(\theta_0)
		=J_{Z,it,h}\Delta\theta+r_{it,h},
		\]
		with $\|r_{it,h}\|\leq M_{it,h}\|\Delta\theta\|^2$.
		For the fixed residual-maker $M_C$, the envelope condition implies
		\[
		\left\|\frac1{NT}\sum_{t,i}(M_Cr)_{it,h}u_{i,t,h}\right\|
		=O_p(\|\Delta\theta\|^2)=O_p(T^{-1}),
		\]
		and the LLN gives
		$(NT)^{-1}\sum_{t,i}(M_CJ_Z)_{it,h}u_{i,t,h}=D_h+o_p(1)$.
		Thus the sample moment equals its oracle value plus
		$D_h\Delta\theta+o_p(T^{-1/2})$. The product rule gives
		\[
		\partial_{\theta'}\mathrm z_{g,t,h}^C
		=e_g'\mathcal J_t+\sum_{s=0}^h
		\{\boldsymbol w_s'J_r\mathcal J_{t+s}
		+(J_r\boldsymbol\varepsilon_{t+s})'W_s\},
		\]
		and hence \eqref{eq:main-internal-D}. For a generated residual-maker,
		the same Taylor bound applies to the complete moment derivative in
		\eqref{eq:D-total}. Full column rank in \eqref{eq:news} gives the
		corresponding result for news composites.
	\end{proof}

	\subsection{Regularity conditions and proofs of the inference results}
	\label{app:theory}

	\begin{assumption}[Time-dimension inference]
		\label{ass:regular}
		For a fixed finite collection of horizons:
		\begin{enumerate}[label=(\roman*),leftmargin=2em]
			\item $\E[g_{t,h}(\vartheta_h,\eta_0)]=\mathbf0$, $\Sigma_h$ has full column
			rank, and $\mathcal W_h$ is positive definite; $M$, $K$, $p_A$, and the
			horizons are fixed.
			\item The first-step estimator solves
			$T^{-1}\sum_tm_{\eta,t}(\widehat\eta)=\mathbf0$ with $A_\eta$ nonsingular, or
			is a smooth function of such a solution, and the composite weight maps have
			locally constant rank. With SVAR shocks as instruments, the VAR is stable
			and $\theta$ is locally identified under a fixed normalization.
			\item $(\widehat\vartheta_h,\widehat\eta)\xrightarrow{p}(\vartheta_h,\eta_0)$.
			The sample moments are differentiable on a neighborhood of
			$(\vartheta_h,\eta_0)$, their Jacobians converge in probability to
			$-\Sigma_h$ and $D_h$ uniformly on that neighborhood, and the remainders of
			their first-order expansions are $o_p(T^{-1/2})$.
			\item $T^{-1/2}\sum_t\mathbf m_{t,h}\xrightarrow{d}N(\mathbf0,\boldsymbol\Omega_h)$
			jointly across the reported horizons, where $\mathbf m_{t,h}$ is
			\eqref{eq:stacked}, $\boldsymbol\Omega_h=\sum_{\ell\in\mathbb Z}\Gamma_h(\ell)$
			with $\Gamma_h(\ell)=\E[\mathbf m_{t,h}\mathbf m_{t-\ell,h}']$, and
			$\sum_{\ell}\|\Gamma_h(\ell)\|<\infty$. For the series and bootstrap results,
			the corresponding functional CLT holds.
			\item After the exact sample residualization,
			\begin{equation}
				\frac1{\sqrt T}\sum_t
				\{g_{N,t,h}(\vartheta_h,\eta_0)-g_{t,h}\}=o_p(1).
				\label{eq:cs-approx}
			\end{equation}
			\item $L_h\mathbf J_h^{-1}\boldsymbol\Omega_h\mathbf J_h^{-1\prime}L_h'$ is
			positive definite, and the first step and the panel moments use the same
			effective dates up to a fixed number of endpoints.
		\end{enumerate}
	\end{assumption}

	Stationarity, strong mixing with summable coefficients, and finite moments
	of sufficiently high order are sufficient for (iv). A sufficient condition for
	\eqref{eq:cs-approx} is that the cross-sectional approximation error,
	conditional on aggregate history, has zero mean and absolutely summable
	temporal covariances of total order $O(N^{-1})$. A persistent
	$O_p(N^{-1/2})$ conditional mean, such as targeting imbalance, violates it;
	Proposition~\ref{prop:targeting} then replaces the time-only limit.

	\begin{proposition}[Sandwich limit]
		\label{prop:limit}
		Under Assumption~\ref{ass:regular},
		\begin{equation}
			\sqrt T(\widehat\vartheta_h-\vartheta_h)
			\xrightarrow{d} N\bigl(\mathbf0,\ V_{\vartheta,h}\bigr),
			\qquad
			V_{\vartheta,h}=S_\vartheta\mathbf J_h^{-1}\boldsymbol\Omega_h
			\mathbf J_h^{-1\prime}S_\vartheta',
			\label{eq:unified-limit}
		\end{equation}
		jointly across the fixed collection of horizons and exposure margins.
	\end{proposition}
	\begin{proof}
		Let $\phi=(\vartheta_h',\eta')'$, and let $\bar{\mathbf m}_h(\phi)$ be the
		stacked sample moment, with the panel block computed after exact sample
		residualization. Under exact identification, the estimator solves
		$\bar{\mathbf m}_h(\widehat\phi)=\mathbf0$. The sample Jacobian converges
		uniformly to $\mathbf J_h$, and the remainder condition and
		\eqref{eq:cs-approx} hold, so a mean-value expansion of the stacked equations
		around $\phi_0$ shows that $\sqrt T(\widehat\phi-\phi_0)$ has the same limit
		as $-\mathbf J_h^{-1}\sqrt T\bar{\mathbf m}_h(\phi_0)$. By the joint CLT of
		Assumption~\ref{ass:regular}(iv), $\sqrt T\bar{\mathbf m}_h(\phi_0)
		\xrightarrow{d}N(\mathbf0,\boldsymbol\Omega_h)$, which proves
		\eqref{eq:unified-limit}. For overidentified panel moments, the first-order
		condition replaces $\mathbf J_h^{-1}$ by
		$(\mathbf J_h'\mathcal W\mathbf J_h)^{-1}\mathbf J_h'\mathcal W$.

		\emph{HELP-IV (Proposition~\ref{thm:help}).}
		Sample residualization removes the unit and time effects and the included
		controls exactly, so
		$\widetilde Y_{i,t+h}=\widetilde{\boldsymbol X}_{it,h}'\boldsymbol\vartheta_h
		+\widetilde\xi_{i,t,h}$, and \eqref{eq:main-estimator} gives
		\[
		\sqrt T(\widehat{\boldsymbol\vartheta}_h-\boldsymbol\vartheta_h)
		=\widehat\Sigma_h^{-1}\frac1{\sqrt T}\sum_t\frac1N\sum_i
		\widetilde{\mathbf Z}_{it,h}\widetilde\xi_{i,t,h}.
		\]
		By Assumption~\ref{ass:regular}(iii), $\widehat\Sigma_h\xrightarrow{p}\Sigma_h$,
		which is nonsingular. By \eqref{eq:cs-approx} and
		Assumption~\ref{ass:regular}(iv), the normalized sum converges in distribution
		to $N(\mathbf0,\Omega_h)$. Since $b_h^{\mathrm{HELP}}=\beta_h^P$ by
		Proposition~\ref{prop:help-identification}, it follows that
		$\sqrt T(\widehat b_h^{\mathrm{HELP}}-\beta_h^P)\xrightarrow{d}N(0,V_h)$.
		The continuous mapping theorem gives $\widehat V_h\xrightarrow{p}V_h$, and
		$V_h>0$ by Assumption~\ref{ass:regular}(vi), so Slutsky's theorem gives
		\eqref{eq:help-clt}.

		\emph{HELP-IV with SVAR shocks (Proposition~\ref{thm:help-S}).}
		The instruments depend on $\theta$, so Proposition~\ref{prop:limit} applies
		with $\eta=\theta$, $m_{\eta,t}=k_t$ of \eqref{eq:main-var-primitive}, and
		$A_\eta=A_\theta$ (Lemma~\ref{lem:svar}). The instrument derivative in
		\eqref{eq:own-shock-diff}, stacked over $t,\ldots,t+h$, gives $D_h$ in
		\eqref{eq:main-QD}, so the Jacobian is \eqref{eq:svar-help-jacobian} and
		$\sqrt T(\widehat b_h^{\mathrm{HELP},\mathrm S}-\beta_h^P)\xrightarrow{d}
		N(0,V_h^{\mathrm S})$. Consistency of $\widehat{\mathbf J}_h^{\mathrm S}$ and
		$\widehat\Omega_h^{\mathrm S}$ gives
		$\widehat V_h^{\mathrm S}\xrightarrow{p}V_h^{\mathrm S}>0$, and Slutsky's
		theorem gives \eqref{eq:help-two-contributions}.

		\emph{Composite HELP-IV with SVAR shocks (Proposition~\ref{thm:composite-S}).}
		Holding the panel coefficient fixed, the product rule applied to
		\eqref{eq:main-composites} gives $d\E[g^C_{t,h}]=(D_h^\epsilon+D_h^w)d\theta$,
		with $D_h^w$ in \eqref{eq:composite-weight-derivative-main} and
		$D_h^\epsilon$ in \eqref{eq:composite-shock-derivative-main}.
		The weight derivatives in \eqref{eq:own-weight-diff} and the shock
		derivatives in \eqref{eq:own-shock-diff} exist on the identified branch.
		Products of the two estimation errors are $O_p(T^{-1})$ in the
		sample moment under the derivative-envelope conditions of
		Lemma~\ref{lem:generated-instrument}. Hence Proposition~\ref{prop:limit}
		applies with $\eta=\theta$, $m_{\eta,t}=k_t$, $A_\eta=A_\theta$, and
		$D_h=D_h^w+D_h^\epsilon$, so the Jacobian is
		\eqref{eq:svar-composite-jacobian} and
		$\sqrt T(\widehat b_h^{\mathrm{HELP\text{-}C},\mathrm S}-\beta_h^P)
		\xrightarrow{d}N(0,V_h^{C,\mathrm S})$; Slutsky's theorem with
		$\widehat V_h^{C,\mathrm S}\xrightarrow{p}V_h^{C,\mathrm S}>0$ gives
		\eqref{eq:main-phi-ic}, and \eqref{eq:sandwich-general} gives the split
		\eqref{eq:composite-variance-split}. Lemma~\ref{lem:normalization-invariance}
		permits evaluation in unit-variance coordinates, where the same total
		derivative is $D_h^{\mathrm{S,C}}=B_h^w+B_h^\epsilon$ in
		\eqref{eq:main-internal-D}, with the covariance blocks transformed
		consistently.

		\emph{Composite HELP-IV (Proposition~\ref{thm:composite}).}
		Write $b=b_h^{\mathrm{HELP\text{-}C}}$, which equals $\beta_h^P$ by
		Proposition~\ref{prop:composite-conf}, $\widehat b=\widehat b_h^{\mathrm{HELP\text{-}C}}$, and
		\[
		\bar g_h(b,\underline{\boldsymbol w})
		=\frac1{NT}\sum_{t,i}\widetilde{s_{ig}\mathrm z_{g,t,h}^C(\underline{\boldsymbol w})}
		\bigl(\widetilde Y_{i,t+h}-b\,\widetilde{s_{ig}G_t}\bigr),
		\]
		so that $\bar g_h(\widehat b,\widehat{\underline{\boldsymbol w}}_h)=0$. The
		composite is affine in the weights and the moment is linear in $b$, so
		$\bar g_h(\widehat b,\widehat{\underline{\boldsymbol w}}_h)
		=\bar g_h(b,\underline{\boldsymbol w}_h)
		+\widehat D_{\boldsymbol w,h}(\widehat{\underline{\boldsymbol w}}_h-\underline{\boldsymbol w}_h)
		-\widehat\Sigma_h^C(\widehat b-b)$ holds exactly, where
		$\widehat D_{\boldsymbol w,h}$ is the sample counterpart of
		$D_{\boldsymbol w,h}$ at $b$ and $\widehat\Sigma_h^C$ is the sample
		instrument--regressor moment at $\widehat{\underline{\boldsymbol w}}_h$.
		Combining this identity with \eqref{eq:weight-exact-identity} gives
		\[
		\begin{pmatrix}-\widehat\Sigma_h^C&\widehat D_{\boldsymbol w,h}\\
			\mathbf0&\widehat{\underline\Sigma}_{rz,h}\end{pmatrix}
		\sqrt T\begin{pmatrix}\widehat b-b\\
			\widehat{\underline{\boldsymbol w}}_h-\underline{\boldsymbol w}_h\end{pmatrix}
		=-\sqrt T\begin{pmatrix}\bar g_h(b,\underline{\boldsymbol w}_h)\\
			\widehat{\underline\Sigma}_{rz,h}\underline{\boldsymbol w}_h
			+\widehat{\underline{\mathbf v}}_h\end{pmatrix}.
		\]
		By Assumption~\ref{ass:regular}(iii), the matrix on the left converges in
		probability to $\mathbf J_h^C$ in \eqref{eq:composite-jacobian}, which is
		nonsingular because $\Sigma_h^C\neq0$ and
		$\det\underline\Sigma_{rz,h}=(\det\Sigma_{rz,0})^{h+1}\neq0$. By
		\eqref{eq:cs-approx}, the first block on the right equals $T^{-1/2}$ times
		the sum over $t$ of the first entry of the vector in
		\eqref{eq:composite-meat}, up to $o_p(1)$; by the proof of
		Lemma~\ref{lem:weight-estimation}, the second block equals $T^{-1/2}$ times
		the sum of its weight block, up to $o_p(1)$. Assumption~\ref{ass:regular}(iv)
		gives their joint convergence to $N(\mathbf0,\Omega_h^C)$, so
		$\sqrt T(\widehat b-b)\xrightarrow{d}N(0,V_h^C)$, with
		$V_h^C=e_1'(\mathbf J_h^C)^{-1}\Omega_h^C(\mathbf J_h^C)^{-1\prime}e_1$ equal to
		\eqref{eq:composite-variance-main}. The sample counterparts of $\Sigma_h^C$,
		$D_{\boldsymbol w,h}$, and $\underline\Sigma_{rz,h}$ are consistent, and
		$\widehat\Omega_h^C\xrightarrow{p}\Omega_h^C$, so
		$\widehat V_h^C\xrightarrow{p}V_h^C$; $V_h^C>0$ by
		Assumption~\ref{ass:regular}(vi). Slutsky's theorem gives
		\eqref{eq:composite-clt}.
		With multiple exposure margins, apply the
		selector. For news instruments, use Lemma~\ref{lem:news-weights}, with
		the sample cross moments as first step, $A_\eta=-I$, $D_h=D_{\mathbf e,h}W^E$, and the
		moments centered at $\beta_h^E$. Stacking a fixed number of horizons and
		exposure margins and applying the joint CLT gives the joint result. Under
		correct specification, the product of the $o_p(1)$ weight-matrix estimation
		error and the $O_p(T^{-1/2})$ sample moment is negligible.
	\end{proof}

	The sandwich is that of exactly identified stacked estimating equations
	\citep{neweymcfadden1994}.

	\subsubsection{Applications with a fixed cross section}
	For a fixed number of units, condition on the exposures and apply the
	time-series CLT to the date-level moment itself, which retains the
	cross-sectional noise; the same sandwich applies. With two regions, at most
	one centered exposure margin remains after differencing.

	\subsection{Cross-covariance and iid structural shocks}
	\label{app:cross-covariance}

	This subsection concerns SVAR shocks as instruments. Partition the long-run
	variance \eqref{eq:svar-help-meat} into the IV-moment block $\Omega_{gg}$, the
	cross block $\Omega_{gk}=\Omega_{kg}'$, and the SVAR block $\Omega_{kk}$, and
	let $V_\theta=A_\theta^{-1}\Omega_{kk}A_\theta^{-1\prime}$. By
	\eqref{eq:sandwich-general}, the covariance in Proposition~\ref{thm:help-S}
	splits into a panel term, a first-step term, and twice a cross term,
	\begin{equation}
		V_h^{\mathrm S}=V_h^{\mathrm{reg}}+V_h^{\mathrm{SVAR}}+2C_h^{\mathrm S},
		\label{eq:help-variance-split}
	\end{equation}
	where
	\begin{align*}
		V_h^{\mathrm{reg}}&=e_1'\Sigma_h^{-1}\Omega_{gg}\Sigma_h^{-1\prime}e_1,\\
		V_h^{\mathrm{SVAR}}&=e_1'\Sigma_h^{-1}D_hV_\theta D_h'\Sigma_h^{-1\prime}e_1,
	\end{align*}
	and the cross term is
	\begin{equation}
		C_h^{\mathrm S}
		=-e_1'\Sigma_h^{-1}\Omega_{gk}A_\theta^{-1\prime}D_h'\Sigma_h^{-1\prime}e_1.
		\label{eq:help-cross-covariance}
	\end{equation}
	For the composite, with $D_h^{\mathrm S}=D_h^w+D_h^\epsilon$,
	\begin{equation}
		V_h^{C,\mathrm S}=(\Sigma_h^C)^{-2}\bigl\{\Omega_{gg}
		-2\,\Omega_{gk}A_\theta^{-1\prime}(D_h^w+D_h^\epsilon)'
		+(D_h^w+D_h^\epsilon)V_\theta(D_h^w+D_h^\epsilon)'\bigr\}.
		\label{eq:composite-variance-split}
	\end{equation}
	Thus $\Omega_{gk}=0$ is sufficient for the panel and first-step variances to
	add. Serial independence of $\boldsymbol\epsilon_t$ does not imply this
	condition.

	\begin{proposition}[An iid Gaussian counterexample]
		\label{prop:iid-counterexample}
		There exists a stable model satisfying the baseline HELP-IV exclusions,
		with iid independent Gaussian structural shocks and strongly relevant
		instruments, for which $C_0^{\mathrm S}=-1$. The corrected limiting variance
		is strictly positive and smaller than the variance $V_0^{\mathrm{reg}}$ that
		treats the shocks as known.
	\end{proposition}
	\begin{proof}
		Let $(\epsilon_{g,t},\epsilon_{r,t})'\stackrel{\mathrm{iid}}{\sim}N(0,I_2)$,
		$G_t=\epsilon_{g,t}$, and $R_t=\epsilon_{r,t}$. Let the exposure be
		independent of the aggregate process, with mean zero and variance one.
		For $0<|\rho|<1$, set
		\begin{equation}
			H_t=\rho H_{t-1}+G_{t-1}
			=\sum_{j=1}^{\infty}\rho^{j-1}G_{t-j},\qquad
			Y_{it}=s_{ig}H_t.
			\label{eq:iid-example-dgp}
		\end{equation}
		This is the baseline dynamic model with $\beta_0=0$, $\beta_1=1$,
		and zero confounder loadings and idiosyncratic errors. At $h=0$,
		$\boldsymbol X_{it,0}=\mathbf Z_{it,0}=s_{ig}(G_t,R_t)'$ and
		$\xi_{i,t,0}=s_{ig}H_t$, so that $\Sigma_0=I_2$ and
		$\beta_0^P=b_0^{\mathrm{HELP}}=c_{0,0}=0$.
		All instrument--residual moments vanish because $(G_t,R_t)$ is
		independent of $H_t$. Estimate the restricted spending equation
		$G_t=aG_{t-1}+\epsilon_{g,t}$, with true $a=0$; the other aggregate
		parameters are known. The estimating equation for $a$ has date-level product
		$k_t=G_tG_{t-1}$ at the true value and Jacobian $A_a=-\E[G_{t-1}^2]=-1$.
		The IV moment is $g_{t,0}=(G_tH_t,R_tH_t)'$, and its derivative with respect
		to $a$ is $D_0=(-\E[G_{t-1}H_t],0)'=(-1,0)'$. The products $G_tH_t$ and
		$G_tG_{t-1}$ are martingale differences with respect to the filtration
		generated by the aggregate shocks through $t$, so all cross-date covariances
		are zero, and
		\begin{align*}
			\Omega_{gg,11}&=\E[G_t^2]\E[H_t^2]=(1-\rho^2)^{-1},\\
			\Omega_{gk,1}&=\E[G_t^2]\E[H_tG_{t-1}]=1,\\
			\Omega_{kk}&=\E[G_t^2]\E[G_{t-1}^2]=1.
		\end{align*}
		Hence $V_0^{\mathrm{reg}}=(1-\rho^2)^{-1}$, $V_0^{\mathrm{SVAR}}=1$, and
		$C_0^{\mathrm S}=-\Omega_{gk,1}A_a^{-1}D_{0,1}=-1$, so that
		$V_0^{\mathrm S}=(1-\rho^2)^{-1}+1-2=\rho^2/(1-\rho^2)>0$. Sample demeaning
		changes the normalized moment sums by $o_p(1)$ under the stated stationarity
		and finite moments.
	\end{proof}

	If the aggregate parameters are instead estimated on an independent
	sample of size $T_A$, with $T/T_A\to\tau_A<\infty$, then
	\[
	V_h^{\mathrm S}=V_h^{\mathrm{reg}}
	+\tau_A\,e_1'\Sigma_h^{-1}D_hV_\theta^AD_h'\Sigma_h^{-1\prime}e_1,
	\]
	where $V_\theta^A$ is the covariance of Lemma~\ref{lem:svar} on that
	sample.
	Independence of the two samples sets the cross term to zero.

	For the composite built from SVAR shocks, \eqref{eq:composite-variance-split}
	contains the cross term between the weight and shock-estimation channels,
	\begin{equation}
		2(\Sigma_h^C)^{-2}D_h^wV_\theta(D_h^\epsilon)'.
		\label{eq:composite-weight-shock-covariance}
	\end{equation}
	An aggregate sample independent of the panel does not remove this term,
	because both channels use the same parameter estimate.

	\subsection{Lead alignment and what it changes}
	\label{app:alignment}

	\subsubsection{Composites built from SVAR shocks}
	Decompose $\mathrm z_{g,t,h}^C=\sum_{s=0}^h v^{(s)}_{t,h}$ with
	\begin{equation}
		v^{(0)}_{t,h}=(e_g'+\boldsymbol w_0'J_r)\boldsymbol\varepsilon_t,\qquad
		v^{(s)}_{t,h}=\boldsymbol w_s'J_r\boldsymbol\varepsilon_{t+s}\quad(s\geq1).
		\label{eq:lead-components}
	\end{equation}
	Let $D_{h,s}$ be \eqref{eq:main-QD} with $Z$ replaced by
	$\mathbf s_{ig} v^{(s)}_{t,h}$. Each component includes both the shock derivative
	and the corresponding weight derivative, and $D_h=\sum_sD_{h,s}$. The SVAR
	equations may then be aligned with each lead: treat the first step as $h+1$
	copies of the same estimating equations, dated $t,\ldots,t+h$, each with its
	own copy of $\theta$ that enters the panel moment through $D_{h,s}$. The
	stacked moment becomes
	\begin{equation}
		\mathbf m^{\mathrm{al}}_{t,h}=(g_{t,h}',k_t',k_{t+1}',\ldots,k_{t+h}')',
		\label{eq:aligned-internal}
	\end{equation}
	and its Jacobian has $-\Sigma_h$ and $h+1$ copies of $A_\theta$ on the
	diagonal and $D_{h,0},\ldots,D_{h,h}$ in the first block row. In population
	the copies coincide, and in the sample they differ only through endpoint
	terms. A similar decomposition is available for full-path instruments built
	from SVAR shocks. It is unnecessary for a composite containing only date-$t$
	news shocks.

	\subsubsection{Composites with weights from sample cross moments}
	In block $\ell$ of the weight equations \eqref{eq:weight-moment-main}, the
	date-$t$ summand of $\widehat\Sigma_{rz,\ell-s}\boldsymbol w_s$ is
	$\mathbf R_{t+\ell-s}\boldsymbol z_{r,t}'\boldsymbol w_s$. It can be re-dated to
	$t-s$, the base date of the panel moment in which the proxy
	$\boldsymbol z_{r,t}$ carries the weight $\boldsymbol w_s$, while the summand
	of $\widehat{\mathbf v}_\ell$ keeps date $t$. The recursion propagates these
	shifts into later weights, and the Jacobian remains
	\eqref{eq:composite-jacobian}.

	\begin{lemma}[Invariance of the first-order sum]
		\label{lem:alignment}
		Let $\mathbf m^{\mathrm{al}}_{t,h}$ denote \eqref{eq:aligned-internal} or the
		stacked moment with the aligned weight equations. For fixed $h$, under
		stationarity and finite second moments,
		$T^{-1/2}\sum_t(\mathbf m^{\mathrm{al}}_{t,h}-\mathbf m_{t,h})=O_p(T^{-1/2})$.
		If the autocovariances of both are absolutely summable, their long-run
		variances, and hence the sandwich covariances, coincide.
	\end{lemma}
	\begin{proof}
		For each fixed $s$ and stationary square-integrable $k_t$,
		\[
		\sum_{t=1}^T(k_{t+s}-k_t)
		=\sum_{t=T+1}^{T+s}k_t-\sum_{t=1}^sk_t=O_p(1).
		\]
		This identity proves the claim for SVAR shocks. For weights built from
		sample cross moments, the aligned and unaligned summands differ in block $\ell$
		only by the terms $\mathbf R_{t+\ell-s}\boldsymbol z_{r,t}'\boldsymbol w_s$ dated
		$t-s$ rather than $t$; each of their sums over $t$ changes by $O_p(1)$ by the
		same identity; premultiplication by a fixed
		matrix preserves this order. Dividing these sums by $\sqrt T$ gives
		$o_p(1)$. The endpoint terms have bounded second moments, so the normalized
		differences also converge in $L^2$. Under covariance summability, the
		coefficient blocks of the two sandwiches therefore coincide.
	\end{proof}

	Alignment changes only how a finite-bandwidth or finite-series estimator
	weights offsetting covariances.

	\subsection{Targeting rules: estimation, exact balance, and finite-sample uncertainty}
	\label{app:targeting}

	Partition confounders into a two-way block $\mathbf R_t^{(1)}$ of dimension
	$K_1$ and a one-way block. Let $\mathbf a_i\in\R^d$ contain
	observed targeting characteristics and a constant, and define the residual
	exposure
	\begin{equation}
		\mathbf s_{ig}^{\perp}=\mathbf s_{ig}-\Pi'\mathbf a_i,
		\label{eq:target-rule}
	\end{equation}
	where $\Pi=\E[\mathbf a_i\mathbf a_i']^{-1}\E[\mathbf a_i\mathbf s_{ig}']$ is
	the population projection coefficient. Its sample analogue
	$\widehat\Pi=(\sum_i \mathbf a_i\mathbf a_i')^{-1}\sum_i \mathbf a_i\mathbf s_{ig}'$
	gives $\widehat{\mathbf s}_{ig}^{\perp}=\mathbf s_{ig}-\widehat\Pi'\mathbf a_i$.
	With $f^{(1)}_{t,h}=(G_t,\mathbf R_t^{(1)\prime},\ldots,
	\mathbf R_{t+h}^{(1)\prime})'$ and the corresponding proxy vector
	$p^{(1)}_{t,h}$, the direct implementation is
	\begin{equation}
		X^T_{it,h}=\mathbf s_{ig}\otimes f^{(1)}_{t,h},\qquad
		\widehat Z^T_{it,h}=\widehat{\mathbf s}_{ig}^{\perp}\otimes p^{(1)}_{t,h}.
		\label{eq:target-design}
	\end{equation}
	Use the joint IV estimator in the main text. With multiple exposures, all
	columns of $\Pi$ are estimated jointly and every margin uses its own
	residual exposure. The exposure relevance matrix is now
	$S_\perp=\E[\mathbf s_{ig}^{\perp}\mathbf s_{ig}^{\perp\prime}]$, which must have full rank in the separable
	case.

	\subsubsection{Identification through exposure balance}
	Suppose the omitted one-way component has the form
	\begin{equation}
		U^{\mathrm{ow}}_{i,t,h}=
		\sum_{q>K_1}\sum_{s=0}^h c_{q,h,s}s_{ir_q}R_{q,t+s}.
		\label{eq:oneway}
	\end{equation}
	The targeting restriction is
	$\E[\mathbf s_{ig}^{\perp} s_{ir_q}]=0$ for every omitted exposure, together with
	independence of exposure characteristics from the aggregate process and
	instrument exclusion for the remaining error conditional on those
	characteristics. In the two-way-effects
	case it implies
	\begin{equation}
		\E[Z^T_{it,h}U^{\mathrm{ow}}_{i,t,h}]
		=\sum_{q>K_1}\E[\mathbf s_{ig}^{\perp} s_{ir_q}]\otimes
		\left\{\sum_{s=0}^h c_{q,h,s}
		\E[p^{(1)}_{t,h}R_{q,t+s}]\right\}=0.
		\label{eq:target-bridge}
	\end{equation}
	Centering is implicit after fixed effects. With additional controls this
	condition must hold for the concentrated moment as well.

	\subsubsection{Implementation through characteristic--instrument controls}
	In a balanced panel with common date weights and only two-way effects,
	adding $W_{it,h}=\mathbf a_i\otimes p^{(1)}_{t,h}$ as included controls and using
	$\mathbf s_{ig}\otimes p^{(1)}_{t,h}$ as instruments is equivalent to using
	\eqref{eq:target-design}. Redundant constant interactions are dropped after
	time demeaning. For every pair of primitive instruments $p_{j,t}$
	and $p_{k,t}$ and every residual exposure margin $\widehat s^\perp_{i\ell}$,
	\begin{equation}
		\sum_{i,t}\mathbf a_i p_{j,t}\widehat s^\perp_{i\ell}p_{k,t}
		=\Big(\sum_i \mathbf a_i\widehat s^\perp_{i\ell}\Big)
		\Big(\sum_t p_{j,t}p_{k,t}\Big)=0.
		\label{eq:target-fwl}
	\end{equation}
	With additional controls $C$, the exact Frisch--Waugh--Lovell (FWL) implementation projects on
	$[W,C]$ jointly, and irregular panels require the weighted projection. When
	the instruments are SVAR shocks, $W$ is generated, so \eqref{eq:D-total} is
	used with the projection rebuilt at each $\theta$.

	The derivative of the generated projection has a closed form.
	Stack all included control columns, after dropping redundant columns,
	in a full-column-rank matrix $\mathcal C(\theta)$, and set
	$M(\theta)=I-\mathcal C(\theta)
	(\mathcal C(\theta)'\mathcal C(\theta))^{-1}\mathcal C(\theta)'$.
	Differentiating the projector gives
	\begin{equation}
		dM=-M(d\mathcal C)(\mathcal C'\mathcal C)^{-1}\mathcal C'
		-\mathcal C(\mathcal C'\mathcal C)^{-1}(d\mathcal C)'M.
		\label{eq:target-projector-derivative}
	\end{equation}
	For any stacked variable $v(\theta)$,
	$d\widetilde v=(dM)v+M\,dv$.
	Inserting these derivatives for $Y,X,Z$ into \eqref{eq:D-total}
	accounts for the dependence of the targeting controls on the estimated
	shocks. Formula \eqref{eq:target-projector-derivative} follows by
	differentiating $\mathcal C(\mathcal C'\mathcal C)^{-1}\mathcal C'$
	and collecting the terms using $M\mathcal C=0$. It requires a
	locally fixed control rank, as does the concentrated-moment expansion.

	\subsubsection{Cross-sectional sampling uncertainty}
	Population exposure balance can fail in the sample. For each one-way confounder $q>K_1$, define the sample imbalance
	\begin{equation}
		\widehat\delta_q=\frac1N\sum_i\widehat{\mathbf s}_{ig}^\perp s_{ir_q}.
		\label{eq:balance-moments}
	\end{equation}
	Let $\gamma_q=\E[\mathbf a_i\mathbf a_i']^{-1}\E[\mathbf a_i s_{ir_q}]$ and
	let $\chi_{q,i}=\mathbf s_{ig}^{\perp}(s_{ir_q}-\mathbf a_i'\gamma_q)$ be the
	product of the two projection residuals. The pair $(\widehat\Pi,\widehat\delta_q)$
	solves an exactly identified cross-sectional system that stacks the
	projection equations $\mathbf a_i(\mathbf s_{ig}'-\mathbf a_i'\Pi)$ and the
	balance moment $(\mathbf s_{ig}-\Pi'\mathbf a_i)s_{ir_q}-\delta_q$. Under
	population balance and a cross-sectional CLT, the sandwich covariance of this
	system gives
	\[
	\sqrt N\,\widehat\delta_q\xrightarrow{d}
	N\bigl(\mathbf0,\ \E[\chi_{q,i}\chi_{q,i}']\bigr).
	\]
	The Jacobian block for $\Pi$ replaces $s_{ir_q}$ by its projection residual
	in $\chi_{q,i}$, so this covariance already accounts for the estimation of
	$\Pi$.

	Set
	\begin{equation}
		m_{q,h}=\sum_{s=0}^h c_{q,h,s}\E[p^{(1)}_{t,h}R_{q,t+s}],\qquad
		\kappa_{i,h}=\sum_{q>K_1}\chi_{q,i}\otimes m_{q,h}.
		\label{eq:target-kappa}
	\end{equation}
	The time-averaged omitted-path moment contains
	$\sum_q\widehat\delta_q\otimes m_{q,h}$, which can survive averaging
	over time. In the separable model just described, assuming that the
	remaining finite-$N$ approximation is negligible and $T/N$ is bounded, the
	coefficient therefore acquires a cross-sectional term of order
	$\sqrt{T/N}$ in addition to the time-series sandwich of
	Proposition~\ref{prop:limit}; Proposition~\ref{prop:targeting} gives the joint
	limit. The time component includes any aggregate first-step moments.

	Three regimes arise. If each omitted exposure lies exactly in
	the span of $\mathbf a_i$, then $\widehat\delta_q=0$ in every sample and the displayed
	imbalance term is absent. If balance holds only in population, root-$N$
	balance and $T/N\to0$ suffice to make it negligible. If $T/N\to\tau>0$,
	it generally contributes at first order. Under independence of the leading
	aggregate time component and the cross-sectional sampling component,
	\begin{equation}
		\Var(\widehat{\boldsymbol\beta}_h)\simeq
		L_h\mathbf J_h^{-1}\Bigl(\frac1T\Omega_{T,h}
		+\frac1N E_g\Omega_{N,h}E_g'\Bigr)\mathbf J_h^{-1\prime}L_h',
		\label{eq:target-variance}
	\end{equation}
	where $\Omega_{T,h}=\boldsymbol\Omega_h$ is the long-run variance of the
	stacked moment in \eqref{eq:stacked}, $E_g$ embeds the panel-moment block in
	the stacked moment, and $\Omega_{N,h}$ is the limit of
	$\Var(N^{-1/2}\sum_i\kappa_{i,h})$. The cross-sectional term enters through
	the panel moment only. Without independence, the covariance also includes the
	cross covariance of the two components (Proposition~\ref{prop:targeting}).

	\begin{proposition}[Joint time and cross-sectional targeting limit]
		\label{prop:targeting}
		Suppose \eqref{eq:oneway} holds, the structural remainder satisfies
		conditional instrument exclusion, Assumption~\ref{ass:regular} holds for the
		remaining moments, $\E[\mathbf a_i\mathbf a_i']$ is nonsingular,
		$T/N\to\tau\in[0,\infty)$, and
		\begin{equation}
			\begin{pmatrix}
				T^{-1/2}\sum_t \mathbf m_{t,h}\\ N^{-1/2}\sum_i\kappa_{i,h}
			\end{pmatrix}
			\ \xrightarrow{d}\
			N\!\left(0,
			\begin{pmatrix}\Omega_{T,h}&\Omega_{TN,h}\\
				\Omega_{TN,h}'&\Omega_{N,h}\end{pmatrix}\right).
			\label{eq:target-joint-clt}
		\end{equation}
		Then
		\begin{equation}
			\sqrt T(\widehat{\boldsymbol\beta}_h-\boldsymbol\beta_h^P)
			\xrightarrow{d}N\bigl(\mathbf0,\ V_h^{\mathrm T}\bigr),
			\label{eq:target-general-limit}
		\end{equation}
		where $V_h^{\mathrm T}=L_h\mathbf J_h^{-1}\{\Omega_{T,h}+\tau E_g\Omega_{N,h}E_g'
		+\sqrt\tau(\Omega_{TN,h}E_g'+E_g\Omega_{TN,h}')\}\mathbf J_h^{-1\prime}L_h'$.
		Under exact sample balance,
		$V_h^{\mathrm T}=L_h\mathbf J_h^{-1}\Omega_{T,h}\mathbf J_h^{-1\prime}L_h'$.
	\end{proposition}
	\begin{proof}
		Let $A_a=\E[\mathbf a_i\mathbf a_i']$ and
		$\widehat A_a=N^{-1}\sum_i\mathbf a_i\mathbf a_i'$.
		The normal equations imply
		\begin{equation}
			\widehat\Pi-\Pi
			=\widehat A_a^{-1}N^{-1}\sum_i\mathbf a_i
			\mathbf s_{ig}^{\perp\prime}
			=A_a^{-1}N^{-1}\sum_i\mathbf a_i
			\mathbf s_{ig}^{\perp\prime}+o_p(N^{-1/2}).
			\label{eq:Pi-proof-expansion}
		\end{equation}
		Consequently,
		\begin{align}
			\widehat\delta_q
			&=N^{-1}\sum_i\mathbf s_{ig}^{\perp}s_{ir_q}
			-(\widehat\Pi-\Pi)'N^{-1}\sum_i\mathbf a_i s_{ir_q}\nonumber\\
			&=N^{-1}\sum_i\mathbf s_{ig}^{\perp}
			(s_{ir_q}-\mathbf a_i'\gamma_q)+o_p(N^{-1/2})
			=N^{-1}\sum_i\chi_{q,i}+o_p(N^{-1/2}).
			\label{eq:balance-proof-expansion}
		\end{align}
		Here $\E\chi_{q,i}=0$ by population balance and projection
		orthogonality. Since $\widehat m_{q,h}=m_{q,h}+O_p(T^{-1/2})$,
		\[
		\sum_q\widehat\delta_q\otimes\widehat m_{q,h}
		=N^{-1}\sum_i\kappa_{i,h}
		+O_p((NT)^{-1/2})+o_p(N^{-1/2}).
		\]
		Hence, with bounded $T/N$, the normalized stacked sample moment at the
		population parameters equals
		\begin{equation}
			T^{-1/2}\sum_t\mathbf m_{t,h}
			+\sqrt{T/N}\,E_gN^{-1/2}\sum_i\kappa_{i,h}+o_p(1).
			\label{eq:target-expansion}
		\end{equation}
		Its joint CLT gives a normal limit with covariance
		$\Omega_{T,h}+\tau E_g\Omega_{N,h}E_g'+\sqrt\tau(\Omega_{TN,h}E_g'
		+E_g\Omega_{TN,h}')$, and the argument of Proposition~\ref{prop:limit} then
		gives \eqref{eq:target-general-limit}. If
		$s_{ir_q}=\mathbf a_i'\gamma_q$, then
		\[
		\widehat\delta_q=N^{-1}\sum_i
		\widehat{\mathbf s}_{ig}^{\perp}\mathbf a_i'\gamma_q=0
		\]
		by the sample normal equations. This proves the exact-balance claim.
	\end{proof}

	When $\Omega_{N,h}$ is estimable and units are independently
	sampled, a covariance estimator is
	\begin{equation}
		\widehat\Omega_{N,h}=
		\frac1N\sum_i(\widehat\kappa_{i,h}-\overline{\widehat\kappa}_h)
		(\widehat\kappa_{i,h}-\overline{\widehat\kappa}_h)',
		\label{eq:target-feasible-cross-cov}
	\end{equation}
	with consistently estimated projections, omitted-path coefficients, and
	aggregate moments in $\kappa_{i,h}$; clustered units require the
	corresponding cluster covariance. The variance added to the coefficient
	covariance is
	$N^{-1}L_h\widehat{\mathbf J}_h^{-1}E_g\widehat\Omega_{N,h}E_g'
	\widehat{\mathbf J}_h^{-1\prime}L_h'$.

	If the omitted exposures or the coefficients in $m_{q,h}$ are not
	identified, \eqref{eq:target-variance} has no sample analogue, and one must
	maintain a negligible-imbalance condition or retain the relevant path
	controls. Time-only heteroskedasticity- and autocorrelation-consistent (HAC),
	equal-weighted cosine (EWC), and time-block bootstrap estimators do not
	recover the imbalance term, which time centering removes.

	\subsection{Feasible covariance estimates and simultaneous inference}
	\label{app:variance}

	Compute the sample Jacobian $\widehat{\mathbf J}_h$ of the stacked moments:
	$\widehat\Sigma_h$, the sample derivative $\widehat D_h$ of the concentrated
	moment at $(\widehat\vartheta_h,\widehat\eta)$, holding $\widehat\vartheta_h$
	fixed, and the first-step Jacobian $\widehat A_\eta$. Form the feasible stacked
	moments
	$\widehat{\mathbf m}_{t,h}=(\widehat g_{t,h}',m_{\eta,t}(\widehat\eta)')'$
	from the fitted residuals and the estimated first step, using, for dated
	composites, the aligned version, and center them,
	$\check{\mathbf m}_{t,h}=\widehat{\mathbf m}_{t,h}
	-T^{-1}\sum_{\tau=1}^T\widehat{\mathbf m}_{\tau,h}$. The reported covariance
	is the coefficient block of the sample sandwich,
	\begin{equation}
		\widehat V_h=L_h\widehat{\mathbf J}_h^{-1}\widehat{\boldsymbol\Omega}_h
		\widehat{\mathbf J}_h^{-1\prime}L_h',
		\label{eq:feasible-score}
	\end{equation}
	where $\widehat{\boldsymbol\Omega}_h$ is a long-run covariance estimator
	applied to $\check{\mathbf m}_{t,h}$, and $L_h$ now selects the treatment
	coefficients from the full parameter vector.
	Numerical derivatives of the entire moment reconstruct shocks, responses,
	weights, and any generated controls at each perturbation.

	\subsubsection{Consistent HAC covariance}
	For a bandwidth $b_T$ and an even kernel $k$ with $k(0)=1$, define
	\begin{align}
		\widehat\Gamma_h(\ell)&=T^{-1}\sum_{t=\ell+1}^T
		\check{\mathbf m}_{t,h}\check{\mathbf m}_{t-\ell,h}',\nonumber\\
		\widehat{\boldsymbol\Omega}_h^{\mathrm{HAC}}&=\widehat\Gamma_h(0)+
		\sum_{\ell=1}^{T-1}k(\ell/b_T)
		\{\widehat\Gamma_h(\ell)+\widehat\Gamma_h(\ell)'\}.
		\label{eq:hac}
	\end{align}
	For Bartlett weights, $k(x)=(1-|x|)_+$. The substitution condition is
	\begin{equation}
		b_T\left\{\frac1T\sum_t
		\|\widehat{\mathbf m}_{t,h}-\mathbf m_{t,h}\|^2\right\}^{1/2}=o_p(1).
		\label{eq:hac-substitution}
	\end{equation}
	Because overlap extends beyond $h$ through aggregate persistence, $b_T=h+1$
	is not generally sufficient.

	\begin{proposition}[Consistency of the feasible HAC covariance]
		\label{prop:hac}
		Let $\mathbf m_t$ be stationary with mean zero, autocovariances
		$\Gamma(\ell)=\E[\mathbf m_t\mathbf m_{t-\ell}']$ satisfying
		$\sum_\ell\|\Gamma(\ell)\|<\infty$, finite fourth moments, and fourth-order
		cumulants that are absolutely summable over the three relative time indices.
		Let $k$ be bounded, supported on $[-1,1]$, and continuous at zero with
		$k(0)=1$, and let $b_T\to\infty$ with $b_T=o(\sqrt T)$. If
		\eqref{eq:hac-substitution} holds, the estimator \eqref{eq:hac} converges in
		probability to $\LRV(\mathbf m_t)$, and \eqref{eq:feasible-score} converges
		in probability to $V_h$ for any $\widehat{\mathbf J}_h\xrightarrow{p}\mathbf J_h$.
	\end{proposition}
	\begin{proof}
		Let $\Gamma_T(\ell)=T^{-1}\sum_{t=\ell+1}^T \mathbf m_t\mathbf m_{t-\ell}'$ and
		let $V_T^o$ be the uncentered oracle version of \eqref{eq:hac}.
		Since $\E\Gamma_T(\ell)=(1-\ell/T)\Gamma(\ell)$, dominated
		convergence and covariance summability give
		\[
		\|\E V_T^o-\LRV(\mathbf m_t)\|\longrightarrow0.
		\]
		The fourth-cumulant decomposition and the assumed summability imply
		\[
		\sup_{\ell<T}\Var\{\Gamma_T(\ell)_{jk}\}\leq C/T,
		\qquad
		\|V_T^o-\E V_T^o\|_{L^2}
		\leq C(1+b_T)/\sqrt T=o(1).
		\]
		Thus $V_T^o\xrightarrow{p}\LRV(\mathbf m_t)$. Put
		\[
		R_T=\left(T^{-1}\sum_t\|\widehat{\mathbf m}_t-\mathbf m_t\|^2\right)^{1/2},
		\qquad A_T=\left(T^{-1}\sum_t\|\mathbf m_t\|^2\right)^{1/2}=O_p(1).
		\]
		Cauchy--Schwarz gives, for the uncentered feasible products,
		\[
		\sup_{\ell<T}\|\widehat\Gamma_T^u(\ell)-\Gamma_T(\ell)\|
		\leq2A_TR_T+R_T^2,
		\qquad
		\|\widehat V_T^u-V_T^o\|
		\leq C(1+b_T)(2A_TR_T+R_T^2)=o_p(1).
		\]
		Finally, $\bar{\mathbf m}=O_p(T^{-1/2})$ and
		$\|\overline{\widehat{\mathbf m}}-\bar{\mathbf m}\|\leq R_T$, so
		\[
		\|\widehat V_T-\widehat V_T^u\|
		\leq C(1+b_T)
		\{2\|\overline{\widehat{\mathbf m}}\|(A_T+R_T)
		+\|\overline{\widehat{\mathbf m}}\|^2\}=o_p(1).
		\]
		The three bounds prove the first claim, and the continuous mapping theorem
		the second.
	\end{proof}

	\subsubsection{Equal-weighted cosine inference}
	For $1\leq\nu<T$ define the vector cosine transforms and EWC covariance
	\begin{equation}
		\widehat\Lambda_{j,h}=\sqrt{\frac2T}\sum_{t=1}^T
		\check{\mathbf m}_{t,h}\cos\!\left(\frac{\pi j(t-1/2)}T\right),
		\label{eq:ewc}
	\end{equation}
	and $\widehat{\boldsymbol\Omega}_h^{\mathrm{EWC}}=\nu^{-1}\sum_{j=1}^\nu
	\widehat\Lambda_{j,h}\widehat\Lambda_{j,h}'$. For a scalar coefficient whose
	row of $L_h\widehat{\mathbf J}_h^{-1}$ is $\widehat\ell_h'$, use $\widehat V_h^{\mathrm{EWC}}=\widehat\ell_h'\widehat{\boldsymbol\Omega}_h^{\mathrm{EWC}}
	\widehat\ell_h$ and
	$\operatorname{se}_{\mathrm{EWC}}=\sqrt{\widehat V_h^{\mathrm{EWC}}/T}$. The
	calibration follows \citet{sun2013} and \citet{lazarusetal2018}. At fixed
	$\nu$, $\widehat V_h^{\mathrm{EWC}}$ has a nondegenerate limit and the
	studentized statistic converges to $t_\nu$; at increasing $\nu_T$, it is
	consistent and $t_{\nu_T}$ quantiles are a finite-sample calibration
	(Proposition~\ref{prop:ewc}). With a nonnegligible targeting cross-sectional
	component, the time-only $t_\nu$ pivot does not apply.

	\begin{proposition}[Fixed-series and increasing-series EWC limits]
		\label{prop:ewc}
		Let $\mathbf m_t\in\R^p$ have positive definite long-run covariance $\Omega$,
		let $\ell$ be a fixed nonzero row of $L_h\mathbf J_h^{-1}$, let
		$\widehat\ell\xrightarrow{p}\ell$, and let $\Lambda_j$ denote \eqref{eq:ewc}
		computed with $\mathbf m_t$ in place of $\check{\mathbf m}_{t,h}$.
		\begin{enumerate}[label=(\roman*),leftmargin=2em]
			\item Suppose $T^{-1/2}\sum_{t\leq\lfloor rT\rfloor}\mathbf m_t$ converges
			weakly to $\Omega^{1/2}W(r)$ on $[0,1]$ and
			$\widehat\Lambda_j-\Lambda_j=o_p(1)$ for $j=1,\ldots,\nu$. For fixed $\nu$,
			the EWC $t$-statistic with $\widehat\ell$ converges in distribution to
			$t_\nu$.
			\item Suppose $\sum_\ell(1+|\ell|)\|\Gamma(\ell)\|<\infty$, the fourth-order
			cumulants satisfy the condition of Proposition~\ref{prop:hac},
			$\nu_T\to\infty$, $\nu_T/T\to0$, and
			\begin{equation}
				\frac1{\nu_T}\sum_{j=1}^{\nu_T}
				\|\widehat\Lambda_j-\Lambda_j\|^2=o_p(1).
				\label{eq:ewc-substitution}
			\end{equation}
			Then $\widehat\Omega^{\mathrm{EWC}}\xrightarrow{p}\Omega$, and the EWC
			$t$-statistic with $\widehat\ell$ converges in distribution to $N(0,1)$.
		\end{enumerate}
	\end{proposition}
	\begin{proof}
		For (i), put $\varphi_j(r)=\sqrt2\cos(\pi jr)$. Integration by
		parts and the functional CLT give
		\begin{equation}
			\left(T^{-1/2}\sum_t\mathbf m_t,\Lambda_1,\ldots,\Lambda_\nu\right)
			\xrightarrow{d}
			\left(\Omega^{1/2}W(1),
			\Omega^{1/2}\int_0^1\varphi_1\,dW,\ldots,
			\Omega^{1/2}\int_0^1\varphi_\nu\,dW\right).
			\label{eq:ewc-brownian-limit}
		\end{equation}
		The identities
		\[
		\int_0^1\varphi_j(r)\,dr=0,\qquad
		\int_0^1\varphi_j(r)\varphi_k(r)\,dr={\bf1}\{j=k\}
		\]
		make the Gaussian vectors independent, each with covariance $\Omega$.
		Feasible substitution and $\widehat\ell\xrightarrow{p}\ell$ preserve this
		limit. For the scalar contrast $\ell$, the studentized limit is
		$Z/\sqrt{U/\nu}$, with independent $Z\sim N(0,1)$ and
		$U\sim\chi_\nu^2$.

		For (ii), put $\varphi_{j,t}=\sqrt2\cos\{\pi j(t-1/2)/T\}$.
		For $1\leq j,k<T$,
		\[
		T^{-1}\sum_t\varphi_{j,t}\varphi_{k,t}={\bf1}\{j=k\},
		\qquad |\varphi_{j,t-\ell}-\varphi_{j,t}|
		\leq Cj|\ell|/T.
		\]
		Weighted covariance summability therefore gives
		\begin{equation}
			\sup_{j,k\leq\nu_T}
			\|\Cov(\Lambda_j,\Lambda_k)-{\bf1}\{j=k\}\Omega\|
			\leq C\nu_T/T.
			\label{eq:ewc-frequency-cov-bound}
		\end{equation}
		Absolute fourth-cumulant summability and
		$\max_{j,t}|\varphi_{j,t}/\sqrt T|\leq\sqrt{2/T}$ imply
		\[
		|\operatorname{cum}(\Lambda_{j_1,a},\Lambda_{j_2,b},
		\Lambda_{j_3,c},\Lambda_{j_4,d})|\leq C/T.
		\]
		For $\Omega_T^o=\nu_T^{-1}\sum_{j\leq\nu_T}\Lambda_j\Lambda_j'$, it follows that
		\begin{equation}
			\|\E\Omega_T^o-\Omega\|=O(\nu_T/T),\qquad
			\Var\{(\Omega_T^o)_{ab}\}
			\leq C\{\nu_T^{-1}+(\nu_T/T)^2+T^{-1}\}=o(1).
			\label{eq:ewc-variance-bound}
		\end{equation}
		Hence $\Omega_T^o\xrightarrow{p}\Omega$. Let
		$R_T^2=\nu_T^{-1}\sum_j\|\widehat\Lambda_j-\Lambda_j\|^2=o_p(1)$.
		Then
		\[
		\|\widehat\Omega_T-\Omega_T^o\|
		\leq2\left(\nu_T^{-1}\sum_j\|\Lambda_j\|^2\right)^{1/2}R_T+R_T^2
		=o_p(1).
		\]
		Feasible consistency, $\widehat\ell\xrightarrow{p}\ell$, and the estimator
		CLT give the stated normal limit.
	\end{proof}

	\subsubsection{Joint restrictions and horizon bands}
	Stack all desired coefficients and their date-level moments on the same
	sample, and form the joint sandwich covariance $\widehat V$. If $R$ defines $r$ linearly independent restrictions, consistent
	HAC gives
	\begin{equation}
		T(R\widehat\beta-r_0)'
		(R\widehat V R')^{-1}(R\widehat\beta-r_0)
		\ \xrightarrow{d}\ \chi_r^2.
		\label{eq:wald}
	\end{equation}
	Under the fixed-$\nu$ EWC conditions and $\nu\geq r$, the corrected
	statistic $(\nu-r+1)/(\nu r)$ times the left-hand side of
	\eqref{eq:wald}, with EWC in place of HAC, converges to
	$F_{r,\nu-r+1}$. This follows because
	$\nu R\widehat V^{\mathrm{EWC}}R'$, with $\widehat V^{\mathrm{EWC}}$ the
	sandwich formed with $\widehat{\boldsymbol\Omega}^{\mathrm{EWC}}$, has an
	independent Wishart limit. Marginal intervals are pointwise; simultaneous
	bands can be calibrated from the maximum absolute studentized statistic in the
	joint bootstrap below.

	\subsection{Resampling inference}
	\label{app:bootstrap}

	The time-dimension bootstrap resamples contiguous time blocks jointly across
	the aggregate data, the proxies, and the panel cross section, keeping unit
	labels and exposures fixed and using the same draws across horizons. With
	block length $\ell_T\to\infty$ and $\ell_T/T\to0$, joint blocks preserve the
	cross covariance between the panel and first-step moments; independent
	aggregate and panel draws remove it. We resample circular blocks of
	preassembled date records, each containing all lags, leads, and panel
	outcomes needed at its base date, with block length larger than the lag--lead
	span.

	Full re-estimation reruns every estimated component, namely the aggregate
	moments or the VAR, the proxy restrictions, the normalization, the estimated
	shocks, the offset weights, the generated controls, the residualization, and
	the panel IV, on the identified branch of the original sample. Alternatively,
	a centered one-step block bootstrap resamples blocks of $\check{\mathbf m}_t$
	and takes one Newton step from the original estimate,
	\begin{equation}
		\beta^*=\widehat\beta
		-L\widehat{\mathbf J}^{-1}\frac1T\sum_{t=1}^T\check{\mathbf m}_t^*,
		\label{eq:linear-bootstrap}
	\end{equation}
	which reproduces the first-order distribution under a valid block-bootstrap
	approximation \citep{kunsch1989,goncalves2011}.

	When the targeting component of Proposition~\ref{prop:targeting} matters, the
	time draw is supplemented, under the independence used in
	\eqref{eq:target-variance}, by a centered cross-sectional draw of
	$L_h\widehat{\mathbf J}_h^{-1}E_g\widehat\kappa_i$ at scale $N^{-1/2}$.

	\subsection{Weak controlled treatment relevance}
	\label{app:weak}
	For HELP-IV with the instruments of Section~\ref{sec:help-estimation}, write
	$\vartheta=(b,c')'$ and partition the sample moments as
	$\bar g(\vartheta)=(\bar g_g(\vartheta),\bar g_r(\vartheta)')'$.
	If the nuisance path block $\Sigma_{rc}$ is strongly identified, solve
	$\bar g_r(b,\widehat c(b))=0$ at each candidate $b$ and form
	\[
	\zeta_t(b)=g_{g,t}(b,c(b))-\Sigma_{gc}\Sigma_{rc}^{-1}g_{r,t}(b,c(b)).
	\]
	With a consistent HAC estimate of its long-run variance,
	\begin{equation}
		\mathcal C_{1-\alpha}
		=\left\{b:\frac{T\bar g_g(b,\widehat c(b))^2}
		{\widehat\Omega_\zeta(b)}
		\leq\chi^2_{1,1-\alpha}\right\}.
		\label{eq:help-ar}
	\end{equation}
	The denominator accounts for estimation of $c(b)$, and the null moment has a
	regular limit without dividing by the controlled first stage. If the path
	block is weak, the joint moment test over $(b,c)$ is inverted and projected
	onto $b$ \citep{dufourtaamouti2005}. With SVAR shocks or estimated offset
	weights, the first-step moments are stacked into $\zeta_t(b)$.

	\FloatBarrier\clearpage
	\section{Monte Carlo details}
	\label{app:mc}

	This appendix reports the designs and results summarized in
	Section~\ref{sec:montecarlo}, in the order of the main text.

	\subsection{Weak proxies}
	\label{app:mc-weak}

	Table~\ref{tab:mc-weak} raises the standard deviation of the measurement error
	of one proxy in the backward-looking environment from $0.5$ to $2$ and $3$,
	and keeps that of the other proxy at $0.5$. The structural shock then accounts
	for 20 and 10 percent of the variance of the weakened proxy, against 80
	percent in Section~\ref{sec:mc-single}.

	The table also reports the median first-stage $F$ statistic of HELP-IV. The
	statistic is the squared ratio
	of the sample covariance between the treatment and the effective instrument,
	which is the denominator of HELP-IV, to the standard error of that
	covariance. The effective instrument is the combination of instruments that is
	uncorrelated in the sample with every path control, and the standard error
	uses the long-run variance estimator of the HELP-IV standard errors. Because
	the statistic divides the covariance by its own standard error, it does not
	depend on the units of the instruments or of the confounders.

	Besides the measures of the main text, this table and the following ones
	report the range of the estimates, which is the difference between their 95th
	and 5th percentiles. Unlike the RMSE, the range is not dominated by the few
	replications with extreme estimates, which occur because a just-identified IV
	estimator has no finite moments \citep{kinal1980}. Comparing the median
	length of the intervals with the range shows whether the standard errors track
	the dispersion of the estimates.

	\begin{table}[!ht]\centering
		\caption{HELP-IV with weak proxies in the backward-looking environment}
		\label{tab:mc-weak}
		{\footnotesize\setlength{\tabcolsep}{5pt}
			\begin{tabular}{cc r rrrrr}
\toprule
Noise & $h$ & First-stage $F$ & Bias & RMSE & Range & Coverage & Length\\
\midrule
\multicolumn{8}{l}{\textit{Panel A. Noise of the treatment proxy}}\\[1pt]
0.5 & 0 & 48 & $\phantom{-}0.00$ & 0.01 & 0.02 & 0.91 & 0.02\\
 & 2 & 40 & $-0.01$ & 0.12 & 0.39 & 0.88 & 0.40\\
 & 4 & 39 & $-0.02$ & 0.17 & 0.53 & 0.87 & 0.53\\
2 & 0 & 16 & $\phantom{-}0.00$ & 0.01 & 0.04 & 0.91 & 0.04\\
 & 2 & 15 & $-0.02$ & 0.21 & 0.69 & 0.89 & 0.67\\
 & 4 & 15 & $-0.03$ & 0.27 & 0.86 & 0.90 & 0.84\\
3 & 0 & 9 & $\phantom{-}0.00$ & 0.02 & 0.06 & 0.92 & 0.06\\
 & 2 & 8 & $-0.02$ & 0.30 & 0.98 & 0.91 & 0.93\\
 & 4 & 8 & $-0.03$ & 0.38 & 1.22 & 0.91 & 1.14\\
\addlinespace
\multicolumn{8}{l}{\textit{Panel B. Noise of the confounder proxy}}\\[1pt]
2 & 0 & 32 & $\phantom{-}0.00$ & 0.01 & 0.03 & 0.91 & 0.03\\
 & 2 & 22 & $-0.01$ & 0.20 & 0.63 & 0.90 & 0.62\\
 & 4 & 19 & $-0.02$ & 1.00 & 1.09 & 0.91 & 0.88\\
3 & 0 & 22 & $\phantom{-}0.00$ & 0.01 & 0.04 & 0.91 & 0.03\\
 & 2 & 13 & $-0.01$ & 0.82 & 0.91 & 0.92 & 0.84\\
 & 4 & 11 & $-0.04$ & 7.31 & 1.91 & 0.94 & 1.26\\
\bottomrule
\end{tabular}
}

		\vspace{0.6em}
		{\footnotesize
			\begin{minipage}{0.8\textwidth}
				\textit{Notes.} Two-way general equilibrium, $N=T=200$, and $2{,}000$
				replications. Noise is the standard deviation of the measurement error
				of the proxy named in the panel title, and the other proxy keeps a
				standard deviation of $0.5$. The first block of Panel A repeats the
				design of Table~\ref{tab:mc-single} with new draws. The target is
				$\beta_h^P$. First-stage $F$ is the median across replications of the
				first-stage $F$ statistic.
		\end{minipage}}
	\end{table}

	Table~\ref{tab:mc-weak} delivers two results. First, the median bias of HELP-IV
	stays at most $0.04$ in absolute value although the median first-stage $F$
	statistic falls from $39$--$48$ with the proxies of the main text to $8$ or
	$9$ with the noisiest treatment proxy, and its intervals cover the target in
	89--94 percent of replications in every cell with a weakened proxy. These
	designs do not establish uniform validity of the Wald intervals under weak
	identification. Second, weak proxies cost precision. When the noise of the treatment
	proxy is $3$, the range of HELP-IV at $h=2$ and $h=4$ is $2.3$ to $2.5$ times
	that of the first block. When the noise of the confounder proxy is $3$, the
	range is up to $3.6$ times as large, and $12$ of the $2{,}000$ replications
	have errors above $10$ in absolute value at $h=4$. Because these replications
	dominate the RMSE, the RMSE equals $7.31$ and exceeds the range of $1.91$.

	\FloatBarrier
	\subsection{Exact identification and the first step in the forward-looking environment}
	\label{app:mc-forward}

	Table~\ref{tab:mc-exact} reports the exactly identified case, in which the
	researcher observes proxies for all six news shocks instead of two.\footnote{In the forward-looking environment, the
		first step of E-HELP-IV and the time-series step of the
		decomposition method control for two lags of $G_t$ and $R_t$.} The proxies of
	the seven date-$t$ shocks that move $R$, namely $z_{r,t}$ and the six news
	proxies, then span the confounder path through $t+6$. The composite instrument
	therefore holds the whole expected path fixed, and the unabsorbed share of
	Section~\ref{sec:forward-portable-section} is zero, against 2.0 percent with
	two news proxies. TWFE-IV and the decomposition method are biased with the
	same signs as in Table~\ref{tab:mc-single}. E-HELP-IV has a median bias of at most $0.04$ in absolute value and an RMSE of
	$0.17$ to $0.47$, and its intervals cover the target in 92--94 percent of
	replications, slightly more often than the nominal 90 percent.

	\begin{table}[!ht]\centering
		\caption{Forward-looking environment under exact identification}
		\label{tab:mc-exact}
		{\small\setlength{\tabcolsep}{4.2pt}
			\begin{tabular}{cc rr rr rrrrr}
\toprule
 & & \multicolumn{2}{c}{TWFE-IV} & \multicolumn{2}{c}{Decomposition} & \multicolumn{5}{c}{HELP-IV}\\
\cmidrule(lr){3-4}\cmidrule(lr){5-6}\cmidrule(lr){7-11}
$h$ & Target & Bias & RMSE & Bias & RMSE & Bias & RMSE & Range & Coverage & Length\\
\midrule
0 & 0.88 & $-0.10$ & 0.12 & $-0.22$ & 0.25 & $\phantom{-}0.00$ & 0.17 & 0.15 & 0.94 & 0.15\\
2 & 0.82 & $\phantom{-}0.58$ & 0.61 & $-0.22$ & 0.37 & $-0.03$ & 0.47 & 0.66 & 0.92 & 0.68\\
4 & 0.27 & $\phantom{-}0.83$ & 0.90 & $-0.09$ & 0.22 & $-0.04$ & 0.28 & 0.79 & 0.92 & 0.79\\
\bottomrule
\end{tabular}
}

		\vspace{0.6em}
		{\footnotesize
			\begin{minipage}{0.86\textwidth}
				\textit{Notes.} $N=T=200$ and $2{,}000$ replications. The target is the
				information-constrained effect $\beta_h^E$, which under exact
				identification equals the forward-looking partial-equilibrium effect.
				HELP-IV is E-HELP-IV, and the decomposition method is the ex
				ante decomposition of \citet{donaldson2026}.
		\end{minipage}}
	\end{table}

	To locate the source of this over-coverage, Table~\ref{tab:mc-fwdcov}
	compares the intervals of Panel B of Table~\ref{tab:mc-single} and of
	Table~\ref{tab:mc-exact} with two alternatives. The first alternative uses the
	population weights, which need no first-step term. The second omits
	the first-step term from the standard error. The first step sets the weights
	from the sample covariances between the confounder path and the date-$t$
	proxies. The composite instrument therefore has zero sample covariance with
	every combination of $R_t,\ldots,R_{t+6}$ that the proxies can offset. As a
	result, the estimated weights lower the range of the estimates by about 45 to
	65 percent relative to the population weights. With the population weights,
	the intervals cover the target in 87--88 percent of replications. Without the
	first-step term, the standard error ignores the lower dispersion, and the
	intervals cover the target in 96--100 percent of replications. With the
	first-step term, the median length of the intervals is close to the range of
	the estimates, but the $t$-statistic remains less dispersed than a standard
	normal variable, so coverage is 92--95 percent.\footnote{The simulations use
		Student-$t_\nu$ critical values for designs whose simulated $t$-statistic is
		more dispersed than a standard normal variable and normal critical values for
		the other designs, including E-HELP-IV. This calibration depends on the
		design and differs from the prespecified EWC procedure of
		Appendix~\ref{app:variance}.}

	\begin{table}[!ht]\centering
		\caption{Coverage of E-HELP-IV with and without the first-step term
			and with population weights}
		\label{tab:mc-fwdcov}
		{\footnotesize\setlength{\tabcolsep}{5pt}
			\begin{tabular}{c r rrr rr rrr}
\toprule
 & \multicolumn{6}{c}{Estimated weights} & \multicolumn{3}{c}{Population weights}\\
\cmidrule(lr){2-7}\cmidrule(lr){8-10}
 & & \multicolumn{3}{c}{With first-step term} & \multicolumn{2}{c}{Without first-step term} & & & \\
\cmidrule(lr){3-5}\cmidrule(lr){6-7}
$h$ & Range & Length & Coverage & \begin{tabular}[b]{@{}c@{}}Standard\\deviation of $t$\end{tabular} & Length & Coverage & Range & Coverage & \begin{tabular}[b]{@{}c@{}}Standard\\deviation of $t$\end{tabular}\\
\midrule
\multicolumn{10}{l}{\textit{Panel A. Under-identification (Panel B of Table~\ref{tab:mc-single})}}\\[1pt]
0 & 0.25 & 0.28 & 0.95 & 0.82 & 0.38 & 0.96 & 0.45 & 0.88 & 1.15\\
2 & 0.65 & 0.69 & 0.93 & 0.92 & 1.17 & 0.98 & 1.35 & 0.88 & 1.08\\
4 & 0.91 & 0.88 & 0.92 & 0.96 & 1.91 & 0.99 & 2.25 & 0.88 & 1.07\\
\addlinespace
\multicolumn{10}{l}{\textit{Panel B. Exact identification (Table~\ref{tab:mc-exact})}}\\[1pt]
0 & 0.15 & 0.15 & 0.94 & 0.89 & 0.48 & 1.00 & 0.44 & 0.88 & 1.15\\
2 & 0.66 & 0.68 & 0.92 & 0.91 & 1.43 & 1.00 & 1.30 & 0.87 & 1.09\\
4 & 0.79 & 0.79 & 0.92 & 0.93 & 2.34 & 1.00 & 2.15 & 0.88 & 1.08\\
\bottomrule
\end{tabular}
}

		\vspace{0.6em}
		{\footnotesize
			\begin{minipage}{0.86\textwidth}
				\textit{Notes.} $N=T=200$ and $2{,}000$ replications. The standard deviation of $t$ is taken across
				replications, where $t$ is the estimation error divided by the standard error. The columns with
				the first-step term repeat the intervals of Panel B of
				Table~\ref{tab:mc-single} and of Table~\ref{tab:mc-exact}. The columns
				without the first-step term build the standard error from the
				date-level IV moment alone. The last two columns report, from separate
				replications, the estimator that uses the composite instrument with the
				population weights and a standard error without a first-step term.
		\end{minipage}}
	\end{table}

	\FloatBarrier
	\subsection{Shocks estimated from an SVAR}
	\label{app:mc-internal}

	This subsection uses as instruments the shocks estimated from an
	SVAR, as in Appendix~\ref{app:svar-inference}, instead of the proxies. The researcher fits a VAR(2) to $(G_t,R_t)$, which
	is correctly specified. The impact column of the treatment shock is identified
	by the covariance between the reduced-form residuals and the treatment proxy,
	and the confounder shock is the remaining structural shock. HELP-IV then uses
	the estimated shocks as instruments, and composite HELP-IV uses a single
	instrument whose weights come from the impulse responses of the fitted SVAR.
	The sandwich covariance stacks the SVAR estimating equations with the IV
	moment, and for composite HELP-IV these first-step moments are aligned with
	the lead of the instrument that they offset (Appendix~\ref{app:alignment}).

	\begin{table}[!ht]\centering
		\caption{Backward-looking environment with proxies and with SVAR shocks
			as instruments}
		\label{tab:mc-internal}
		\resizebox{\textwidth}{!}{\footnotesize\setlength{\tabcolsep}{3.4pt}%
			\begin{tabular}{cc rrrr rrrr rrrr}
\toprule
 & & \multicolumn{4}{c}{HELP-IV, external proxies} & \multicolumn{4}{c}{HELP-IV, SVAR shocks} & \multicolumn{4}{c}{Composite HELP-IV, SVAR shocks}\\
\cmidrule(lr){3-6}\cmidrule(lr){7-10}\cmidrule(lr){11-14}
$h$ & Target & Bias & Range & Coverage & Length & Bias & Range & Coverage & Length & Bias & Range & Coverage & Length\\
\midrule
0 & 0.62 & $\phantom{-}0.00$ & 0.02 & 0.89 & 0.02 & $\phantom{-}0.00$ & 0.02 & 0.90 & 0.02 & $\phantom{-}0.00$ & 0.02 & 0.90 & 0.02\\
2 & 0.66 & $-0.01$ & 0.40 & 0.88 & 0.40 & $-0.01$ & 0.34 & 0.87 & 0.34 & $-0.01$ & 0.35 & 0.87 & 0.34\\
4 & 0.21 & $-0.02$ & 0.56 & 0.88 & 0.53 & $-0.02$ & 0.47 & 0.87 & 0.47 & $\phantom{-}0.00$ & 0.58 & 0.90 & 0.59\\
\bottomrule
\end{tabular}
}

		\vspace{0.6em}
		{\footnotesize
			\begin{minipage}{0.9\textwidth}
				\textit{Notes.} Two-way general equilibrium, $N=T=200$, and the
				$2{,}000$ replications of Panel A of Table~\ref{tab:mc-single}. The first
				block repeats HELP-IV of that panel.
		\end{minipage}}
	\end{table}

	Table~\ref{tab:mc-internal} shows that HELP-IV and composite HELP-IV with
	SVAR shocks both have a median bias of at most $0.02$ in absolute
	value. Their intervals cover the target in 87--90 percent of replications,
	against 88--89 percent with the proxies as instruments. The median length of the
	intervals is within $0.02$ of the range of the estimates at every horizon, so
	the standard errors with the first-step term track the dispersion of the
	estimates.

	\FloatBarrier
	\subsection{One-way general equilibrium}
	\label{app:mc-don}

	To show that the bias of the decomposition method in Panel~A of
	Table~\ref{tab:mc-single} comes from two-way general equilibrium on impact
	(Section~\ref{sec:donaldson-connection}),
	Table~\ref{tab:mc-oneway} removes $R_t$ and $R_{t-1}$ from the equation for
	$G_t$, so that general equilibrium is one-way. Because the
	impact response of the treatment under the fixed path then equals its
	unrestricted impact response, the decomposition method recovers $\beta_h^P$
	in population.\footnote{\citet[Appendix~C.4]{donaldson2026} considers a
		specification similar to that of our cross-sectional step, which adds
		exposure times the lagged treatment and the lagged confounder shock.} TWFE-IV
	is biased by the exposure-correlation component alone, which is $-0.18$ on
	impact and at most $0.04$ in absolute value thereafter. HELP-IV and the
	decomposition method both have a median bias of at most $0.02$ in absolute
	value, and their RMSEs differ by at most $0.02$.

	\begin{table}[!ht]\centering
		\caption{Backward-looking environment under one-way general equilibrium}
		\label{tab:mc-oneway}
		\resizebox{\textwidth}{!}{\small\setlength{\tabcolsep}{3.4pt}%
			\begin{tabular}{cc rr rrrrr rrrrr}
\toprule
 & & \multicolumn{2}{c}{TWFE-IV} & \multicolumn{5}{c}{Decomposition} & \multicolumn{5}{c}{HELP-IV}\\
\cmidrule(lr){3-4}\cmidrule(lr){5-9}\cmidrule(lr){10-14}
$h$ & Target & Bias & RMSE & Bias & RMSE & Range & Coverage & Length & Bias & RMSE & Range & Coverage & Length\\
\midrule
0 & 0.62 & $-0.18$ & 0.18 & $\phantom{-}0.00$ & 0.02 & 0.07 & 0.97 & 0.10 & $\phantom{-}0.00$ & 0.01 & 0.02 & 0.88 & 0.02\\
2 & 0.66 & $-0.04$ & 0.09 & $-0.01$ & 0.12 & 0.37 & 0.84 & 0.33 & $-0.01$ & 0.12 & 0.39 & 0.88 & 0.40\\
4 & 0.21 & $\phantom{-}0.00$ & 0.10 & $-0.02$ & 0.16 & 0.52 & 0.80 & 0.41 & $-0.02$ & 0.17 & 0.55 & 0.87 & 0.53\\
\bottomrule
\end{tabular}
}

		\vspace{0.6em}
		{\footnotesize
			\begin{minipage}{0.95\textwidth}
				\textit{Notes.} Same design as Panel A of Table~\ref{tab:mc-single},
				except that the equation for $G_t$ omits $R_t$ and $R_{t-1}$.
		\end{minipage}}
	\end{table}

	The two methods differ in the coverage of their intervals. The intervals of
	HELP-IV cover the target in 87--88 percent of replications, whereas those of
	the decomposition method cover it in 97 percent at $h=0$ but in only 84 and
	80 percent at $h=2$ and $h=4$. The difference comes from the covariance
	estimator of the decomposition method.
	\citet[Appendix~A.7]{donaldson2026} writes her estimator as a just-identified
	system of generalized method of moments (GMM) conditions that stacks the panel
	and the time-series moments over all horizons. The baseline covariance
	estimator, specification~(ii) of that appendix, sets to zero the covariances
	between the panel and the time-series moments and the covariances across
	horizons. The resulting standard error is distorted in both directions: the
	baseline intervals are longer than the range of the estimates at $h=0$ and
	about 10 to 20 percent shorter at $h=2$ and $h=4$.

	\FloatBarrier
	\subsection{Two exposure margins}
	\label{app:mc-margins}

	To evaluate HELP-IV with multiple exposure margins (Section~\ref{sec:multiple-exposure-margins}), this
	subsection lets the treatment reach each unit through two exposure margins. The first margin is the exposure
	$\sig$ of Section~\ref{sec:mc-single-dgp}. The second, $s_{ig,2}$, equals
	$0.5\,\sig$ plus an independent normal term with standard deviation $0.8$, so
	that the two margins have a correlation of $0.53$. The outcome depends on the
	second margin through $s_{ig,2}(0.35\,G_t-0.12\,G_{t-1}+0.08\,G_{t-2})$.
	Confounder exposure is $\sir=0.65\,\sig+0.30\,s_{ig,2}$ plus an independent
	standard normal term, so it rises with both margins. Because
	$0.65+0.30\times0.5=0.8$, the projection of $\sir$ on $\sig$ alone keeps the
	coefficient $0.8$ of Section~\ref{sec:mc-single-dgp}. HELP-IV interacts the
	treatment, the path controls, and the instruments with both margins, whereas
	TWFE-IV is estimated margin by margin. Both panel regressions control for the
	lags of $G_t$ and $R_t$ interacted with both margins.

	\begin{table}[!ht]\centering
		\caption{Two exposure margins}
		\label{tab:mc-margins}
		{\small\setlength{\tabcolsep}{4.6pt}
			\begin{tabular}{cc rr rrrrr}
\toprule
 & & \multicolumn{2}{c}{TWFE-IV} & \multicolumn{5}{c}{HELP-IV}\\
\cmidrule(lr){3-4}\cmidrule(lr){5-9}
$h$ & Target & Bias & RMSE & Bias & RMSE & Range & Coverage & Length\\
\midrule
\multicolumn{9}{l}{\textit{Panel A. Margin 1}}\\[1pt]
0 & 0.62 & $-0.14$ & 0.15 & $\phantom{-}0.00$ & 0.01 & 0.03 & 0.89 & 0.03\\
2 & 0.66 & $\phantom{-}0.35$ & 0.37 & $-0.03$ & 0.13 & 0.40 & 0.88 & 0.40\\
4 & 0.21 & $\phantom{-}0.59$ & 0.62 & $-0.03$ & 0.18 & 0.59 & 0.85 & 0.53\\
\addlinespace
\multicolumn{9}{l}{\textit{Panel B. Margin 2}}\\[1pt]
0 & 0.35 & $-0.07$ & 0.07 & $\phantom{-}0.00$ & 0.01 & 0.03 & 0.89 & 0.03\\
2 & 0.23 & $\phantom{-}0.13$ & 0.13 & $-0.01$ & 0.05 & 0.16 & 0.87 & 0.16\\
4 & 0.10 & $\phantom{-}0.22$ & 0.23 & $-0.01$ & 0.07 & 0.22 & 0.85 & 0.20\\
\bottomrule
\end{tabular}
}

		\vspace{0.6em}
		{\footnotesize
			\begin{minipage}{0.8\textwidth}
				\textit{Notes.} The design is otherwise that of Panel A of
				Table~\ref{tab:mc-single}. HELP-IV denotes HELP-IV with multiple exposure
				margins. The targets are $\beta_{1,h}^P$ in Panel A and
				$\beta_{2,h}^P$ in Panel B.
		\end{minipage}}
	\end{table}

	Table~\ref{tab:mc-margins} shows that HELP-IV recovers both margin-specific
	partial-equilibrium effects, $\beta_{1,h}^P$ and $\beta_{2,h}^P$. Its median
	bias is at most $0.03$ in absolute value, and its intervals cover the target
	in 85--89 percent of replications on both margins. TWFE-IV is biased on both
	margins, because confounder exposure rises with both. On the second margin,
	its bias reaches $0.22$ against a target of $0.10$.

	\FloatBarrier
	\subsection{The two-region model}
	\label{app:mc-model}

	This subsection describes the two-region model of
	\citet[Appendix~D]{donaldson2026}, its calibration, and the design of the
	simulations. Symbols specific
	to her model follow her paper and are local to this subsection and the next.

	The model is a monetary union of two regions, Home ($H$) and Foreign ($F$),
	each of size $0.5$. The model extends \citet{nakamurasteinsson2014} by letting the
	regions differ in their exposure to the aggregate spending shock and in their
	intertemporal elasticity of substitution. We log-linearize the model around
	the symmetric zero-inflation steady state and solve the linear model with
	Dynare. Lower-case
	variables are log deviations from the steady state, and spending is measured
	in units of steady-state output, so a spending shock of one is one percent
	of steady-state output.

	Households in each region consume both goods, with a share $\phi_j$ of the
	good of their own region $j\in\{H,F\}$. Home and Foreign consumer-price
	inflation are therefore $\pi^{C}_{H,t}=\phi_H\pi_{H,t}+(1-\phi_H)\pi_{F,t}$ and
	$\pi^{C}_{F,t}=\phi_F\pi_{F,t}+(1-\phi_F)\pi_{H,t}$, where $\pi_{H,t}$ and
	$\pi_{F,t}$ are the producer-price inflation rates. The Home household
	satisfies the Euler equation
	\[
	c_{H,t}=\E_tc_{H,t+1}-\sigma_H\bigl(i_t-\E_t\pi^{C}_{H,t+1}\bigr),
	\]
	where $i_t$ is the nominal interest rate of the union and $\sigma_H$ is the
	Home elasticity of intertemporal substitution. Complete markets link
	consumption across regions through
	$c_{H,t}/\sigma_H-c_{F,t}/\sigma_F=(\phi_H+\phi_F-1)\tau_t$, where
	$\tau_t=\tau_{t-1}+\pi_{F,t}-\pi_{H,t}$ is the relative price of Foreign
	goods. The Foreign Euler equation follows from the Home Euler equation and
	the risk-sharing condition. With a
	steady-state spending share of $0.2$ and a trade elasticity $\eta$, goods
	markets clear when
	\begin{align*}
		y_{H,t}&=0.8\Bigl[\phi_H\bigl(c_{H,t}+\eta(1-\phi_H)\tau_t\bigr)
		+(1-\phi_F)\bigl(c_{F,t}+\eta\phi_F\tau_t\bigr)\Bigr]+g_{H,t},\\
		y_{F,t}&=0.8\Bigl[(1-\phi_H)\bigl(c_{H,t}-\eta\phi_H\tau_t\bigr)
		+\phi_F\bigl(c_{F,t}-\eta(1-\phi_F)\tau_t\bigr)\Bigr]+g_{F,t},
	\end{align*}
	where $g_{H,t}$ and $g_{F,t}$ are regional government purchases.

	Firms keep their price with probability $\alpha$ in each quarter, so each
	region has the Phillips curve
	$\pi_{j,t}=\varkappa\,mc_{j,t}+\beta\E_t\pi_{j,t+1}$, with real marginal cost
	\begin{align*}
		mc_{H,t}&=\frac{1/\nu+1-a}{a}\,y_{H,t}+\frac{c_{H,t}}{\sigma_H}+(1-\phi_H)\tau_t,\\
		mc_{F,t}&=\frac{1/\nu+1-a}{a}\,y_{F,t}+\frac{c_{F,t}}{\sigma_F}-(1-\phi_F)\tau_t,
	\end{align*}
	where $\beta$ is the discount factor, $\nu$ is the Frisch elasticity of labor
	supply, and $a$ is the elasticity of output with respect to labor. The slope is
	$\varkappa=\zeta(1-\alpha)(1-\alpha\beta)/\alpha$, where
	$\zeta=1/[1+((1+1/\nu)/a-1)\vartheta]$ is the strategic-complementarity factor
	of \citet{nakamurasteinsson2014} and $\vartheta=7$ is the elasticity of
	substitution between varieties.

	The nominal rate follows the policy rule
	\[
	i_t=\rho_i\,i_{t-1}+(1-\rho_i)\bigl(\phi_\pi\pi^{agg}_t+\phi_y\,y^{agg}_t\bigr)+u_t,
	\]
	where $\pi^{agg}_t=0.5\,(\pi^{C}_{H,t}+\pi^{C}_{F,t})$ and
	$y^{agg}_t=0.5\,(y_{H,t}+y_{F,t})$ are inflation and output in the union. The
	policy shock $u_t=\rho_u u_{t-1}+\epsilon_{i,t}+\sum_{k\ge1}\epsilon^{n}_{k,t-k}$
	combines a contemporaneous shock $\epsilon_{i,t}$ with news shocks
	$\epsilon^{n}_{k,t}$, which are announced at $t$ about date $t+k$. The
	confounder is the real rate $r_t=i_t-\E_t\pi^{agg}_{t+1}$, and the
	regressions use its annualized value $R_t=4r_t$.

	The persistence $\rho_u$ of the policy shock defines the two scenarios of her
	paper. In the aligned scenario, $\rho_u=0.85$, so a policy shock moves the
	real rate persistently, as a spending shock does, and a few date-$t$ policy
	shocks can reproduce the path of the real rate that a spending shock
	generates. In the misaligned scenario, $\rho_u=0$, so a policy shock moves the
	real rate only briefly, as the last rows of Table~\ref{tab:mc-repl} show.
	Our simulations use the aligned
	scenario, and the misaligned scenario serves only to check our implementation
	against her figures.

	Regional spending follows $g_{j,t}=\rho_g g_{j,t-1}+\ell_j x_t+e_{j,t}$, where
	$x_t$ is the common spending impulse, the loading $\ell_j$ is the exposure of
	region $j$, and $e_{j,t}$ is a regional shock. The loadings satisfy
	$0.5\,\ell_H+0.5\,\ell_F=1$, so aggregate spending
	$G_t=0.5\,(g_{H,t}+g_{F,t})$ moves one for one with the impulse. In her model,
	$x_t=\epsilon_{g,t}$, the spending shock, so spending does not respond to the
	real rate and general equilibrium is one-way. In our two-way version,
	$x_t=\epsilon_{g,t}-\psi\,r_{t-1}$, so spending responds to the real rate with
	a one-quarter lag. Because of the lag, $G_t$ contains no contemporaneous
	policy shock, as Proposition~\ref{lemma:raw-fails} requires.

	The calibration follows Table~A12 of \citet{donaldson2026}, with $\beta=0.99$,
	$\nu=1$, $\eta=2$, $a=0.63$, $\phi_H=\phi_F=0.85$, $\alpha=0.75$, $\rho_i=0.8$,
	$\phi_\pi=1.5$, $\rho_g=0.85$, and no hand-to-mouth households. Four settings
	depart from that table, because its printed values do not reproduce her
	Figures~3 and~A3. First, the slope includes the factor $\zeta$, which gives
	$\varkappa=0.0053$. Second, the output coefficient inside the bracket of the
	policy rule is $\phi_y=0.5$, so its effective value is $0.1$. Third, the
	elasticities $\sigma_H=1.2$ and $\sigma_F=0.8$ and the loadings $\ell_H=0.8$ and
	$\ell_F=1.2$ are the values of her Table~3. The region more exposed to
	spending thus has the lower elasticity of intertemporal substitution, so its
	consumption responds less to the real rate, and exposure to the treatment and
	exposure to the real rate are negatively correlated. Fourth, we choose
	$\rho_u=0.85$ and $\rho_u=0$ to reproduce her figures, because she does not
	report the values behind them.\footnote{Her Table~3, which serves another
		exercise, sets $\rho_u=0.75$ and a hand-to-mouth share of $0.2$.}
	Table~\ref{tab:mc-repl} shows that our implementation reproduces her
	Figures~3 and~A3 with the four settings.

	\begin{table}[!ht]\centering
		\caption{Figures 3 and A3 of \citet{donaldson2026} against our implementation}
		\label{tab:mc-repl}
		\adjustbox{max width=\textwidth}{%
			\footnotesize\setlength{\tabcolsep}{4pt}%
			\begin{tabular}{l ccccc}
				\toprule
				& \multicolumn{5}{c}{Quarter}\\
				\cmidrule(lr){2-6}
				Object & 0 & 1 & 2 & 3 & 4\\
				\midrule
				Partial-equilibrium multiplier $\mathcal B_h^P$ & .955 (.954) & .935 (.932) & .91 (.911) & .89 (.891) & .875 (.872)\\
				TWFE & 1.36 (1.36) & 1.345 (1.34) & 1.33 (1.33) & 1.31 (1.31) & 1.295 (1.29)\\
				Ex ante, one shock, misaligned & 1.22 (1.23) & 1.225 (1.23) & 1.22 (1.23) & 1.215 (1.22) & 1.205 (1.21)\\
				Ex ante, four shocks, misaligned & 1.13 (1.14) & 1.12 (1.13) & 1.105 (1.12) & 1.095 (1.11) & 1.09 (1.10)\\
				Ex ante, one shock, aligned & 1.015 (1.02) & 1.0 (1.01) & .99 (.992) & .975 (.976) & .96 (.96)\\
				Ex ante, four shocks, aligned & --- (.955) & --- (.933) & --- (.913) & --- (.893) & --- (.874)\\
				$R$ after a spending shock, relative to impact & 1 (1) & 1.33 (1.34) & 1.44 (1.45) & 1.40 (1.43) & 1.30 (1.34)\\
				$R$ after a policy shock, misaligned & .20 (.195) & .125 (.123) & .08 (.078) & .05 (.049) & .033 (.031)\\
				$R$ after a policy shock, aligned & .20 (.195) & .19 (.192) & .18 (.180) & .16 (.163) & .145 (.146)\\
				\bottomrule
		\end{tabular}}

		\vspace{0.6em}
		{\footnotesize
			\begin{minipage}{0.9\textwidth}
				\textit{Notes.} Each cell gives the value read from her figure and, in
				parentheses, the value of our implementation. Her horizons 1 to 5 are
				quarters 0 to 4. Multipliers are cumulative. Responses of $R$ are
				annualized, in percentage points. The policy shock in the last two rows
				is scaled so that its impact on $R$ equals the impact of a unit
				spending shock. Her figure places the ex ante decomposition with four
				shocks in the aligned scenario at about the partial-equilibrium multiplier.
		\end{minipage}}
	\end{table}

	We set the feedback coefficient $\psi$ from a statistic of annual US data. We
	estimate local projections of the federal funds rate and of the annual change
	in national defense spending relative to GDP on the monetary policy shock of
	\citet{romerromer2004} over 1971 to 2007. The projections control for two lags
	of the two variables and of the shock. The statistic divides the cumulative
	spending response over years 0 to 4 by the cumulative funds-rate response.
	The statistic equals $-0.21$, so the point estimate implies that spending falls after a
	monetary tightening. We compute the same statistic on annual averages of a
	long simulation of the model, with the spending shock and the contemporaneous
	policy shock as the only shocks, and $\psi=0.456$ reproduces it.

	Each replication simulates $T=200$ quarters after a burn-in of 200 quarters.
	The regressions use 194 of the 200 quarters, because the first two quarters
	provide the lagged controls and the last four provide the leads up to $h=4$.
	The spending shock has standard deviation $1$. The contemporaneous policy
	shock and the news shocks about the policy shock one, two, and three quarters
	ahead have standard deviation $0.25$, and the regional spending shocks have
	standard deviation $0.5$. The other news shocks serve only to compute
	population values. The researcher observes the spending shock, the
	contemporaneous policy shock, and the three news shocks without error, as in
	the simulations of \citet[Section~3.3]{donaldson2026}.\footnote{Her text does
		not state whether the shocks in her simulations carry measurement error.}

	With two regions, the panel regressions with unit and time effects reduce to
	time-series regressions on the difference between the regions, so exposure
	enters only through the difference $\ell_H-\ell_F=-0.4$. The outcome and the
	treatment are the cumulative variables
	\[
	\mathcal Y_{j,t+h}=\sum_{k=0}^{h}\frac{y_{j,t+k}-y_{j,t-1}}{100+y_{j,t-1}},
	\qquad
	\mathcal G_{j,t+h}=\sum_{k=0}^{h}\frac{g_{j,t+k}-g_{j,t-1}}{100+y_{j,t-1}},
	\]
	where $100+y_{j,t}$ is the output of region $j$ in levels. The regressions of
	the outcome on the treatment control for the lagged growth of the output difference, as in
	\citet{donaldson2026}. The first step of E-HELP-IV controls for two lags of
	$G_t$ and $R_t$, whereas the time-series step of the decomposition has no
	controls and uses Newey--West standard errors with four lags.

	\FloatBarrier
	\subsection{The cumulative fiscal multiplier in the two-region model}
	\label{app:mc-transformed}

	This subsection evaluates E-HELP-IV for the transformed target of
	Section~\ref{sec:generalized-target} in the model of
	Appendix~\ref{app:mc-model}. The target $\mathcal B_h^P$ is the partial-equilibrium
	multiplier, the cumulative fiscal multiplier when the expected path of the real
	rate is held fixed; \citet{donaldson2026} calls it the portable multiplier. Because the model is forward-looking, the
	exercise draws on Sections~\ref{sec:forward-portable-section}
	and~\ref{sec:generalized-target}. The experiment behind the target combines
	the spending shock with the sequence of date-$t$ policy shocks, the
	contemporaneous shock and 40 news shocks, that holds the path of the real
	rate fixed over quarters 0 to 40. The target divides the cumulative response
	of the output difference to the experiment by the cumulative response of the
	spending difference. Under two-way general equilibrium, the spending response
	itself changes when the path of the real rate is held fixed, so the
	denominator of the target differs from the unrestricted spending response.

	We compare three estimators. The benchmark
	is the shift-share design of Section~\ref{sec:generalized-target}, which
	instruments the cumulative treatment by exposure times the one-quarter change
	in aggregate spending relative to lagged output, as in her fiscal
	application. The decomposition is her ex ante decomposition with the four
	date-$t$ policy shocks. E-HELP-IV is the transformed design of
	Section~\ref{sec:generalized-target} with the composite instrument of
	Section~\ref{sec:forward-portable-section}, built from the same four shocks.
	The decomposition and E-HELP-IV thus use the four shocks to offset the path
	of the real rate over quarters 0 to 4 as closely as possible. The standard
	error of E-HELP-IV includes the first-step term for the estimated weights, and
	the decomposition uses her baseline covariance estimator.

	\begin{table}[!ht]\centering
		\caption{Cumulative fiscal multiplier in the two-region model of
			\citet{donaldson2026}}
		\label{tab:mc-dsge}
		\resizebox{\textwidth}{!}{\footnotesize\setlength{\tabcolsep}{3.6pt}%
			\begin{tabular}{cc rr rrrrr rrrrr}
\toprule
 & & \multicolumn{2}{c}{Shift-share} & \multicolumn{5}{c}{Decomposition} & \multicolumn{5}{c}{E-HELP-IV}\\
\cmidrule(lr){3-4}\cmidrule(lr){5-9}\cmidrule(lr){10-14}
$h$ & Target & Bias & RMSE & Bias & RMSE & Range & Coverage & Length & Bias & RMSE & Range & Coverage & Length\\
 & & (1) & (2) & (3) & (4) & (5) & (6) & (7) & (8) & (9) & (10) & (11) & (12)\\
\midrule
\multicolumn{14}{l}{\textit{Panel A. One-way general equilibrium (the model of Donaldson, 2026)}}\\[1pt]
0 & 0.95 & $\phantom{-}0.48$ & 0.51 & $\phantom{-}0.01$ & 0.58 & 1.66 & 0.94 & 1.70 & $-0.01$ & 0.49 & 1.30 & 0.91 & 1.15\\
2 & 0.91 & $\phantom{-}0.51$ & 0.55 & $\phantom{-}0.01$ & 0.65 & 1.83 & 0.95 & 1.97 & $-0.01$ & 0.74 & 1.46 & 0.92 & 1.26\\
4 & 0.87 & $\phantom{-}0.52$ & 0.59 & $\phantom{-}0.03$ & 0.77 & 2.16 & 0.97 & 2.37 & $-0.01$ & 2.91 & 1.75 & 0.94 & 1.44\\
\addlinespace
\multicolumn{14}{l}{\textit{Panel B. Two-way general equilibrium, feedback matched to data}}\\[1pt]
0 & 0.95 & $\phantom{-}0.48$ & 0.51 & $\phantom{-}0.00$ & 0.56 & 1.47 & 0.93 & 1.50 & $\phantom{-}0.00$ & 0.39 & 1.15 & 0.92 & 1.01\\
2 & 0.91 & $\phantom{-}0.58$ & 0.61 & $\phantom{-}0.03$ & 0.71 & 1.62 & 0.96 & 1.82 & $-0.01$ & 0.56 & 1.33 & 0.92 & 1.12\\
4 & 0.87 & $\phantom{-}0.66$ & 0.71 & $\phantom{-}0.07$ & 1.06 & 1.97 & 0.98 & 2.31 & $\phantom{-}0.00$ & 1.29 & 1.60 & 0.93 & 1.30\\
\bottomrule
\end{tabular}
}

		\vspace{0.6em}
		{\footnotesize
			\begin{minipage}{0.9\textwidth}
				\textit{Notes.} Aligned scenario, $T=200$ quarters, two regions, and
				$2{,}000$ replications. The target is $\mathcal B_h^P$, and $h$ is in quarters.
				Panel B sets the feedback coefficient to $\psi=0.456$, which matches the
				data statistic of Appendix~\ref{app:mc-model}.
		\end{minipage}}
	\end{table}

	Table~\ref{tab:mc-dsge} delivers three results. First, the shift-share design,
	which corrects neither for the HEDGE bias nor for the omitted-history bias of
	Proposition~\ref{lemma:raw-fails}, is biased by $0.48$ to $0.66$ in both panels.
	Second, under one-way general equilibrium (Panel A), the decomposition method
	and E-HELP-IV both have a median bias of at most $0.03$ in absolute value,
	which replicates the finding of \citet{donaldson2026} that her ex ante
	decomposition with four shocks recovers the partial-equilibrium multiplier in
	the aligned scenario. E-HELP-IV is more precise, with a range 19--22 percent narrower and intervals 32--39 percent
	shorter.

	Third, under two-way general equilibrium (Panel B), E-HELP-IV keeps a median
	bias of at most $0.01$ in absolute value, and its intervals cover the target in
	92--93 percent of replications. The bias of the decomposition method grows
	with the horizon and reaches $0.07$ at $h=4$, 8 percent of the target. Its
	intervals nevertheless cover the target in 93--98 percent of replications,
	because they are 1.5 to 1.8 times as long as those of E-HELP-IV; coverage does
	not reveal the bias.

	In both panels, single replications dominate the RMSE at long horizons. In
	Panel A, a single replication accounts for 46 and 90 percent of the sum of
	squared errors of E-HELP-IV at $h=2$ and $h=4$, so its RMSE exceeds that of
	the decomposition method at these horizons although its range is narrower.
	In Panel B, the largest error at $h=4$ accounts for 61 percent of the sum of
	squared errors of E-HELP-IV and for 55 percent of that of the decomposition
	method, while the range of E-HELP-IV is 19 percent narrower.

	The population limits, which we compute from the impulse responses of the
	model, show that the bias of the decomposition method in Panel B is not a
	finite-sample effect. The limit of E-HELP-IV stays within $0.002$ of the
	target, whereas the limit of the decomposition method exceeds the target by 1 to 8 percent at $h=1$ to $4$. Because the fall in
	spending that the higher real rate causes accumulates, the excess keeps
	growing beyond $h=4$ and reaches 19 percent at $h=8$, two years after the
	shock.

	\FloatBarrier
	\subsection{Multiple aggregate confounders}
	\label{app:mc-multi}

	To support the comparison of the three HELP-IV designs in
	Section~\ref{sec:mc-multiple-results}, Table~\ref{tab:mc-multiapp} reports the
	median bias and the range of each design.\footnote{Because the outcome
		equation of Section~\ref{sec:mc-multiple-dgp} contains no lags of the
		aggregates, the panel regressions of Section~\ref{sec:mc-multiple} control
		only for the lagged outcome.} In Panel A, the range of
	multiple-confounder HELP-IV does not exceed that of composite HELP-IV up to
	$h=6$ but is larger at $h=8$, $2.74$ against $1.93$. In Panel B, by contrast,
	the range of composite HELP-IV is $1.4$ to $1.5$ times that of
	multiple-confounder HELP-IV at every horizon. In the same panel, the median
	length of the intervals of multiple-confounder HELP-IV falls from 96 percent
	of the range of its estimates at $h=2$ to 79 percent at $h=8$, whereas the
	ratio for composite HELP-IV stays between 93 and 102 percent. The standard
	error of multiple-confounder HELP-IV thus understates the dispersion of its
	estimates at long horizons.

	\begin{table}[!ht]\centering
		\caption{Median bias and range with multiple aggregate confounders}
		\label{tab:mc-multiapp}
		{\small\setlength{\tabcolsep}{7pt}
			\begin{tabular}{c rrr rrr}
\toprule
 & \multicolumn{3}{c}{Median bias} & \multicolumn{3}{c}{Range}\\
\cmidrule(lr){2-4}\cmidrule(lr){5-7}
$h$ & Multiple & Composite & Targeting rule & Multiple & Composite & Targeting rule\\
\midrule
\multicolumn{7}{l}{\textit{Panel A. Large measurement error, no within-period response}}\\[1pt]
2 & $-0.05$ & $-0.05$ & $-0.02$ & 0.63 & 0.74 & 0.51\\
4 & $-0.09$ & $-0.07$ & $-0.05$ & 1.00 & 1.22 & 0.71\\
6 & $-0.10$ & $-0.08$ & $-0.06$ & 1.50 & 1.53 & 0.84\\
8 & $-0.08$ & $-0.07$ & $-0.07$ & 2.74 & 1.93 & 0.93\\
\addlinespace
\multicolumn{7}{l}{\textit{Panel B. Small measurement error, within-period response}}\\[1pt]
2 & $-0.06$ & $-0.04$ & $-0.02$ & 0.56 & 0.80 & 0.53\\
4 & $-0.09$ & $-0.08$ & $-0.05$ & 0.78 & 1.12 & 0.67\\
6 & $-0.10$ & $-0.09$ & $-0.07$ & 0.92 & 1.38 & 0.77\\
8 & $-0.09$ & $-0.07$ & $-0.06$ & 1.12 & 1.61 & 0.80\\
\bottomrule
\end{tabular}
}

		\vspace{0.6em}
		{\footnotesize
			\begin{minipage}{0.7\textwidth}
				\textit{Notes.} Multiple, Composite, and Targeting rule denote
				multiple-confounder, composite, and targeting-rule HELP-IV. Panels A
				and B follow the panels of Table~\ref{tab:mc-multi}.
		\end{minipage}}
	\end{table}

	In Panel A, the extreme estimates of multiple-confounder HELP-IV come from
	the replications with a weak first stage. Its first-stage $F$ statistic is
	below one in $0.8$ and $4.2$ percent of replications at $h=6$ and $8$. Without
	these replications, its median bias changes by at most $0.02$, and its RMSE
	falls from $1.38$ to $0.49$ at $h=6$ and from above $900$ to $0.78$ at $h=8$.
	In Panel B, by contrast, the statistic is never below one.

	\begin{table}[!ht]\centering
		\caption{Multiple aggregate confounders under the other two combinations of
			measurement error and within-period response}
		\label{tab:mc-multibase}
		\resizebox{\textwidth}{!}{\footnotesize\setlength{\tabcolsep}{3.1pt}%
			\begin{tabular}{cc rr rrrr rrr rrr}
\toprule
 & & \multicolumn{2}{c}{TWFE-IV} & \multicolumn{4}{c}{Multiple-confounder} & \multicolumn{3}{c}{Composite} & \multicolumn{3}{c}{Targeting rule}\\
\cmidrule(lr){3-4}\cmidrule(lr){5-8}\cmidrule(lr){9-11}\cmidrule(lr){12-14}
$h$ & Target & Bias & RMSE & Bias & RMSE & Coverage & Length & RMSE & Coverage & Length & RMSE & Coverage & Length\\
 & & (1) & (2) & (3) & (4) & (5) & (6) & (7) & (8) & (9) & (10) & (11) & (12)\\
\midrule
\multicolumn{14}{l}{\textit{Panel A. Small measurement error, no within-period response}}\\[1pt]
2 & 1.29 & $\phantom{-}1.03$ & 1.11 & $-0.05$ & 0.17 & 0.85 & 0.52 & 0.20 & 0.86 & 0.62 & 0.15 & 0.88 & 0.49\\
4 & 0.99 & $\phantom{-}1.01$ & 1.16 & $-0.09$ & 0.26 & 0.82 & 0.72 & 0.31 & 0.86 & 0.90 & 0.21 & 0.85 & 0.65\\
6 & 0.67 & $\phantom{-}0.77$ & 1.00 & $-0.11$ & 0.32 & 0.79 & 0.83 & 0.38 & 0.86 & 1.12 & 0.25 & 0.84 & 0.74\\
8 & 0.44 & $\phantom{-}0.52$ & 0.84 & $-0.09$ & 0.38 & 0.80 & 0.94 & 0.43 & 0.88 & 1.29 & 0.27 & 0.84 & 0.78\\
\addlinespace
\multicolumn{14}{l}{\textit{Panel B. Large measurement error, within-period response}}\\[1pt]
2 & 1.21 & $\phantom{-}1.12$ & 1.21 & $-0.06$ & 0.20 & 0.88 & 0.64 & 0.28 & 0.91 & 0.91 & 0.17 & 0.89 & 0.55\\
4 & 0.89 & $\phantom{-}1.12$ & 1.31 & $-0.08$ & 0.30 & 0.89 & 0.95 & 0.45 & 0.93 & 1.54 & 0.22 & 0.86 & 0.69\\
6 & 0.60 & $\phantom{-}0.86$ & 1.18 & $-0.10$ & 0.70 & 0.92 & 1.34 & 0.59 & 0.95 & 2.09 & 0.25 & 0.86 & 0.78\\
8 & 0.39 & $\phantom{-}0.58$ & 1.05 & $-0.08$ & 90.35 & 0.96 & 2.11 & 0.75 & 0.96 & 2.61 & 0.27 & 0.87 & 0.84\\
\bottomrule
\end{tabular}
}

		\vspace{0.6em}
		{\footnotesize
			\begin{minipage}{0.9\textwidth}
				\textit{Notes.} Same statistics as
				Table~\ref{tab:mc-multi}. Panel A sets
				$(\sigma_\eta,\varpi)=(0.5,0)$ and Panel B $(1,0.5)$.
		\end{minipage}}
	\end{table}

	Table~\ref{tab:mc-multibase} reports the other two combinations of
	measurement error and within-period response, which
	Section~\ref{sec:mc-multiple-results} summarizes. With a small measurement
	error and no within-period response, the first-stage $F$ statistic of
	multiple-confounder HELP-IV is never below one, and the RMSE of composite
	HELP-IV is $1.1$ to $1.2$ times that of multiple-confounder HELP-IV. With a
	large measurement error and a within-period response, the statistic is below
	one in $0.7$ and $5.8$ percent of replications at $h=6$ and $8$. The errors of
	multiple-confounder HELP-IV then exceed $10$ in absolute value in $18$
	replications at $h=8$, whereas the largest error of composite HELP-IV is
	only $6.6$.

	The standard error of targeting-rule HELP-IV adds the cross-sectional part of
	Section~\ref{sec:help-inference}, because the orthogonality of
	Assumption~\ref{ass:targeting} holds in a sample of $N$ units only up to an
	error of order $N^{-1/2}$. With $N=T=200$, this part raises the coverage of
	targeting-rule HELP-IV by at most $0.8$ percentage points at the reported
	horizons.

	\FloatBarrier\clearpage
	\section{Inference for the empirical applications}
	\label{app:se-donaldson}

	This appendix describes the standard errors in
	Tables~\ref{tab:app-don} and~\ref{tab:app-cns-main}. The HELP-IV instruments
	are generated from an estimated aggregate model, so inference accounts jointly
	for the estimation of that model and for the panel regression. The HELP-IV rows
	use the covariance formulas for SVAR shocks in
	Propositions~\ref{thm:help-S} and~\ref{thm:composite-S}
	(Appendix~\ref{app:svar-inference}), with the transformed outcome
	and treatment in the fiscal application. The decomposition row uses the covariance
	estimator described in Appendix~\ref{app:se-don-spec2}, and the shift-share IV
	row uses a time-block bootstrap (Appendix~\ref{app:boot}). The targeting-rule
	entries contain only the time-series component of uncertainty
	(Appendix~\ref{app:se-fiscal}).

	\subsection{Analytic standard errors}
	For a common effective sample, the analytic variance is the coefficient
	element of the sample sandwich,
	\[
	\widehat{\Var}(\widehat\beta_h)
	=\frac1T\,L_h\widehat{\mathbf J}_h^{-1}\widehat{\boldsymbol\Omega}_h
	\widehat{\mathbf J}_h^{-1\prime}L_h',
	\]
	where $\widehat{\mathbf J}_h$ is the sample Jacobian
	\eqref{eq:svar-help-jacobian} or \eqref{eq:svar-composite-jacobian} and
	$\widehat{\boldsymbol\Omega}_h$ is the long-run covariance of the stacked
	date-level moments, the panel IV moment and the SVAR estimating equations
	(Appendix~\ref{app:svar-inference}). For full-path estimation the treatment
	selector picks the coefficient on the treatment from the full parameter
	vector. The derivative holds the panel coefficient fixed, differentiates the
	complete concentrated moment, and is evaluated analytically
	(Appendix~\ref{app:svar}).

	\subsubsection{Fiscal multipliers}
	\label{app:se-fiscal}
	Backward-looking HELP-IV uses the bivariate proxy-SVAR described in
	Section~\ref{sec:app-donaldson}. Its first-step moments cover the VAR
	coefficients, residual covariance, proxy-identification moments, and shock
	normalizations, as in \eqref{eq:main-var-primitive} and Lemma~\ref{lem:svar}.
	For targeting-rule HELP-IV, interactions between targeting characteristics
	and estimated shocks are generated controls. Their derivatives enter
	through \eqref{eq:D-total} and \eqref{eq:target-projector-derivative}.
	For E-HELP-IV, the four-variable SVAR also determines the news
	offset. Its derivative includes the residual-path term in
	\eqref{eq:explicit-news-W}, because the path is not exactly spanned. We use
	Bartlett weights with bandwidth $\lfloor4(T/100)^{2/9}\rfloor=3$.

	The targeting-rule standard errors in Tables~\ref{tab:app-don} and
	\ref{tab:app-linboot} contain the time-series component only. They are
	complete standard errors under exact sample balance or under imbalance that is
	negligible at the time-series scale. Balance in the population alone adds the
	term in \eqref{eq:target-variance}, and with 46 states and 39 annual
	observations, the approximation $T/N\to0$ does not justify omitting it. The
	sample analogue \eqref{eq:target-feasible-cross-cov} requires the omitted
	exposures and their path loadings; when these are not identified, the formula
	cannot be evaluated. The entries therefore do not establish coverage under
	population balance alone, which would require an estimable augmented model or
	a justified joint sampling procedure. The controlled first stages of the
	targeting-rule specifications are strong, with first-stage $F$ statistics above
	$48$ at every reported horizon.

	\subsubsection{Stock-wealth responses}
	The five-variable SVAR is common to all six outcome specifications. County
	observations are population weighted; county effects, state-by-quarter
	effects, lagged interactions, and the Bartik control are projected out
	jointly. The relevant $\Sigma_h^C$ and $D_h$ are those of this weighted,
	residualized design. Path neutrality of the composite in the aggregate data
	implies orthogonality of these panel moments only if the residualized county
	exposures are independent of the aggregate process; we maintain this
	condition, which does not follow from aggregate covariances alone.

	For the dated composite, decompose the complete derivative as
	$D_h=\sum_{s=0}^hD_{h,s}$, including both shock and weight derivatives in
	each component, and use the aligned stacked moment and Jacobian of
	\eqref{eq:aligned-internal}. They give the same sandwich covariance as the
	unaligned version, up to fixed endpoint terms, but their finite-bandwidth
	estimates can differ, in either direction. We use bandwidth $h+1=8$ quarters
	at $h=7$. A difference between two outcome coefficients stacks the moments
	of both outcome regressions and uses the difference of their coefficient
	rows of the inverse Jacobian.

	If the panel and SVAR use different date sets $\mathcal I_P$ and
	$\mathcal I_A$, with sizes $T_P$ and $T_A$, the panel moment is averaged over
	$\mathcal I_P$ and the SVAR estimating equations over $\mathcal I_A$, and the
	sandwich must retain the covariance of the two averages over their overlap. On a common calendar of length
	$T_*$, the two summands receive weights
	$(T_*/T_P){\bf1}\{t\in\mathcal I_P\}$ and
	$(T_*/T_A){\bf1}\{t\in\mathcal I_A\}$ before their joint covariance is
	formed, with normalization $T_*^{-1}$. The common-sample formulas apply
	directly when the date sets agree up to fixed endpoints; other overlap
	patterns require their corresponding joint limit.

	The \emph{CNS benchmark} and \emph{No confounder controls} rows retain the
	county-and-quarter clustered standard errors of the original specification.
	They do not estimate structural shocks or offset weights.

	\subsubsection{Direction of the first-step correction}
	The cross-covariances in \eqref{eq:help-variance-split} and
	\eqref{eq:composite-variance-split} can be negative.
	For backward-looking HELP-IV in the fiscal application, treating the
	estimated shocks as known gives standard errors of $0.35$, $0.50$, $0.52$,
	and $0.67$, against $0.38$, $0.40$, $0.36$, and $0.53$ after the joint
	correction. For nontradable payroll, the corresponding values are $1.09$ and
	$0.74$. Accounting for the first step lowers the standard error in four of
	these five cases.

	\subsection{The decomposition method}
	\label{app:se-don-spec2}

	We implement the procedure of \citet{donaldson2026}. Her aggregate local
	projection of the federal funds rate controls for lagged realizations of the
	funds rate and of both shocks, log real per capita output, the average federal
	tax rate, CPI inflation, and log real per capita defense spending. Her
	correction cumulates the effects of the offsetting monetary policy shocks from
	the impact date onward.

	The decomposition row uses the baseline covariance estimator, specification~(ii)
	of \citet[Appendix~A.7]{donaldson2026}. At each horizon $k=0,\ldots,h$, her
	estimator combines two sets of moments. The panel moments come from the
	regressions of the outcome and of spending on $\sig G_t$ and $\sig RR_t$.
	The time-series moments come from the local projection of the federal funds
	rate at lead $k$. The baseline estimator treats the moments of different
	horizons as uncorrelated and sets to zero the covariance between the panel and
	the time-series moments. Within a horizon, the panel moments are clustered by
	year, and the time-series moments use Newey--West weights with
	bandwidth $3$. The standard error of the multiplier follows by the delta
	method. With bandwidth $4$ or $h+1$ instead, the standard errors change by at
	most $0.01$. The block bootstrap of Appendix~\ref{app:boot} gives standard
	errors that are 18 to 46 percent larger, because the baseline estimator omits
	the covariances across horizons (Appendix~\ref{app:mc-don}).

	\subsection{Block bootstrap}
	\label{app:boot}

	A circular block draw reweights every panel observation dated in the
	corresponding period. Let $c_t$ denote the number of sampled blocks containing
	date $t$. All panel observations at date $t$ receive weight $c_t$, and dates
	with $c_t=0$ are omitted. We use circular blocks of four annual observations,
	which cover the longest reported response window, and $1{,}000$
	replications.

	Within each replication, we:

	\begin{enumerate}[label=(\roman*),leftmargin=2.2em,itemsep=0.3ex]
		\item re-estimate the panel regressions using the block weights; and
		\item recompute the ratio that defines the shift-share IV multiplier.
	\end{enumerate}

	The bootstrap standard error is the interquartile range of the replication
	estimates divided by $1.349$, which is robust to draws in which the
	estimated first-stage denominator is close to zero.

	The shift-share IV row contains no generated shock. A Bartlett HAC estimator
	based on the cross-sectionally averaged IV moment gives standard errors slightly
	smaller than the block bootstrap. Three- or six-year blocks instead of
	four-year blocks change the bootstrap standard errors little.

	For the HELP-IV rows, Table~\ref{tab:app-linboot} compares the analytic
	standard errors with block-bootstrap standard errors. The block draw is applied
	to the stacked date-level moments of the analytic sandwich, mapped through the
	inverse sample Jacobian, so the panel moment and the first-step moments of the same
	dates are drawn together. The block
	length is the bandwidth plus one, four years in the fiscal application and nine
	quarters in the stock-wealth application, which matches the triangular
	Bartlett weights asymptotically, up to circular and endpoint conventions
	\citep{kunsch1989}. The standard error is the standard deviation of $10{,}000$
	draws. The two sets of standard errors differ by less than four percent in the
	fiscal application and by less than three percent in the stock-wealth
	application. Because both use the same sample Jacobian and stacked moments,
	the comparison checks the long-run variance estimator.

	\begin{table}[!ht]\centering
		\caption{Analytic and block-bootstrap standard errors}
		\label{tab:app-linboot}
		\begin{tabular}{lcccccc}
			\toprule
			\multicolumn{7}{l}{\emph{Panel A. Fiscal-multiplier application}}\\
			& Year 1 & Year 2 & Year 3 & Year 4 & & \\
			\midrule
			HELP-IV, analytic & 0.38 & 0.40 & 0.36 & 0.53 & & \\
			\quad block bootstrap & 0.39 & 0.39 & 0.35 & 0.51 & & \\
			\addlinespace
			E-HELP-IV, analytic & 0.41 & 0.59 & 0.75 & 1.03 & & \\
			\quad block bootstrap & 0.42 & 0.61 & 0.77 & 1.03 & & \\
			\addlinespace
			Targeting-rule HELP-IV, analytic & 0.31 & 0.26 & 0.24 & 0.32 & & \\
			\quad block bootstrap & 0.31 & 0.26 & 0.24 & 0.32 & & \\
			\midrule
			\multicolumn{7}{l}{\emph{Panel B. Stock-wealth application, $h=7$}}\\
			& (1) & (2) & (3) & (4) & (5) & (6)\\
			\midrule
			Composite HELP-IV, analytic & 0.43 & 0.54 & 0.63 & 0.74 & 0.69 & 0.93 \\
			\quad block bootstrap & 0.43 & 0.54 & 0.63 & 0.75 & 0.70 & 0.94 \\
			\bottomrule
		\end{tabular}

		\vspace{0.6em}
		{\footnotesize
			\begin{minipage}{0.92\textwidth}
				\textit{Notes.} Standard errors of the HELP-IV rows of
				Tables~\ref{tab:app-don} and~\ref{tab:app-cns-main}. The analytic
				rows repeat the values reported in those tables. The block
				bootstrap uses circular blocks of four years in Panel A and nine
				quarters in Panel B, with $10{,}000$ draws. Columns (1) to (6) of Panel B are the columns of
				Table~\ref{tab:app-cns-main}.
		\end{minipage}}
	\end{table}

	\subsection{Scope of the reported inference}

	The standard errors and intervals are pointwise in the response horizon. The
	analytic construction accounts for the estimation of the aggregate model and
	of the generated HELP-IV objects, with the targeting qualification above. It
	does not address weak identification, for which Appendix~\ref{app:weak}
	applies; the controlled first stages reported here do not indicate
	weakness.

\end{document}